\documentclass[aps,prl,twocolumn,superscriptaddress,showpacs,floatfix,nobibnotes]{revtex4}
\usepackage{epsfig}
\usepackage{epstopdf}

\usepackage{graphicx}
\usepackage{float}
\usepackage{longtable}
\usepackage{CJK}
\usepackage{color}

\usepackage{mathptmx, courier, pifont}
\usepackage[scaled=0.92]{helvet}
\usepackage[T1]{fontenc}
\usepackage{textcomp}

\begin{document}

\title{Probing Quadruple Deformation in Transitional Nuclei via Angular Momentum Projection}

\author{Xian-Zhi Zhao}
\affiliation{Department of Physics, Liaoning Normal University,
Dalian 116029, P. R. China}

\author{Sheng-Nan Wang}
\affiliation{Department of Physics, Liaoning Normal University,
Dalian 116029, P. R. China}

\author{Yu Zhang }\email{dlzhangyu_physics@163.com}
\affiliation{Department of Physics, Liaoning Normal University,
Dalian 116029, P. R. China}

\date{\today}

\begin{abstract}
Within the interacting boson model (IBM), a geometric analysis of transitional nuclei is carried out through angular momentum projection of the intrinsic coherent state. The results indicate that $K$-mixing effects in the calculations are typically negligible, validating the use of $K$-fixed projection for semiclassical analyses of the spin dependence of quadrupole deformation in the IBM. Further analysis indicates that in a rotating transitional system, quadrupole deformation is stretched with increasing angular momentum, providing a geometrically intuitive perspective on the commonly observed Jacobi-type transitions, as exemplified by the case studies of $^{160}$Gd and $^{162}$Dy. The method is additionally applied to the yrast states of $^{170}$Os to probe quadrupole deformation changes linked to the observed low-spin $B(E2)$ anomaly behavior, demonstrating the capability of IBM-based angular momentum projection in interpreting exotic collective phenomena.
\end{abstract}

\pacs{21.60.Fw, 21.60.-n, 21.10.Re, 27.70.+q}

\maketitle

\begin{center}
\vskip.2cm\textbf{I. Introduction}
\end{center}\vskip.2cm

Nuclear deformation is a central concept for understanding nuclear many-body problems~\cite{Bohrbook}. Theoretically, mean-field methods are widely used to extract deformation properties from both microscopic and phenomenological nuclear models. Among these, the interacting boson model (IBM), combined with the coherent-state (mean-field) approximation~\cite{IachelloBook87}, offers a convenient framework for studying quadrupole deformations (shapes) and associated collective modes. Over the past two decades, the IBM has been widely employed to describe shape phase transitions~\cite{CJ2009,CJC2010} and shape coexistence~\cite{HW2011,Frank2004,Ramos2014,Ramos2015,Nomura2012,Nomura2016}.
Recently, two distinctive collective phenomena, the Jacobi-type transition~\cite{Zhang2017,Zhang2021} and the $B(E2)$ anomaly~\cite{Grahn2016,Saygi2017,Cederwall2018,Goasduff2019,Zhang2017,Zhang2021,Zhang2022,Zhang2024,Pan2024,Wang2020,Zhang2025,Teng2025,Teng2025II,Teng2025III,Fu2025}, have been linked to spin-dependent deformation evolution. Jacobi-type transitions are prevalent in transitional nuclei, particularly in the rare-earth region~\cite{Zhang2017,Zhang2021}, and manifests as characteristic non-monotonic behavior in E-GOS (E-Gamma Over Spin) curves~\cite{Regan2003} along the yrast line. The $B(E2)$ anomaly is characterized by a depressed $B(E2)$ ratio, $B(E2;4_1^+\rightarrow2_1^+)/B(E2;2_1^+\rightarrow0_1^+)<1.0$, along with $E(4_1^+)/E(2_1^+)\geq2.0$, and has been observed in several neutron-deficient nuclei~\cite{Grahn2016,Saygi2017,Cederwall2018,Goasduff2019}. Both phenomena have been successfully described within the IBM using specific model Hamiltonian~\cite{Zhang2017,Zhang2021,Zhang2022,Zhang2024,Pan2024,Wang2020,Zhang2025,Teng2025,Teng2025II,Teng2025III,Fu2025}, and their underlying deformations can be extracted via the mean-field method. However, mean-field solutions typically violate the rotational symmetry inherent in the Hamiltonian, making it difficult to compare with physical spectra that preserve the symmetry. To reveal influences of spins on deformation, it is necessary to restore rotational symmetry, using e.g., angular momentum projection (AMP)~\cite{Ringbook}, which is the central objective of this work.

Prior AMP applications within the IBM have focused primarily on obtaining approximate solutions of the Hamiltonian~\cite{Ewart1979,Kuyucak1987,Kuyucak1987II} or validating ground-state deformations~\cite{Dobes1985,Dobes1990,Otsuka1987}. A systematic investigation into how angular momentum (spin) governs the quadrupole deformation evolution remains lacking.
More broadly, AMP is widely recognized as a powerful technique~\cite{Ringbook} and extensively employed in various sophisticated beyond-mean-field calculations~\cite{HS1995,Sun2016,Sun2021,Otsuka2001,Bender2003,Schmid2004,Niksic2006,Bender2008,Gao2009,Rodriguez2010,Yao2010,Bally2014,Shimizu2021,Wang2022}. Recent shell-model based studies~\cite{Gao2022,Lu2025} indicated that $K$-mixing effects in AMP may be negligible for certain nuclear systems, which offers a promising route to simplify AMP implementations in quantum many-body calculations. However, whether this simplification remains valid for nuclei with different mean-field deformations needs to be examined in other models. Another key objective of this work is therefore to assess $K$-mixing effects within the IBM across different transitional regimes. This is particularly motivated by the IBM's exact solvability: AMP calculations under various conditions can be rigorously validated against exact solutions~\cite{IachelloBook87}, in contrast to other microscopic approaches, which typically require huge model spaces.

The article is organized as follows. Section II introduces the model Hamiltonian and outlines the AMP formalism.
Section III presents a systematic analysis of $K$-mixing in the AMP calculations, together with a discussion of the spin-dependent quadrupole deformation in transitional systems.
Section IV provides two illustrative examples demonstrating how AMP uncovers the underlying quadrupole geometry associated with exotic spectral phenomena. A summary is given in Section V.

\begin{center}
\vskip.2cm\textbf{II. Theoretical Framework}
\end{center}\vskip.2cm

In the IBM~\cite{IachelloBook87}, the building blocks are two types of boson (operators): the monopole $s$ boson with $J^\pi=0^+$ and the
quadrupole $d$ boson with $J^\pi=2^+$. All physical operators are constructed utilizing the creation and annihilation operators of these bosons.
One key advantage of the IBM is its ability to robustly describe collective modes associated with diverse quadrupole deformations,
thereby providing a convenient framework for assessing standard many-body techniques across varying structural regimes.

\begin{center}
\vskip.2cm\textbf{A. Model Hamiltonian and Its Mean-Field Geometry}
\end{center}\vskip.2cm

\begin{figure}
\begin{center}
\includegraphics[scale=0.28]{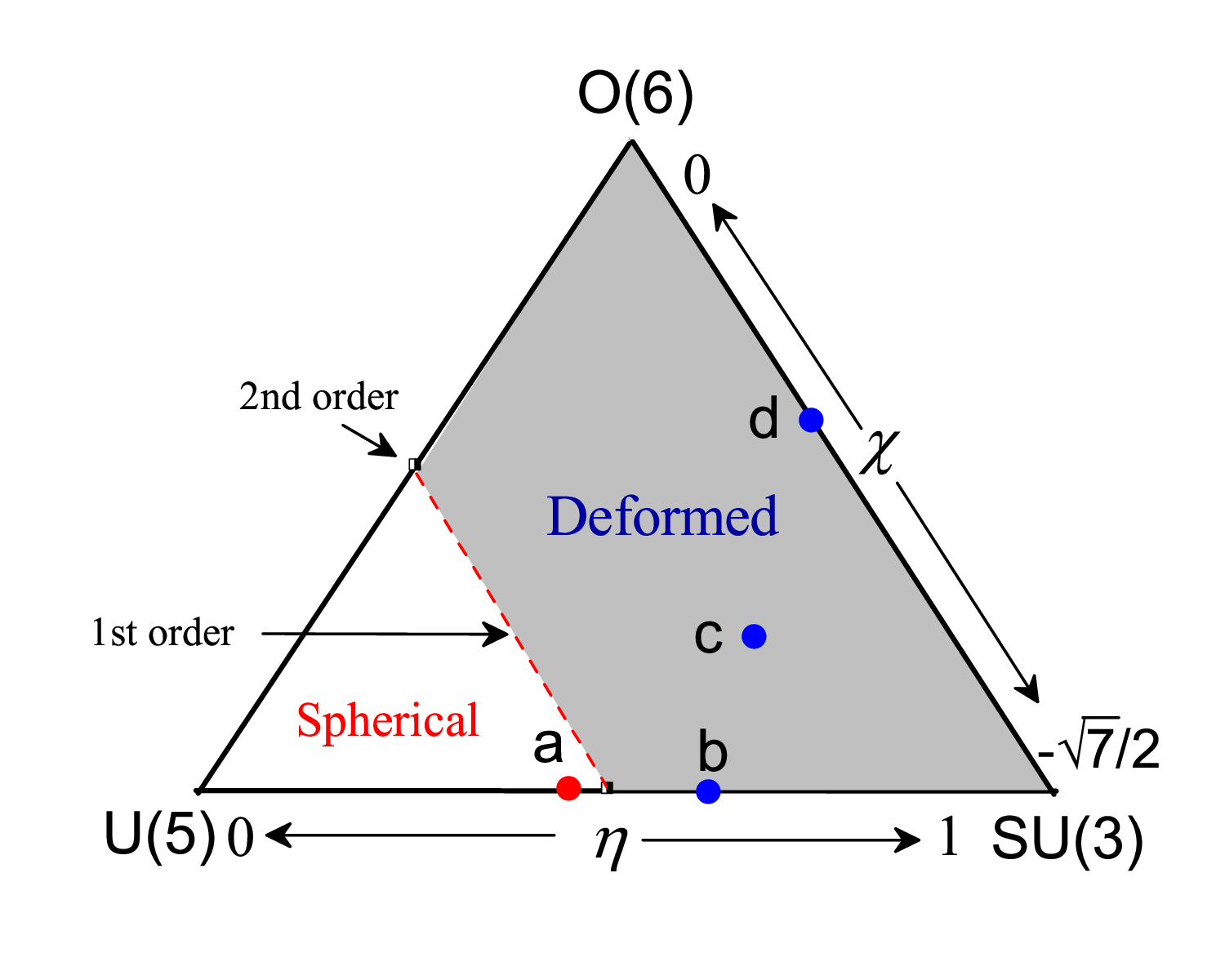}
\caption{(Color online) The schematic shape phase diagram of the IBM is shown, where "a", "b", "c"
and "d" denote the representative parameter points selected to illustrate different transitional
situations. First-order phase transitions are indicated by the dashed line, which partitions the triangle phase
diagram into the spherical and deformed regions. \label{F1}}
\end{center}
\end{figure}

To comprehensively account for all typical quadrupole deformations, the model Hamiltonian is designed as follows~\cite{Teng2025}:
\begin{eqnarray}\label{H}
\hat{H}=a_1\hat{n}_d+a_2\hat{Q}^\chi\cdot\hat{Q}^\chi+a_3\hat{V}_3+b_1(\hat{J}\times \hat{Q}^\chi\times \hat{J})^{(0)}+b_2\hat{J}^2\,
\end{eqnarray}
with $a_i(i=1,2,3)$, $b_k(k=1,2)$ and $\chi$ representing the adjustable parameters, and
\begin{eqnarray}
&&\hat{n}_d=d^\dag\cdot\tilde{d},\\
&&\hat{J}_u=\sqrt{10}(d^\dag\times\tilde{d})_u^{(1)},\\
&&\hat{Q}_u^\chi=(d^\dag\times\tilde{s}+s^\dag\times\tilde{d})_u^{(2)}+\chi(d^\dag\times\tilde{d})_u^{(2)},\\ \label{V3}
&&\hat{V}_3=(d^\dag\times d^\dag\times d^\dag)^{(3)}\cdot(\tilde{d}\times\tilde{d}\times\tilde{d})^{(3)}\, ,
\end{eqnarray}
where the index $u$ denotes generally the $z$-component of a tensor operator. If we set $a_3=b_1=b_2=0$, the Hamiltonian given in (\ref{H}) reduces to the consistent-$Q$ form~\cite{Warner1983}, which
was widely adopted to describe the shape phase diagram of the IBM (see Fig.~\ref{F1}). Within the IBM, typical quadrupole deformations are realized as distinct dynamical symmetry limits: U(5) for the spherical, SU(3) for the axially-deformed, and O(6) for the $\gamma$-soft. However, triaxial deformations at the mean-field level and the associated triaxial rotor modes~\cite{Zhang2024,Zhang2025} cannot be accommodated within the consistent-$Q$ form; they emerge only upon inclusion of high-order interactions, such as the cubic $\hat{V}_3$ term~\cite{VC1981} defined in (\ref{V3}). Furthermore, the final two terms in (\ref{H}) are explicitly introduced~\cite{Zhang2025,Teng2025,Teng2025II} to account for the $B(E2)$ anomaly phenomenon, which will serve as a representative case for illustrating the AMP analysis in this work.  Collectively, the Hamiltonian form in (\ref{H}) is sufficient to address all characteristic quadrupole modes on an equal footing.
To describe $B(E2)$ transitions and spectroscopic quadrupole moments, the $E2$ operator is adopted herein
\begin{eqnarray}
\hat{T}(E2)=e\hat{Q}^\chi\, ,
\end{eqnarray}
where $e$ denotes the effective charge, and $\hat{Q}^\chi$ is taken to be identical to the quadrupole operator appearing in the Hamiltonian (\ref{H}). The $B(E2)$ transitions are then evaluated via
\begin{eqnarray}
B(E2;J_i\rightarrow J_f)=\frac{|\langle\alpha_f J_f \parallel\hat{T}(E2)\parallel\alpha_i J_i\rangle|^2}{2J_i+1}\,
\end{eqnarray}
with $\alpha$ representing all quantum numbers other than $J$. Similarly, the electric quadrupole moments are evaluated via
\begin{eqnarray}\label{Q}
Q(J)=\sqrt{\frac{16\pi}{5}}\langle\alpha J M|\hat{T}(E2)|\alpha J M\rangle_{M=J}\, .
\end{eqnarray}

To identify the mean-field geometry, one must determine the classical limit of the Hamiltonian. This is accomplished
by employing the coherent (intrinsic) state defined as~\cite{IachelloBook87}
\begin{eqnarray}\label{coherent}
|\beta, \gamma, N\rangle=G[s^\dag + \beta \mathrm{cos} \gamma~
d_0^\dag\ + \frac{1}{\sqrt{2}} \beta \mathrm{sin} \gamma (d_2^\dag +
d_{ - 2}^\dag)]^N |0\rangle\, ,
\end{eqnarray}
where $\beta$ and $\gamma$ denote the quadrupole deformation parameters, and the normalization factor is given by $G=1/\sqrt{N!(1+\beta^2)^N}$, with $N$ representing the total boson number. Note that the deformation parameters in the IBM are related to those introduced by Bohr and Mottelson in the geometric model via $\gamma_{\mathrm{BM}}=\gamma$ and $\beta_{\mathrm{BM}}=t\beta$, with $t$ being a dimensionless scaling factor~\cite{IachelloBook87}. It has been established~\cite{IachelloBook87} that the intrinsic state defined in (\ref{coherent}) can be used to derive a wide range of mean-field properties of nuclei within the IBM framework. Within this state, the full-order classical potential function corresponding to the Hamiltonian (\ref{H}) is derived as~\cite{Teng2025II}
\begin{eqnarray}\label{V}
V(\beta,\gamma,N)&=&\langle\beta, \gamma,
N|\hat{H}|\beta, \gamma,N\rangle\\\nonumber
&=&a_1\frac{N\beta^2}{(1+\beta^2)}+a_2\Big[N\frac{(\beta^2+\chi^2\beta^2+5)}{1+\beta^2}\\ \nonumber
&~&+N(N-1)\frac{(4\beta^2-4\sqrt{\frac{2}{7}}\chi\beta^3\mathrm{cos}3\gamma+\frac{2}{7}\chi^2\beta^4)}{(1+\beta^2)^2}\Big]\\
\nonumber
&~&+a_3N(N-1)(N-2)\frac{(\beta^6\mathrm{cos}^23\gamma-\beta^6)}{7(1+\beta^2)^3}\\ \nonumber
&~&+b_1\Big[N(N-1)\frac{(2\sqrt{105}\chi\beta^4-14\sqrt{30}\beta^3\mathrm{cos}3\gamma)}{35(1+\beta^2)^2}\\ \nonumber
&~&+N\frac{\sqrt{105}\chi\beta^2}{5(1+\beta^2)}\Big]+b_2N\frac{6\beta^2}{1+\beta^2}\, .
\end{eqnarray}
The mean-field geometry of the Hamiltonian is subsequently identified by minimizing
the classical potential function with respect to the deformation parameters ($\beta,~\gamma$), yielding the ground-state energy $E_g=V(\beta_\mathrm{e},\gamma_\mathrm{e})$, where $\beta_\mathrm{e}$ and $\gamma_\mathrm{e}$ denote the equilibrium values that characterize the intrinsic shape of the system.

For clarity, we firstly set the Hamiltonian parameters to $a_3=b_1=b_2=0$, while choosing $a_1=\varepsilon(1-\eta)$ and $a_2=-\varepsilon\frac{\eta}{4N}$, so as to explore distinct dynamical symmetries and shape-transitional regimes within the IBM. In this parametrization, the Hamiltonian reduces to the consistent-$Q$ form,
\begin{eqnarray}\label{CQ}
\hat{H}_{\mathrm{CQ}}=\varepsilon\Big[(1-\eta)\hat{n}_d-\frac{\eta}{4N}\hat{Q}^\chi\cdot\hat{Q}^\chi\Big]\, ,
\end{eqnarray}
where $\varepsilon$ denotes a scaling parameter and, for convenience, is set to unity in the subsequent discussions.
The resulting phase diagram in the $\eta\times\chi$ parameter space is displayed in Fig.~\ref{F1}, where the three symmetry limits of the IBM reside at the vertices of the triangle, parameterized as $\eta=0$ for U(5), $(\eta,~\chi)=(1,~0)$ for O(6), and $(\eta,~\chi)=(1,~-\sqrt{7}/2)$ for SU(3). The associated mean-field geometry (shape) in each limit can be inferred from the classical potential function (\ref{V}) evaluated under the corresponding parameter sets: a $\gamma$-independent potential with $\beta_\mathrm{e}=0$ characterizes U(5); a $\gamma$-independent potential with $\beta_\mathrm{e}>0$ corresponds to O(6); and an axially symmetric potential with $\beta_\mathrm{e}>0$ and $\gamma_\mathrm{e}=0$ describes SU(3). It has been rigorously established that the first-order shape phase transitions between spherical and deformed shapes occur within the triangle phase diagram~\cite{Iachello2004}, and the critical line is analytically described (in the large-$N$ limit) by
\begin{equation}
\eta_\mathrm{c}=\frac{14}{28+\chi^2}\,
\end{equation} with $\chi\in[-\sqrt{7}/2,0)$. This critical line cuts the triangle into two distinct regions, the spherical and the deformed, as evident from Fig.~\ref{F1}. Additionally, a second-order transition occurs along the U(5)-O(6) leg ($\chi=0$) at the critical point $\eta_\mathrm{c}=0.5$, corresponding to the spherical to $\gamma$-soft shape evolution. In contrast, the SU(3)-O(6) boundary ($\eta=1$), which connects the axially symmetric prolate and $\gamma$-unstable limits, is found to represent a crossover in the large-$N$ limit. Consequently, the triangle phase diagram provides a comprehensive representation of all characteristic quadrupole shapes and their transitional behaviors.

To illustrate representative cases for subsequent numerical analysis,
in addition to the symmetry limits, we select several parameter points within the triangle phase diagram: $(\eta,~\chi)=(0.45,~-1.323),~(0.65,~-1.323),~(0.7,~-0.8),~(1.0,~-0.5)$. These points are labeled "a", "b", "c" and "d", respectively, in Fig.~\ref{F1}. They correspond to: (a) a weakly deformed (nearly spherical) configuration; (b) a system lying on the U(5)-SU(3) leg; a deformed configuration located inside the triangle; and (d) a system lying on the SU(3)-O(6) leg.

\begin{center}
\vskip.2cm\textbf{B. Angular Momentum Projection}
\end{center}\vskip.2cm

As discussed above, the mean-field potential given in (\ref{V}) is derived from the coherent state
defined in the intrinsic frame, which explicitly breaks the rotational
symmetry~\cite{IachelloBook87}. To construct an IBM potential
with well-defined angular momentum number $J$, the
AMP technique is employed here to restore the SO(3) rotational symmetry in the coherent
state. The AMP operator is defined as
\begin{equation}\label{projection}
\hat{P}_{MK}^J=\frac{(2J+1)}{8\pi^2}\int~D_{MK}^{J*}(\Omega)\hat{R}(\Omega)d\Omega\, ,
\end{equation}
where $\hat{R}(\Omega)=\mathrm{e}^{-i\theta_1 \hat{J}_z}\mathrm{e}^{-i\theta_2 \hat{J}_y}\mathrm{e}^{-i\theta_3 \hat{J}_z}$ denotes the rotational operator in terms of the Euler angles $\Omega=\{\theta_1,\theta_2,\theta_3\}$, and $D_{MK}^{J}(\Omega)=\mathrm{e}^{-iM\theta_1}d_{MK}^J(\theta_2)\mathrm{e}^{-iK\theta_3}$ is the Wigner function. Here, $J$ represents the total angular momentum, with
$M$ and $K$ denoting its projection onto the laboratory-fixed $z$-axis and the intrinsic $z$-axis, respectively. The projection operator defined in (\ref{projection}) satisfies~\cite{HS1995}
\begin{eqnarray}
\hat{P}_{MK}^{J\dag}=\hat{P}_{KM}^J,~~\hat{P}_{K^\prime M^\prime }^{J^{\prime}}\hat{P}_{MK}^J=\delta_{JJ^\prime}\delta_{MM^\prime}\hat{P}_{K^\prime K}^J\, .
\end{eqnarray}
Using the AMP operator, the projected energy function for given $J$ and $K$ is constructed as
\begin{eqnarray}\label{AMPI}
&&V(\beta,\gamma,N)_{K}^J=\frac{\langle\beta,\gamma,N|\hat{H}\hat{P}_{KK}^J|\beta,\gamma,N\rangle}{\langle\beta,\gamma,N|\hat{P}_{KK}^J|\beta,\gamma,N\rangle}\\ \nonumber
&&=\frac{\int\int\int~d\theta_1d\theta_3d\theta_2\mathrm{sin}\theta_2 D_{KK}^{J*}(\Omega)
\langle\beta, \gamma, N|\hat{H}R(\Omega)|\beta,
\gamma, N\rangle}{\int\int\int~d\theta_1d\theta_3~d\theta_2\mathrm{sin}\theta_2
D_{KK}^{J*}(\Omega)\langle\beta, \gamma, N|R(\Omega)|\beta,
\gamma, N\rangle}\,
\end{eqnarray}
with $\theta_1\in[0,2\pi]$, $\theta_2\in[0,\pi]$ and $\theta_3\in[0,2\pi]$.
In constructing the projected potential, a fundamental step involves applying rotational transformation to the boson operator $b_{JM}^\dag$ ($J=0,2$), given by $R(\Omega)b_{JM}^\dag R^{-1}(\Omega)=\sum_{M^\prime}D_{M^\prime M}^J(\Omega)b_{JM^\prime}^\dag$.

For an axially-deformed system such as the SU(3) limit in the IBM, $K=0$ is usually assumed in the AMP calculations. In this case, the projected potential surface for the ground-state band ($K=0$) can be further simplified
as \begin{eqnarray}\label{Vg}
&&V(\beta, \gamma,N)_g^J=\frac{\langle\beta, \gamma,
N|\hat{H}\hat{P}_{00}^J|\beta, \gamma, N\rangle}
{\langle\beta, \gamma, N|\hat{P}_{00}^J|\beta, \gamma, N\rangle} \\
\nonumber &&=\frac{\int~d\theta_2\mathrm{sin}\theta_2 d_{00}^J(\theta_2)
\langle\beta, \gamma, N|\hat{H}\mathrm{e}^{-i\theta_2 \hat{J}_y}|\beta,
\gamma, N\rangle}{\int~d\theta_2\mathrm{sin}\theta_2
d_{00}^J(\theta_2)\langle\beta, \gamma, N|\mathrm{e}^{-i\theta_2 \hat{J}_y}|\beta,
\gamma, N\rangle}\, ,
\end{eqnarray}
where it is assumed that only rotational transformation about the $y$ axis, $\hat{R}_y=e^{-i\theta_2\hat{J}_y}$, is retained, consistent with the axial symmetry of system.
To give a brief illustration of the $K=0$ projection for axially symmetric situation, we examine the SU(3) limit $(\eta=1,~\chi=\frac{-\sqrt{7}}{2})$ with boson number $N=10$. Minimizing the potential function given in Eq.~(\ref{Vg}) yields $V(\beta_\mathrm{e},\gamma_\mathrm{e})_g^J=-5.75,~-5.6938,~-5.5625$ for $J=0,~2,~4$, respectively, with corresponding equilibrium deformations all stabilized with $\beta_\mathrm{e}\approx1.4$ and $\gamma_\mathrm{e}=0$. Notably, these results derived from Eq.~(\ref{Vg}) are exactly reproduced by AMP calculations using Eq.~(\ref{AMPI}) for the same SU(3) Hamiltonian.
Moreover, the computed energy values actually coincide with the exact SU(3) solutions obtained from full diagonalization of the SU(3) Hamiltonian derived from (\ref{CQ}), which yields $E(0_1^+)=-5.75$, $E(2_1^+)=-5.6938$, and $E(4_1^+)=-5.5625$. In this work, the diagonalization of the IBM Hamiltonian is performed within the SU(3) basis, characterized by $|N,(\lambda,\mu),\alpha,JM\rangle_{\mathrm{su3}}$, where $(\lambda,\mu)$ are the quantum numbers used to label the irreducible representations of the SU(3) group, and $\alpha$ is the additional quantum number in the reduction of SU(3)$\supset$SO(3). Accordingly, the eigenvector can be expanded in terms of the orthogonal SU(3) basis~\cite{Xiu2021},
\begin{eqnarray}
|\Psi\rangle_{\xi J M}=\sum_{(\lambda,\mu),\alpha}C_{(\lambda,\mu),\alpha}^\xi|N,(\lambda,\mu),\alpha,JM\rangle_{\mathrm{su3}}\, ,
\end{eqnarray}
where $C_{(\lambda,\mu),\alpha}^\xi$ are the expansion coefficients, with $\xi$ representing all quantum numbers other than $J$ and $M$. While such exact diagonalization is applicable to arbitrary IBM Hamiltonian, analytical solutions for dynamical symmetry limits can be derived rigorously using the group-algebra method~\cite{IachelloBook87}. For example, the ground-state band energies in the SU(3) limit described by (\ref{CQ}) can be analytically expressed as \begin{eqnarray}\label{EJ}
E(J_g^+)_{\mathrm{SU(3)}}=-\frac{2N+3}{4}+\frac{3}{32N}J(J+1)\,
\end{eqnarray}
with $J=0,~2,~4,\cdots,~2N$. Anyway, the calculations suggest that AMP onto the coherent state works equivalently well as the method of exact diagonalization for the SU(3) limit. The resulting $\gamma_\mathrm{e}=0$ is fully consistent with the $K=0$ projection assumption, as is expected from the axial symmetry inherent in the SU(3) system. The geometry of the SU(3) limit is traditionally understood from its mean-field potential, which yields $\beta_\mathrm{e}=\sqrt{2}$ and $\gamma_\mathrm{e}=0^\circ$ in the classical ($N\rightarrow\infty$) limit~\cite{IachelloBook87}. This result, together with the excitation energy rule $E(J_g^+)\propto J(J+1)$ shown in (\ref{EJ}), establishes a prolate-rotor picture for the SU(3) limit in the IBM. The similar methodology can also be applied to the U(5) (spherical vibrator) and O(6) ($\gamma$-unstable rotor) symmetry limits to illustrate their respective geometric modes~\cite{IachelloBook87}.

For systems exhibiting triaxial or $\gamma$-unstable deformation, i.e., those deviating from axial deformation, there is no a priori justification for assuming that Eq.~(\ref{AMPI}) with $K=0$ remains valid for deriving the deformation. In such cases, a more general implementation of the AMP method, incorporating $K$-mixing, should be employed. Using the AMP operator $\hat{P}_{MK}^J$, one can extract the states of definite angular momentum from the coherent state via
\begin{eqnarray}\label{PS}
|\Psi_M^J\rangle=\sum_K g_K \hat{P}_{MK}^J|\beta, \gamma,
N\rangle\, ,
\end{eqnarray}
where the coefficients $g_K$ represent the expansion weights. To get the energies $E^J$, one needs to minimize the energy function with respect to the coefficients $g_K$, leading to
the generalized eigenvalue equation~\cite{Ringbook}
\begin{eqnarray}\label{HWE}
\sum_{K^\prime} [h_{KK^\prime }^J-E^Ln_{KK^\prime}^J]g_{K^\prime}=0\, ,
\end{eqnarray}
where the Hamiltonian kernel and norm kernel are defined, respectively, as
\begin{eqnarray}
&&h_{KK^\prime}^J=\langle\beta, \gamma,
N|\hat{H}\hat{P}_{KK^\prime}^J|\beta, \gamma,N\rangle,\\
&&n_{KK^\prime}^J=\langle\beta, \gamma,
N|\hat{P}_{KK^\prime}^J|\beta, \gamma,N\rangle\, .
\end{eqnarray}
Because the projected states (\ref{PS}) are non-orthogonal, we follow the the standard procedure~\cite{Ringbook}
to solve Eq.~(\ref{HWE}). First, the norm kernel is diagonalized,
\begin{eqnarray}
\sum_{K^\prime}n_{KK^\prime}^J f_{a,K^\prime}^{J}=n_a^J f_{a,K}^J\, .
\end{eqnarray}
The orthonormalized bases are then constructed as~\cite{Wang2022}
\begin{eqnarray}\label{basis}
|\phi_a\rangle=\frac{1}{\sqrt{n_a^J}}\sum_Kf_{a,K}^{J}\hat{P}_{MK}^J|\beta, \gamma,
N\rangle\, .
\end{eqnarray}
Solving the generalized eigenvalue equation (\ref{HWE}) is now equivalent to diagonalizing the Hamiltonian within these orthonormalized bases.
This yields a set of eigenvalues $E^J$, ordered by energy: the lowest defines the yrast state, whose associated $\beta_\mathrm{e}$ and $\gamma_\mathrm{e}$
characterize the quadrupole deformation at given $J$; the next-lowest corresponds to the yare state, and so on. In the diagonalization procedure, the quantum number $K$ is no longer a fixed value. Consequently, the resulting eigenvector are generally superpositions of different $K$ components, as indicated in (\ref{basis}). This actually defines the meaning of $K$-mixing in the present AMP calculations. In what follows, we compare  two distinct approaches for computing yrast-level energies within the AMP framework: the first (denoted by AMP$^\mathrm{I}$) minimizes the projected potential function for a fixed $K$, as defined in Eq.~(\ref{AMPI}); the second (denoted by AMP$^\mathbf{II}$) performs full AMP calculations using the $K$-mixing bases, as specified in Eq.~(\ref{basis}). Obviously, for $J=0$, there are no differences between the two AMP methods. The validity of the AMP calculations can be assessed by comparing the results with exact solutions obtained from Hamiltonian diagonalization.

\begin{center}
\vskip.2cm\textbf{III. AMP Analysis of Quadrupole Deformation}
\end{center}\vskip.2cm

From the above example for the SU(3) limit, it is evident that the AMP method not only enables the calculation of energy spectra but also serves as a valuable tool for characterizing quadrupole deformation in the IBM. Traditionally, distinct collective modes, which are manifested by their characteristic spectral features, are defined and often labeled according to the ground-state quadrupole deformations identified at the mean-field level. Prototypical examples include the spherical vibrator (U(5) limit) and the axially-deformed rotor (SU(3) limit). In contrast to those with rigid shapes, transitional systems may exhibit spin-dependent deformation, implying the presence of $\beta$- or $\gamma$-softness already in ground states. The AMP method discussed herein thus provides a natural framework for investigating the spin dependence of quadrupole deformation and its associated spectroscopic signatures.

\begin{center}
\vskip.2cm\textbf{A. Ground-state Deformation}
\end{center}\vskip.2cm

\begin{figure}[!h]
\begin{center}
\includegraphics[scale=0.12]{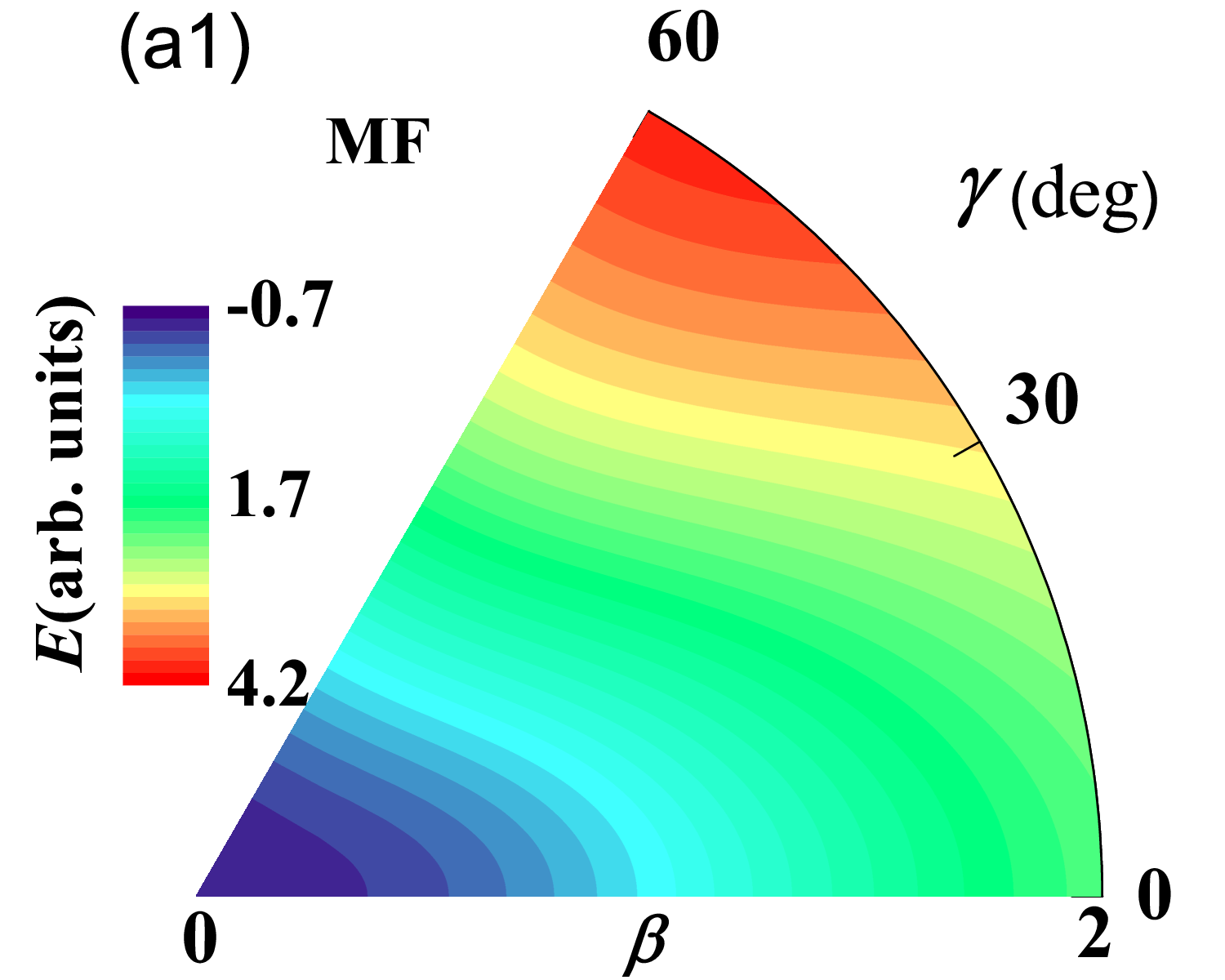}
\includegraphics[scale=0.12]{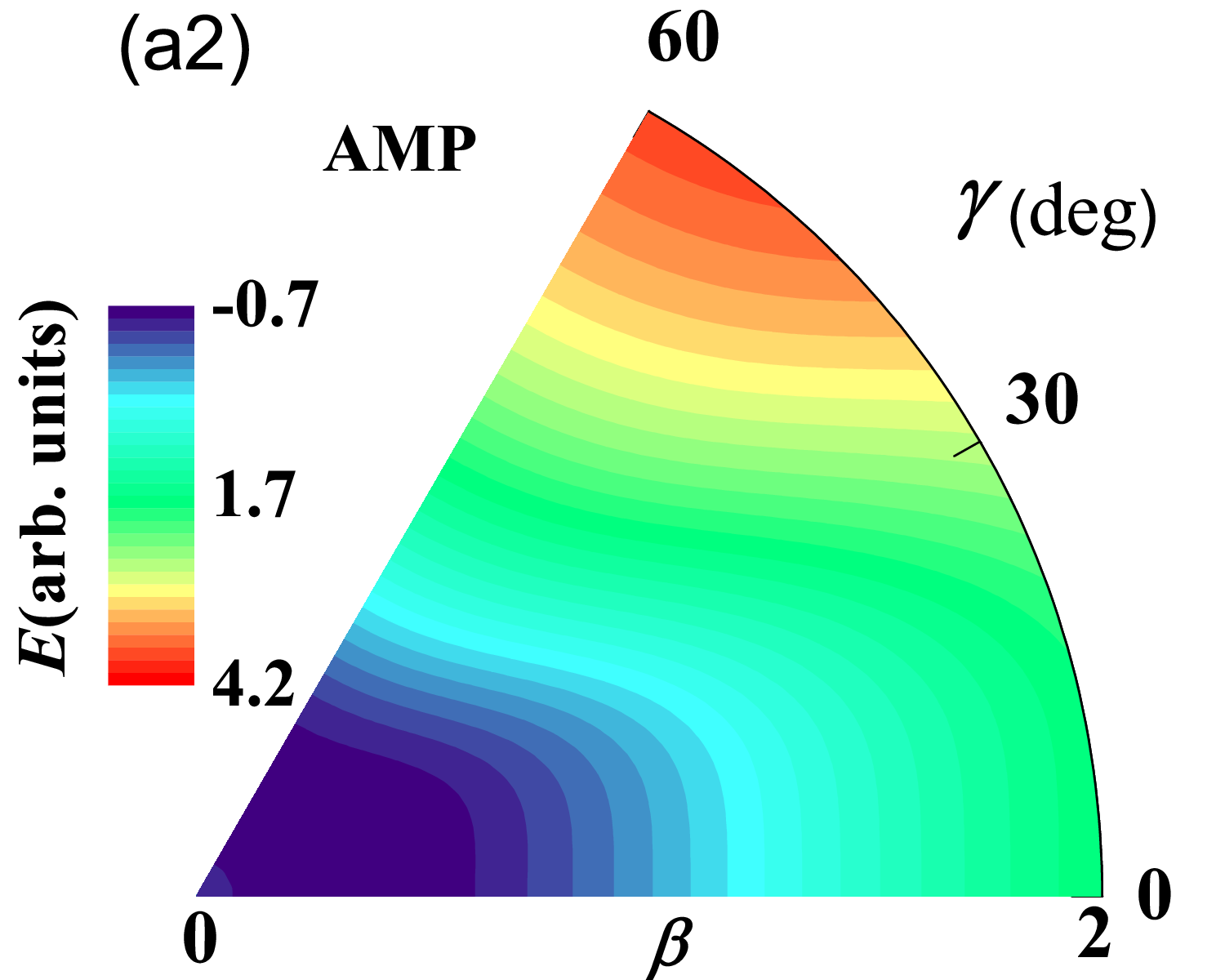}
\includegraphics[scale=0.12]{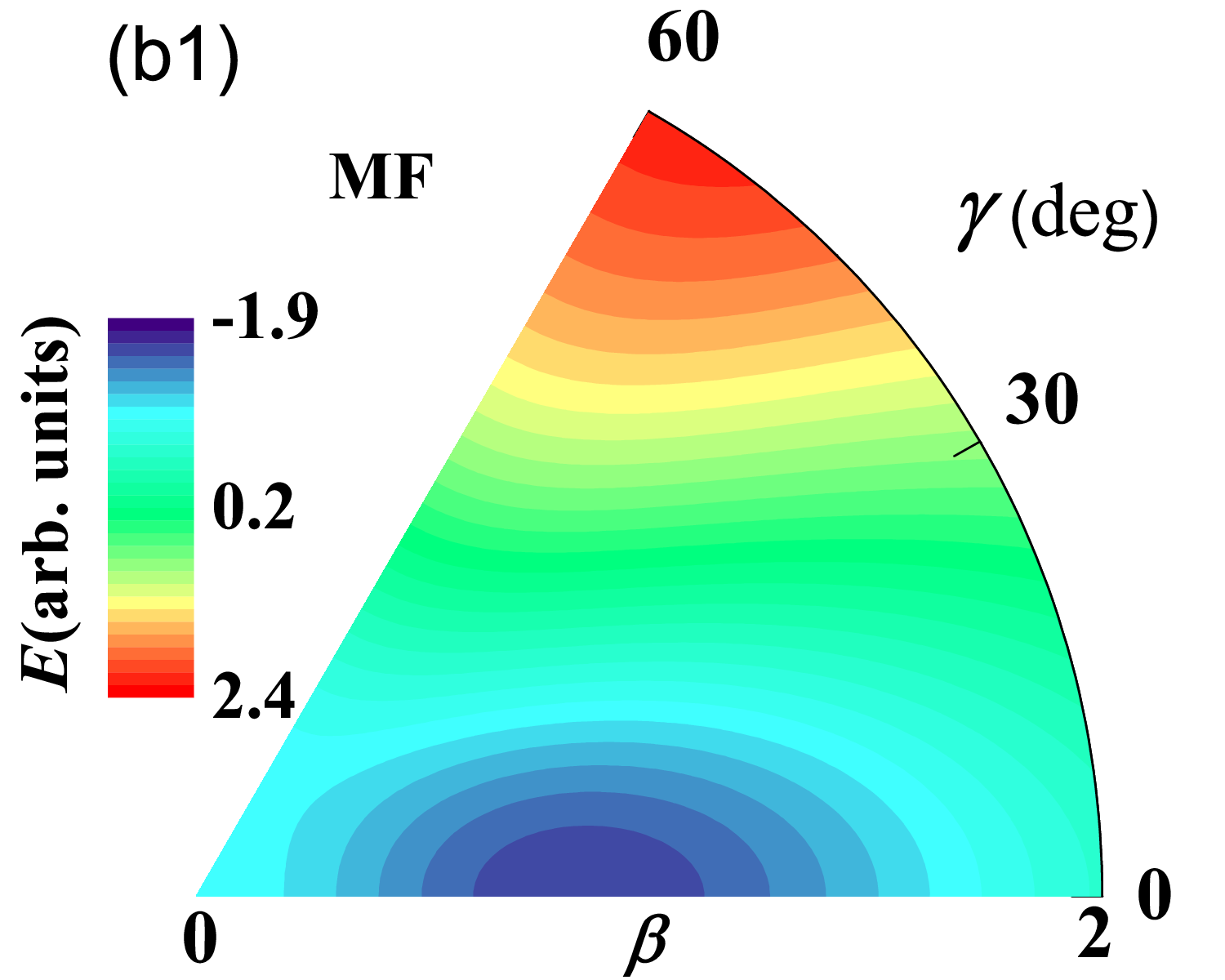}
\includegraphics[scale=0.12]{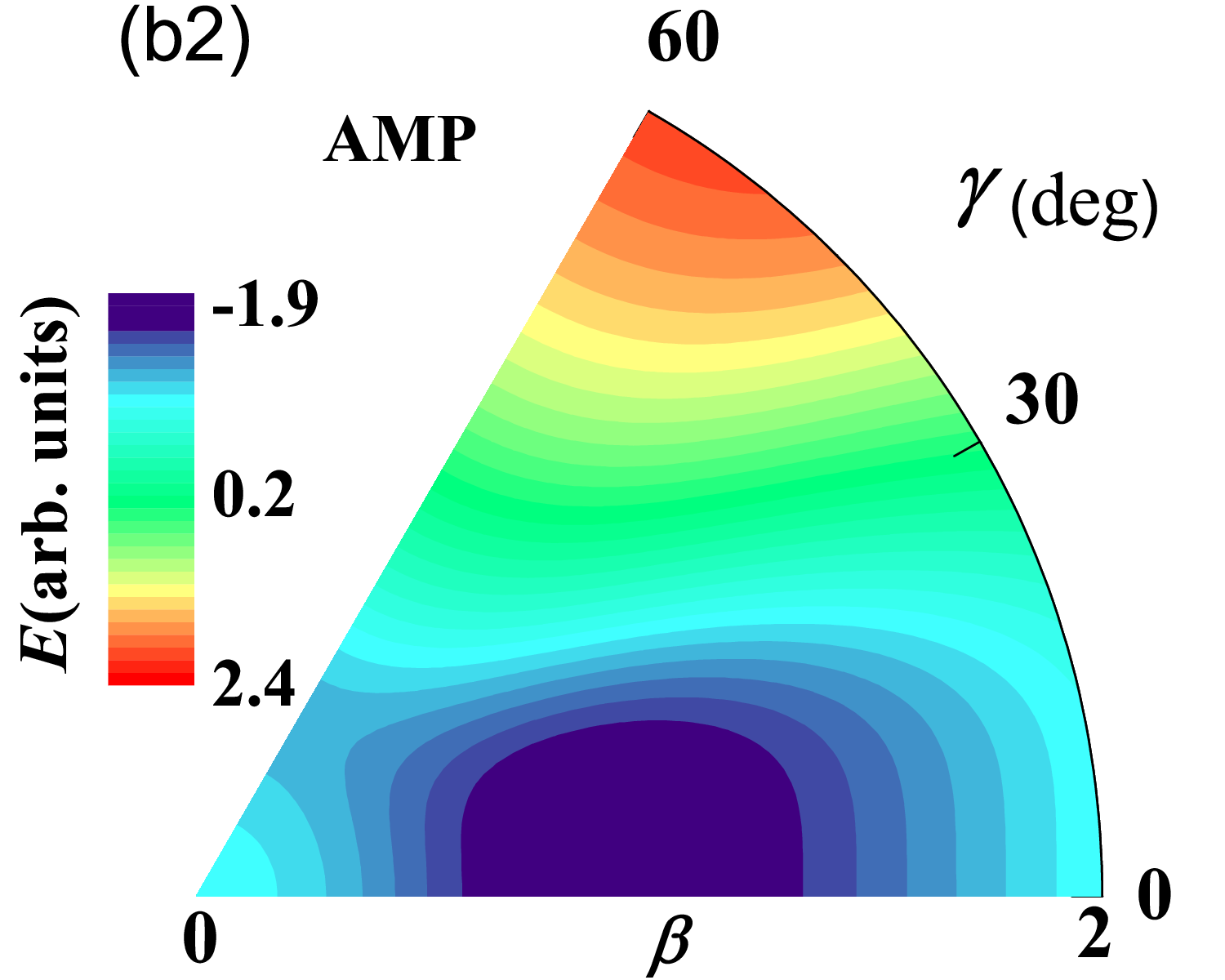}
\includegraphics[scale=0.12]{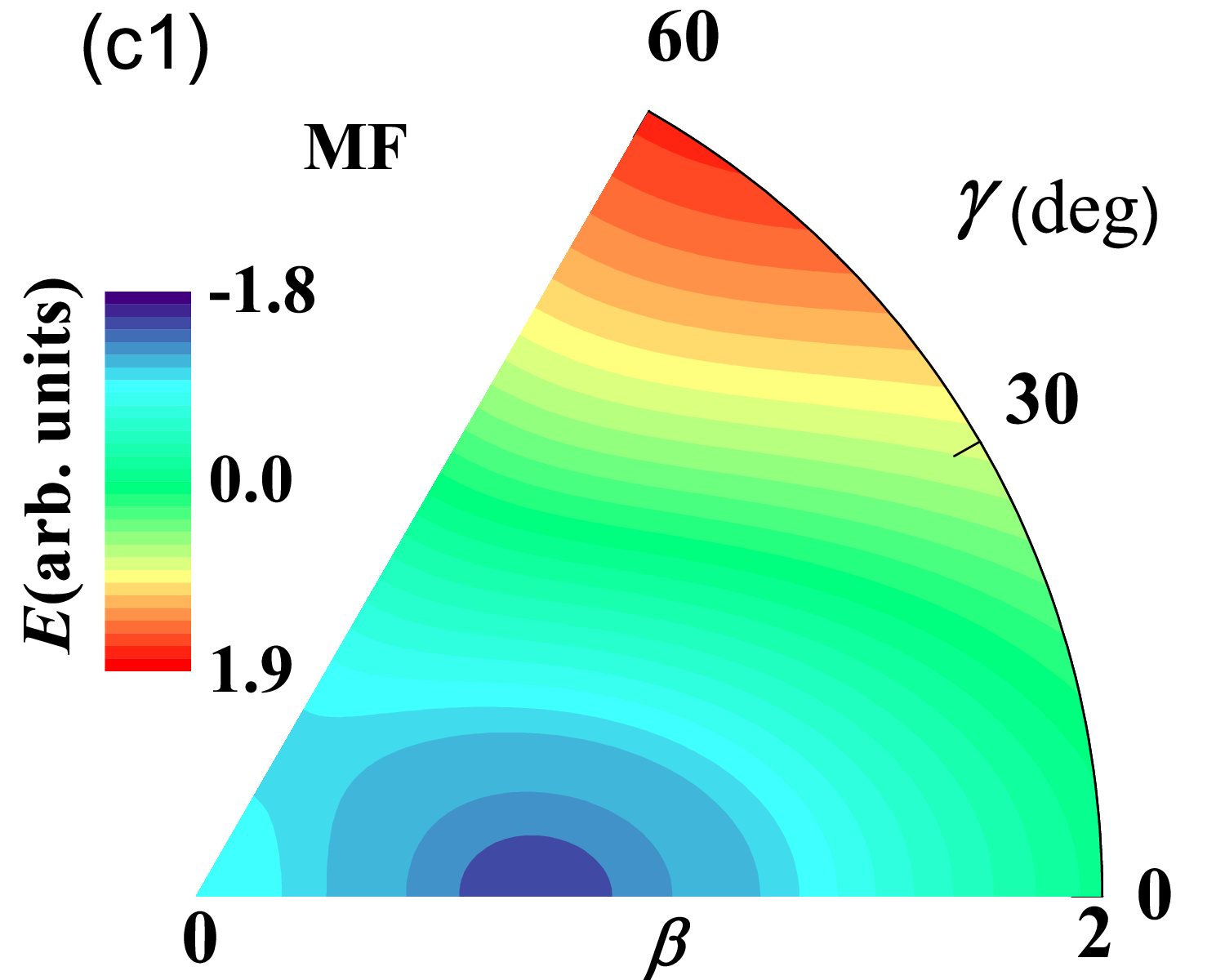}
\includegraphics[scale=0.12]{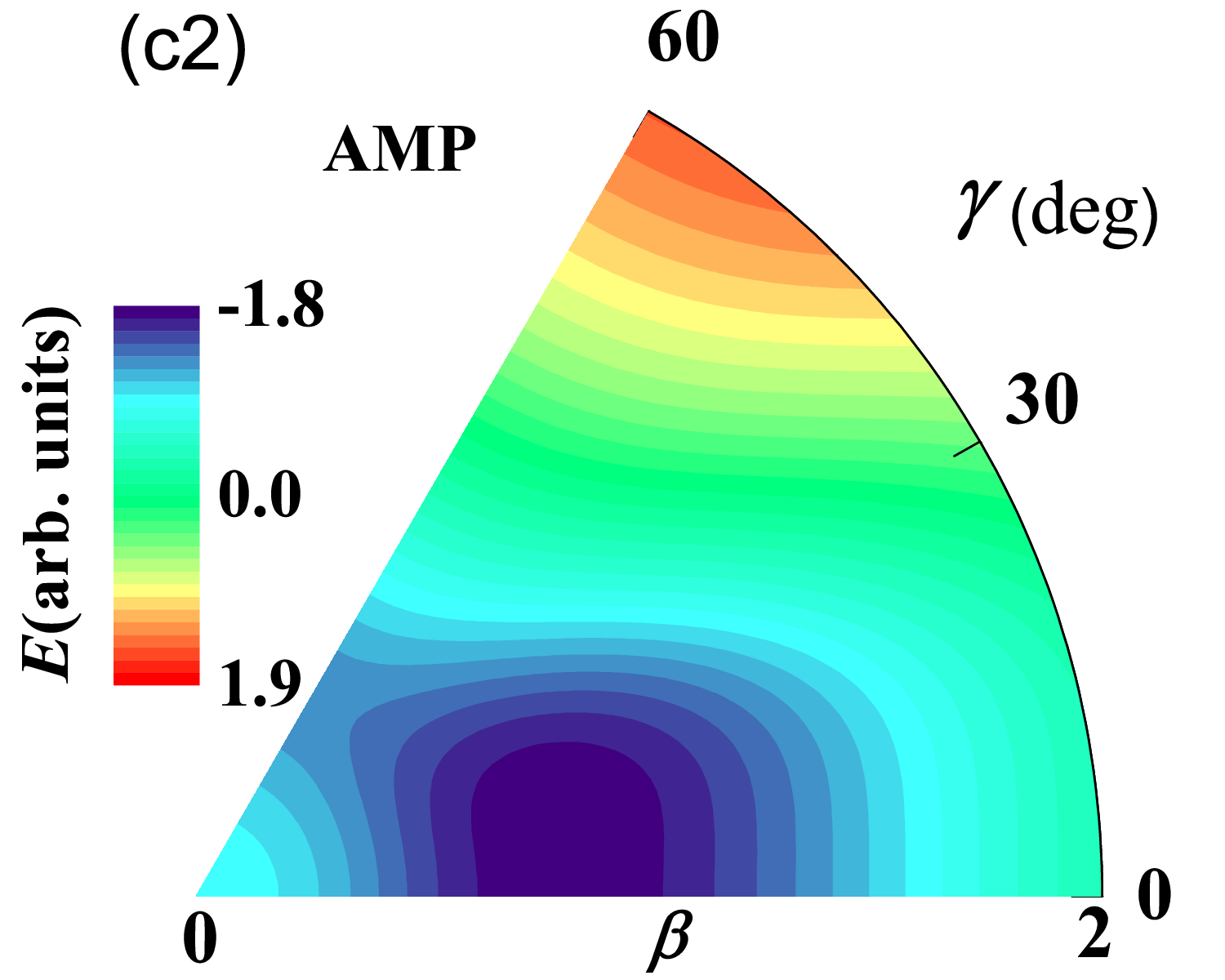}
\includegraphics[scale=0.12]{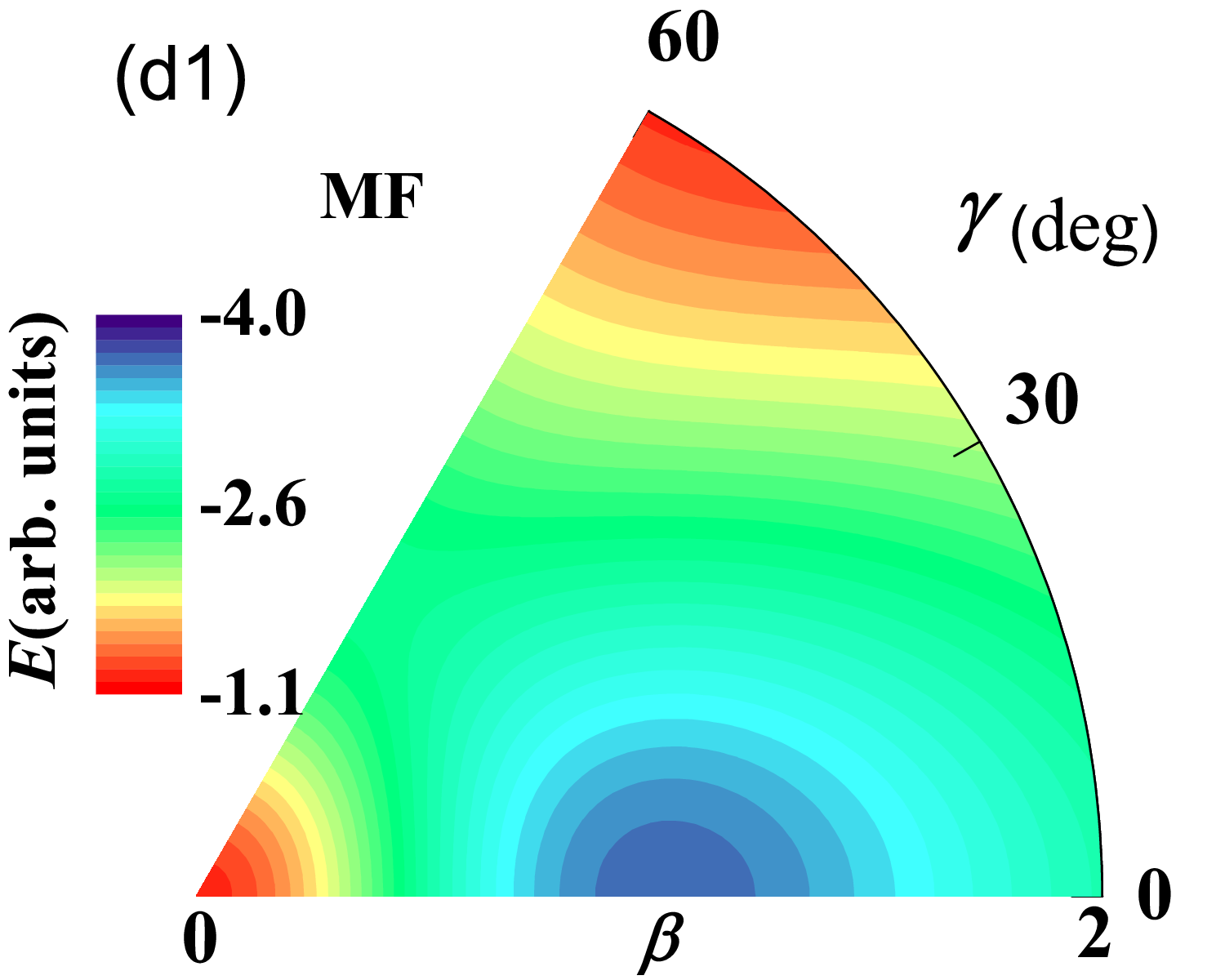}
\includegraphics[scale=0.12]{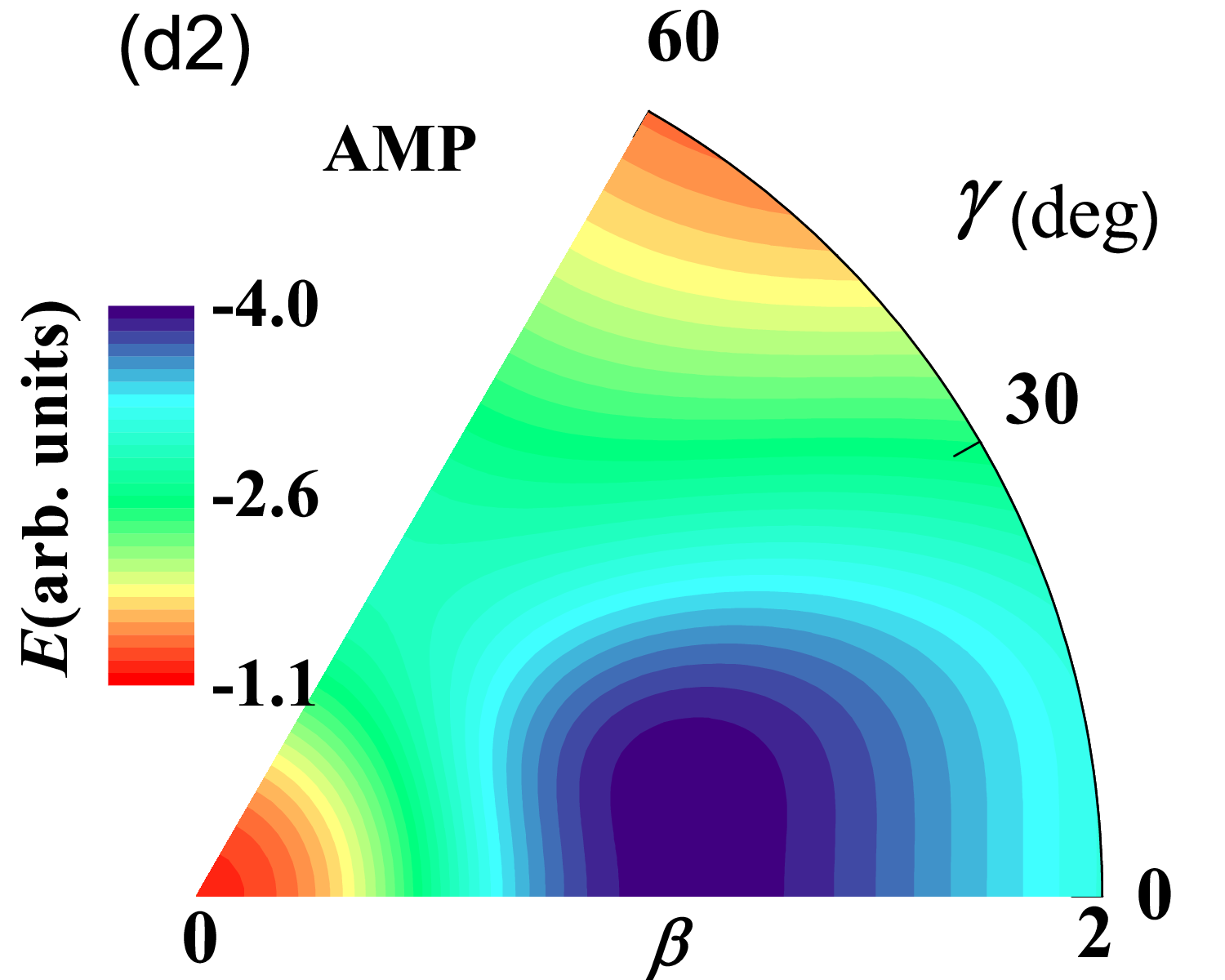}
\includegraphics[scale=0.12]{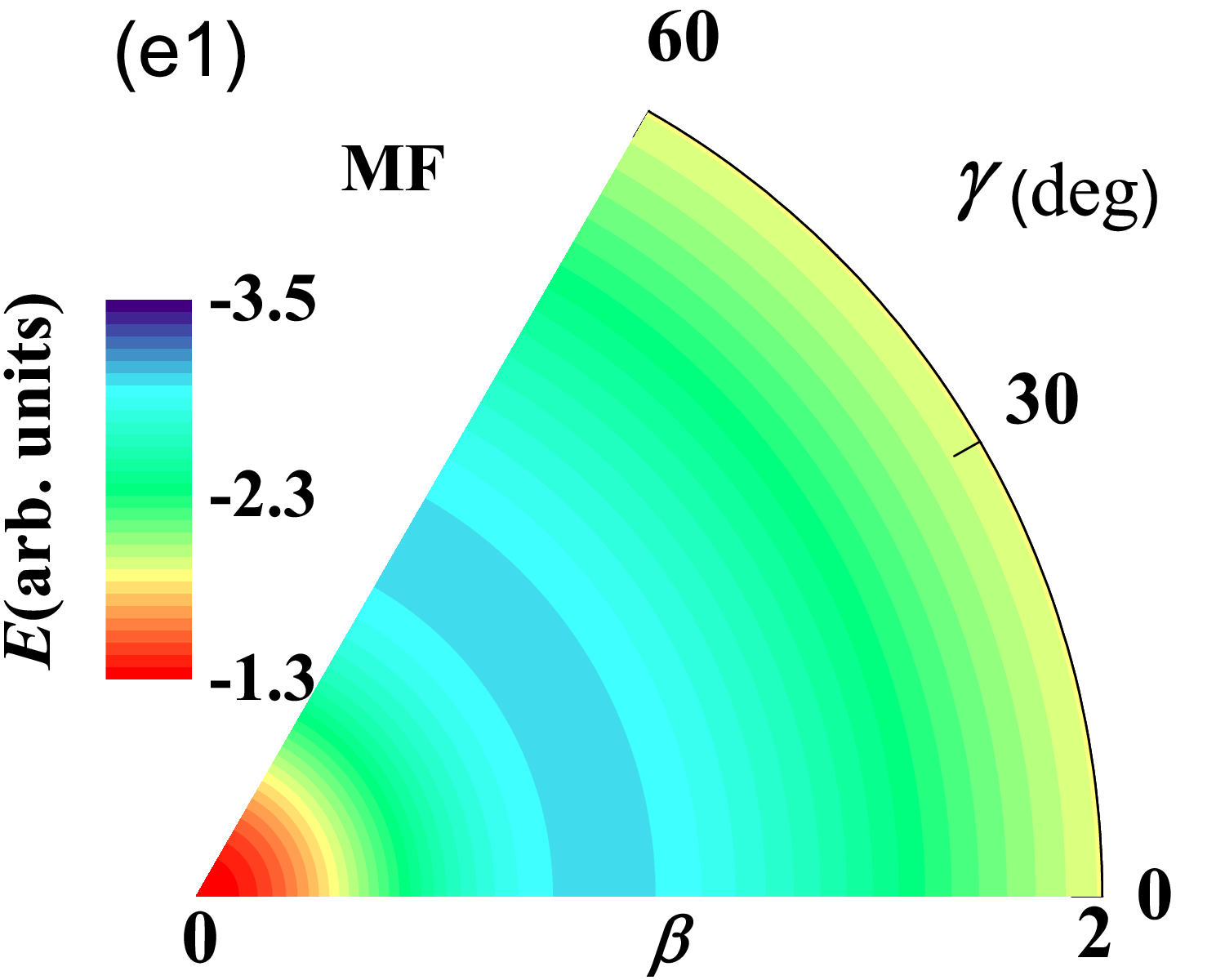}
\includegraphics[scale=0.12]{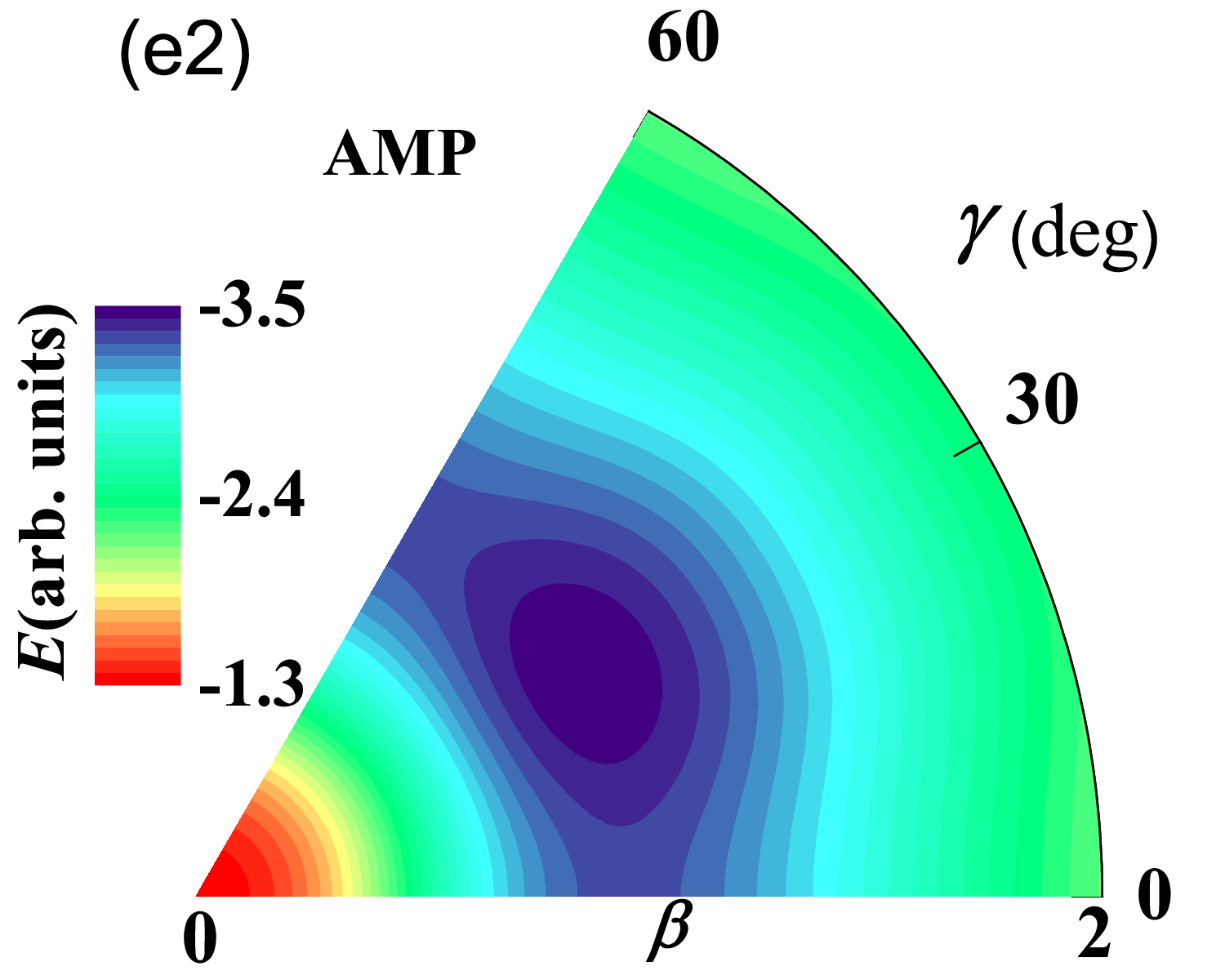}
\includegraphics[scale=0.12]{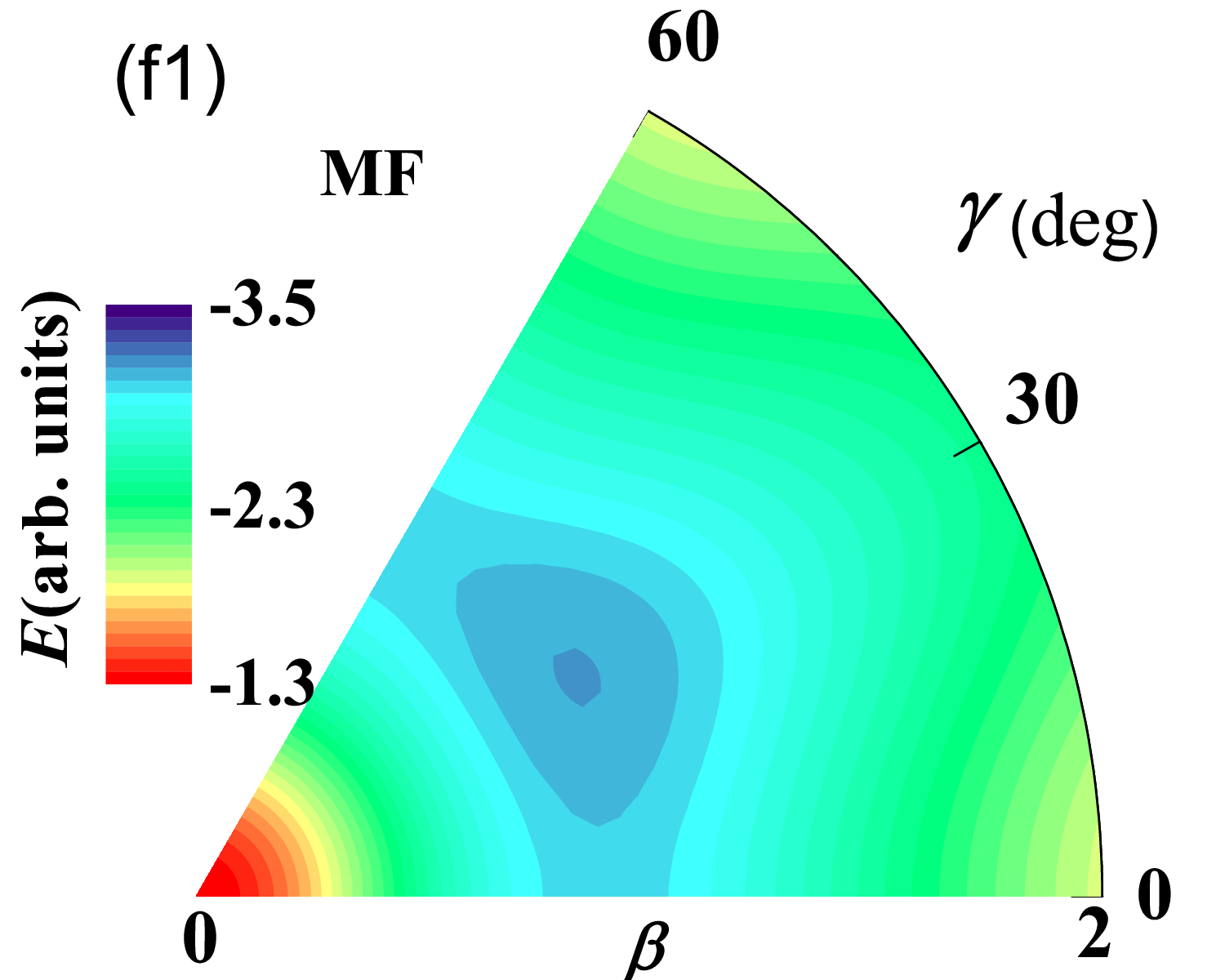}
\includegraphics[scale=0.12]{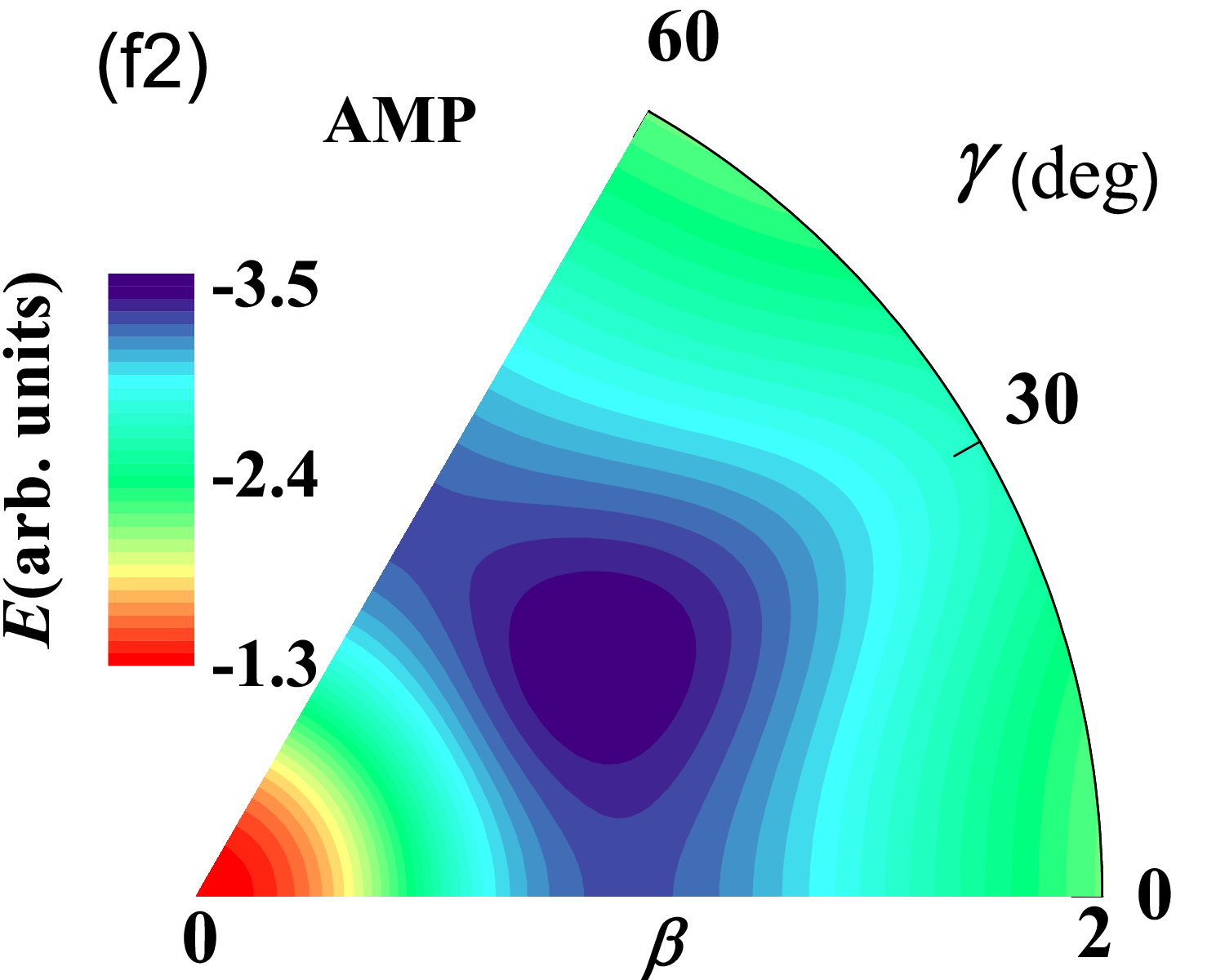}
\includegraphics[scale=0.12]{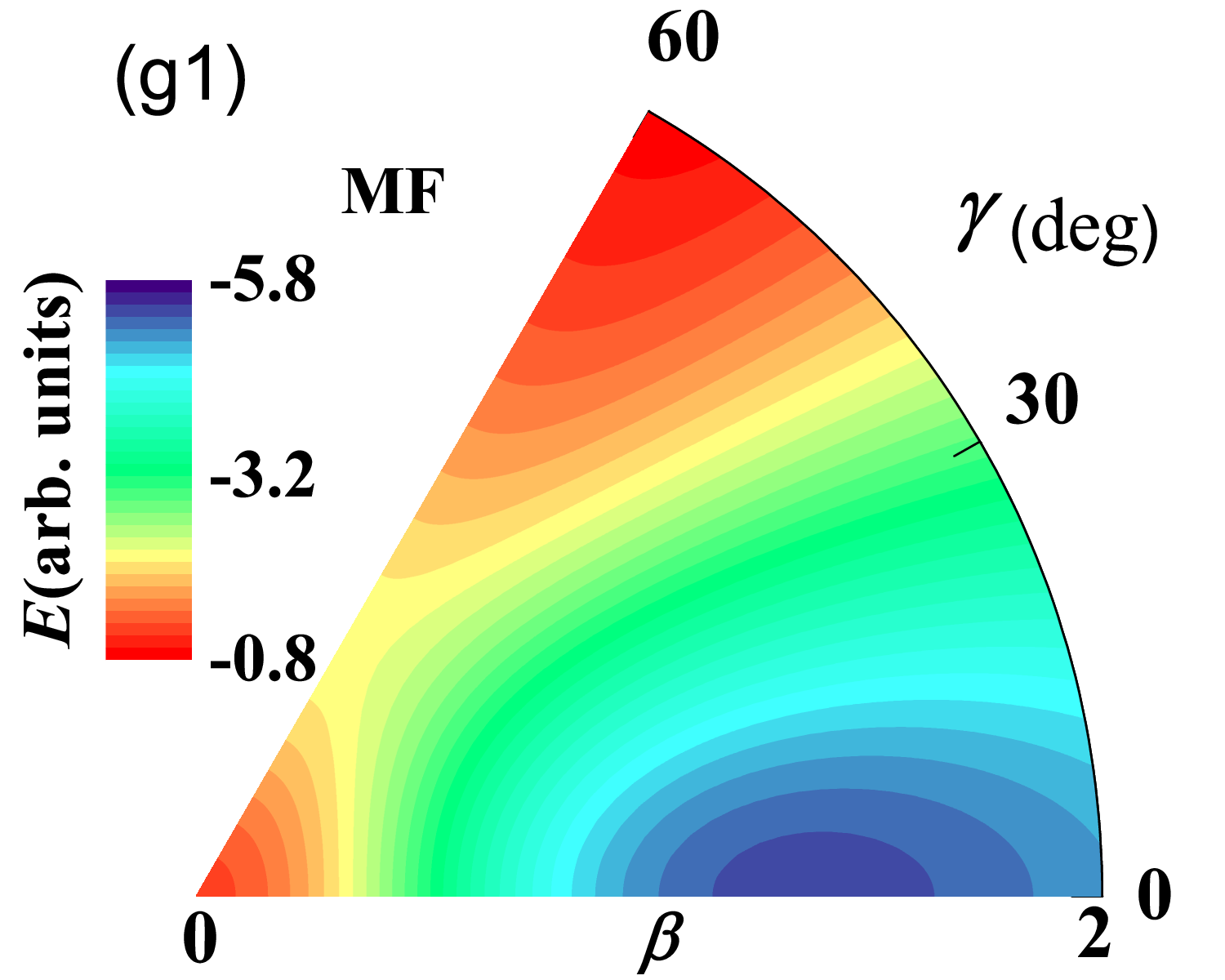}
\includegraphics[scale=0.12]{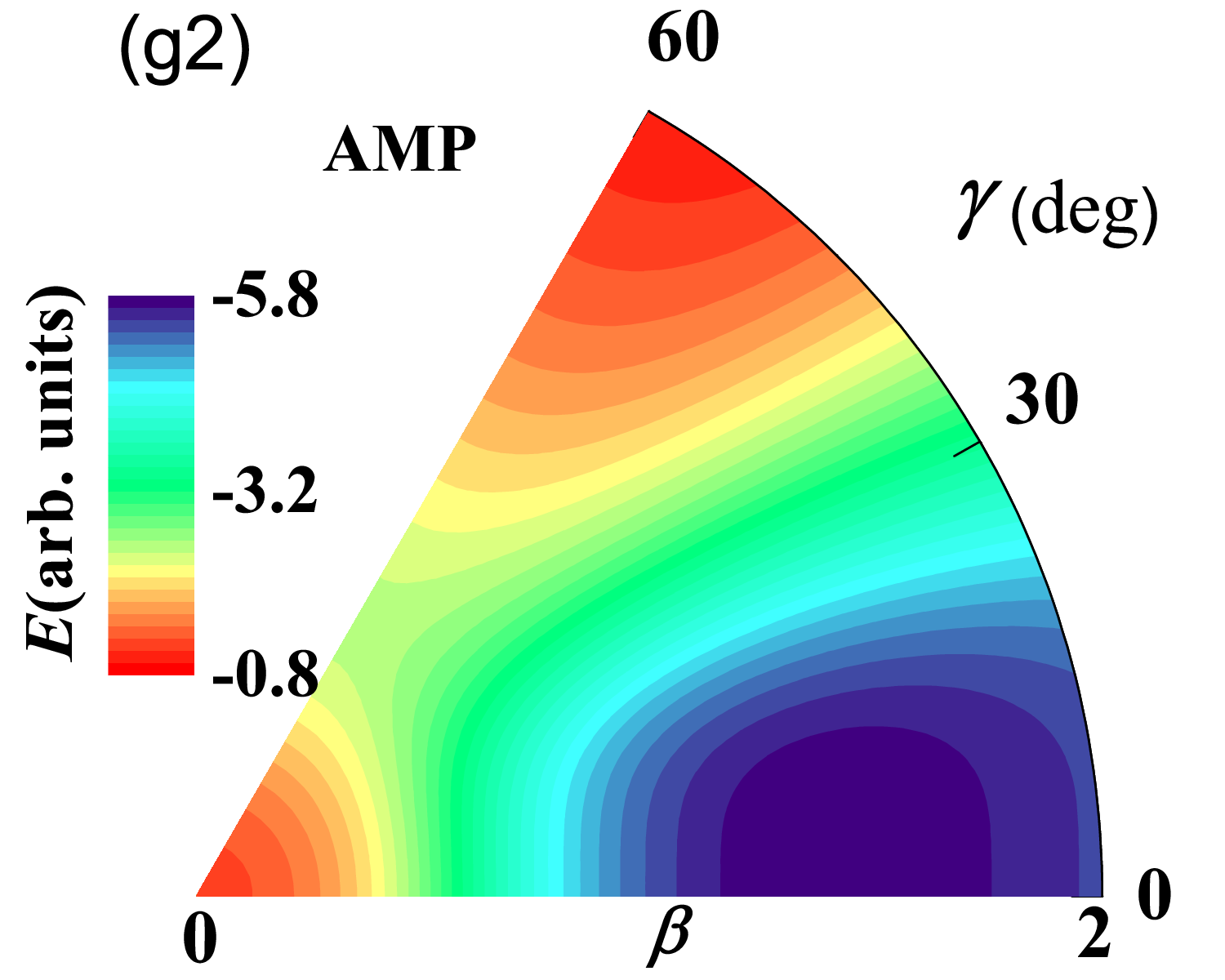}
\caption{(Color online) The mean-field (MF) potentials (in arbitrary units) at the selected parameter points "a", "b", "c" and "d" (see Fig.~\ref{F1}) are presented for comparison with
the corresponding $J=0$ potentials obtained from AMP. In addition, results for the O(6) limit, a triaxial configuration built based on O(6) and the SU(3) limit (see text for the parameter illustrations) are also provided, as shown in panels (e), (f) and (g), respectively.
\label{F2}}
\end{center}
\end{figure}

The ground-state quadrupole deformation of an IBM system governed either by the Hamiltonian in (\ref{H}) or by its consistent-$Q$ reduction in (\ref{CQ}) can be extracted from the mean-field potential given in (\ref{V}). Since the coherent state introduced in (\ref{coherent}) explicitly breaks rotational symmetry, the deformation obtained at the mean-field level requires further refinement through restoration of rotational symmetry. This is accomplished using the AMP technique applied to the coherent state, yielding the projected potential function via Eq.~(\ref{AMPI}) with $J=0$. To illustrate the AMP corrections on the mean-field descriptions, contour plots of both the mean-field potentials and their AMP-corrected counterparts are presented in Fig.~\ref{F2} for selected parameter points ("a", "b", "c", "d") indicated in the triangle phase diagram (see Fig.~\ref{F1}).

As shown in Fig.~\ref{F1}, the parameter point "a", corresponding to $(\eta,~\chi)=(0.45,~-1.323)$, lies within the spherical region and represents a weakly deformed configuration at the mean-field level. The parameter point "b", with $(\eta,~\chi)=(0.65,~-1.323)$, resides in the deformed region along the U(5)-SU(3) leg, thus characterizing a transitional case in the U(5)-SU(3) transitional region. Similarly, parameter point "c", defined by $(\eta,~\chi)=(0.7,~-0.8)$, represents a mixed transitional case involving U(5), SU(3) and O(6) modes; in contrast, parameter point "d", corresponding to $(\eta,~\chi)=(1.0,~-0.5)$, lies on the SU(3)-O(6) leg and therefore denotes a deformed configuration in the SU(3)-O(6) transitional region. Note that the classification of deformation types in the phase diagram is based on the coherent-state (mean-field) calculations~\cite{Iachello2004}, whereby representative parameter points are selected to characterize distinct mean-field deformation regimes.
Additionally, we present the corresponding potential surfaces for three further parameter points ("e", "f", "g"). Parameter point "e", with $(\eta,~\chi)=(1.0,~0)$, corresponds exactly to the O(6) limit ($\gamma$-unstable), which coincides with one vertex of the triangle phase diagram. Parameter point "g", specified by $(\eta,~\chi)=(1.0,~-\sqrt{7}/2)$, realizes the SU(3) limit (axially deformed) and corresponds to another vertex of the triangle. Collectively, these selected parameter points span a representative set of mean-field deformations across the triangle phase diagram, except for triaxial deformation, which cannot be generated solely by the consistent-$Q$ Hamiltonian (\ref{CQ}). To address this limitation, we also include parameter point "f", located outside the triangle but constructed as a perturbation around the O(6) limit ("e"). Specifically, parameter point "f" is described by the Hamiltonian $\hat{H}_{\mathrm{O(6)}}+0.01\hat{V}_3$, where $\hat{H}_{\mathrm{O(6)}}$ denotes the O(6) Hamiltonian with $(\eta,~\chi)=(1.0,~0)$ and $\hat{V}_3$ is the cubic $\hat{V}_3$ term defined in (\ref{V3}). This type of Hamiltonian was suggested to yield a $\gamma$-stable triaxial shape at the mean-field level~\cite{VC1981,Sorgunlu2008}. In all calculations, the total boson number is fixed at $N=10$.

As shown in Fig.~\ref{F2}, the projected potentials (AMP) generally exhibit deeper potential wells than the unprojected (MF) counterparts, while preserving the overall qualitative features of the potential energy surfaces derived from mean-field calculations. This consistency suggests that mean-field calculations provide a sufficiently accurate description of the ground-state deformations in these IBM systems. Nevertheless, projection yield clear quantitative improvements, as evidenced by the energy scales displayed in each panel of Fig.~\ref{F2}. For instance, at the point "a", the energy value, corresponding to the minimum of the projected potential (also referred as minimizing-after-projection (MAP)), is $V_{\mathrm{min}}^{J=0}=-0.74$. This value closely reproduces the exact ground-state energy, $E(0_\mathrm{g})=-0.75$, derived from Hamiltonian diagonalization, whereas the mean-field calculation yields only $V_{\mathrm{min}}=-0.58$.

A further notable change concerns the O(6) limit. As illustrated in Fig.~\ref{F1}(e1), the unprojected potential displays perfect $\gamma$-independence, with degenerate energy surfaces spanning the range $\gamma=0^\circ$ to $\gamma=60^\circ$. Such $\gamma$-softness underpins the conventional characterization of the O(6) limit as a $\gamma$-unstable (soft) rotor~\cite{IachelloBook87}. In contrast, the projected potential surface shown in Fig.~\ref{F2}(e2) exhibits a well-defined minimum at $(\beta_\mathrm{e},~\gamma_\mathrm{e})=(1.0,~30^\circ)$, implying a triaxial deformation for the finite-$N$ O(6) system~\cite{Otsuka1987}, which is consistent with the earlier analysis by Dobe\v{s}~\cite{Dobes1985}. A demonstration of the equivalence between $\gamma$ instability and triaxiality in O(6), based on AMP, was provided by Otsuka and Sugita~~\cite{Otsuka1987}.
The triaxial character becomes even more pronounced in case (f) depicted in Fig.~\ref{F2}(f1). In this case, the mean-field potential surface already suggests a collective mode characteristic of a triaxial rotor, and projection further strengthens the recognition of the triaxial geometry, as revealed in Fig.~\ref{F2}(f2). These results are also consistent with the spectral analysis of the O(6)-related triaxial case presented in \cite{Sorgunlu2008}. By contrast, the pictures presented in Fig.~\ref{F2}(g1)-(g2) confirm that the SU(3) limit corresponds to a prolate rotor.

\begin{center}
\vskip.2cm\textbf{B. $K$-Mixing Effects}
\end{center}\vskip.2cm

For $J>0$, multiple $K$ components contribute to a given $J$. In such cases, the AMP calculations can be carried out in two ways: AMP$^\mathrm{I}$, which employs the $K$-fixing projection using Eq.~(\ref{AMPI}), and AMP$^\mathrm{II}$, which incorporates $K$-mixing, as specified by the orthonormalized bases defined in (\ref{basis}). To assess $K$-mixing effects, we take $J=4$ as a representative example to compare the results obtained from the $K=0$ AMP calculations (AMP$^\mathrm{I}$) and those from $K$-mixing AMP calculations (AMP$^\mathrm{II}$). First, the $J=4$ energies $E(4_1^+)$ computed via both projection schemes are summarized in Table~\ref{T1} for all cases shown in Fig.~\ref{F2}, alongside the exact diagonalization results. As evident from Table~\ref{T1}, the exact (unnormalized) energies of the $4_1^+$ state, obtained via full Hamiltonian diagonalization, are reproduced with high fidelity by both AMP methods, with deviations typically below $1\%$ for most cases. Although $\mathrm{AMP}^\mathrm{II}$ consistently yields improved agreement with the exact results compared to $\mathrm{AMP}^\mathrm{I}$, the quantitative improvements from including $K$-mixing are generally modest. In particular, the equilibrium deformations, $\beta_\mathrm{e}$ and $\gamma_\mathrm{e}$, extracted from the two methods are found to be in close agreement. This indicates that the $K=0$ approximation ($\mathrm{AMP}^\mathrm{I}$) suffices to capture the essential spin-dependent quadrupole deformations within the IBM, even in the triaxial systems conventionally expected to exhibit pronounced $K$-mixing effects from the perspective of the triaxial rotor model~\cite{Bohrbook}. Comparatively, the deviation of AMP calculations from exact solutions tends to increase slightly as the system moves in the spherical region, as partially demonstrated by the results for case "a" listed in the Table. Subsequent discussions will focus primarily on cases located within the deformed region of the phase diagram shown in Fig.~\ref{F1}.
\begin{table}
\caption{The exact energies (in arbitrary units) of the $4_1^+$ state, obtained by diagonalizing the Hamiltonian across all parameter points indicated in Fig.~\ref{F2}, are tabulated for comparison with the corresponding results from the $K=0$ projection (AMP$^\mathrm{I}$) and the $K$-mixed projection (AMP$^\mathrm{II}$).To quantify the deviation of the AMP results, the relative errors, defined as $\Delta\mathrm{AMP}=\mid(\mathrm{Exact}-\mathrm{AMP})/\mathrm{Exact}\mid$, are also reported in the Table.
}\label{T1}
\begin{tabular}{cccccccc}\hline\hline
$E(4_1^+)$&(a)&(b)&(c)&(d)&(e)&(f)&(g) \\
\hline
Exact&-0.084&-1.672&-1.571&-3.795&-3.250&-3.304&-5.5625\\
AMP$^\mathrm{I}$&-0.080&-1.671&-1.570&-3.794&-3.192&-3.268&-5.5625\\
AMP$^\mathrm{II}$&-0.082&-1.672&-1.571&-3.794&-3.232&-3.303&-5.5625\\
$\Delta\mathrm{AMP}^\mathrm{I}$&4.8\%&<1\%&<1\%&<1\%&1.8\%&1.1\%&<1\%\\
$\Delta\mathrm{AMP}^\mathrm{II}$&2.4\%&<1\%&<1\%&<1\%&<1\%&<1\%&<1\%\\
\hline\hline
\end{tabular}
\end{table}

\begin{figure}
\begin{center}
\includegraphics[scale=0.15]{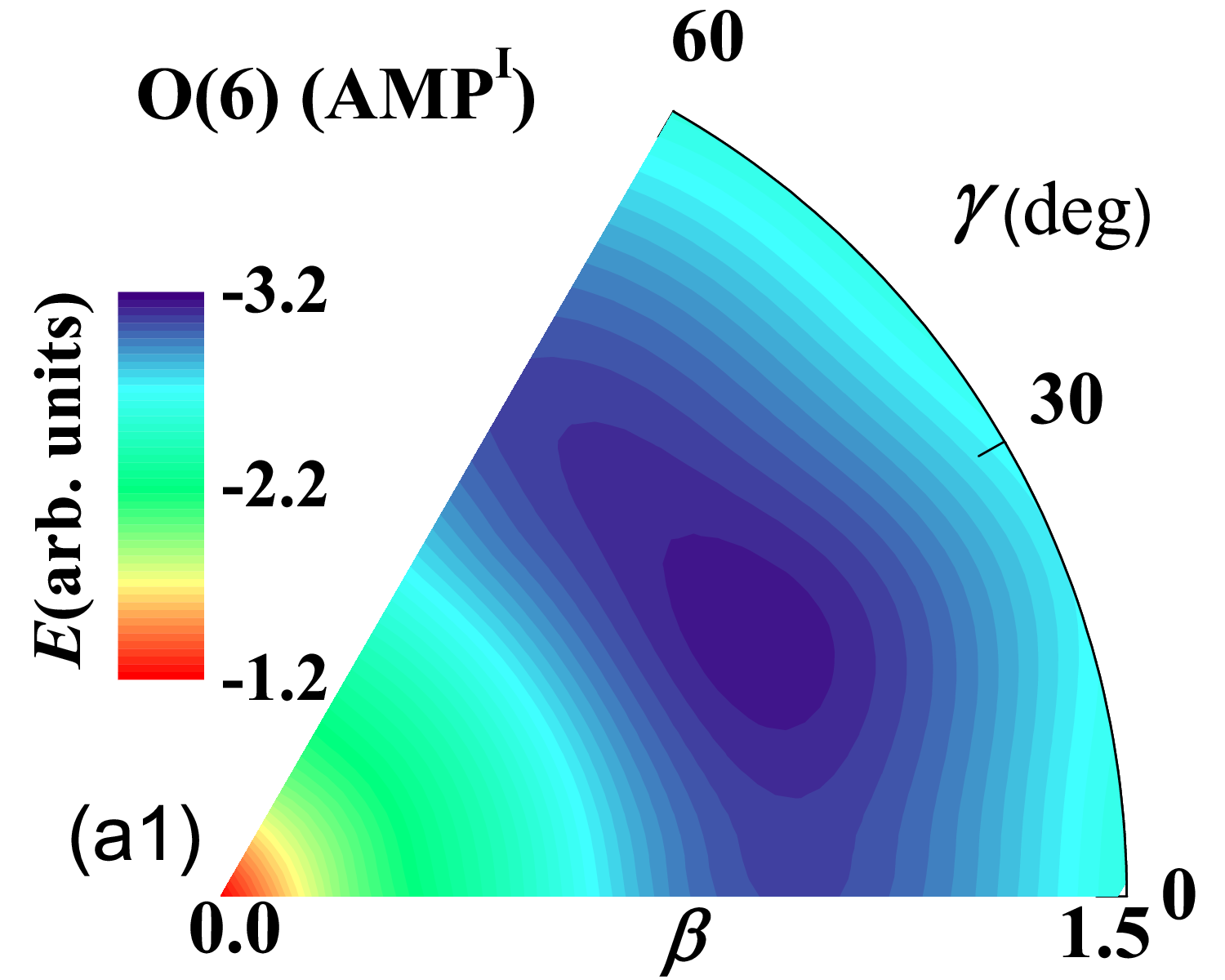}
\includegraphics[scale=0.15]{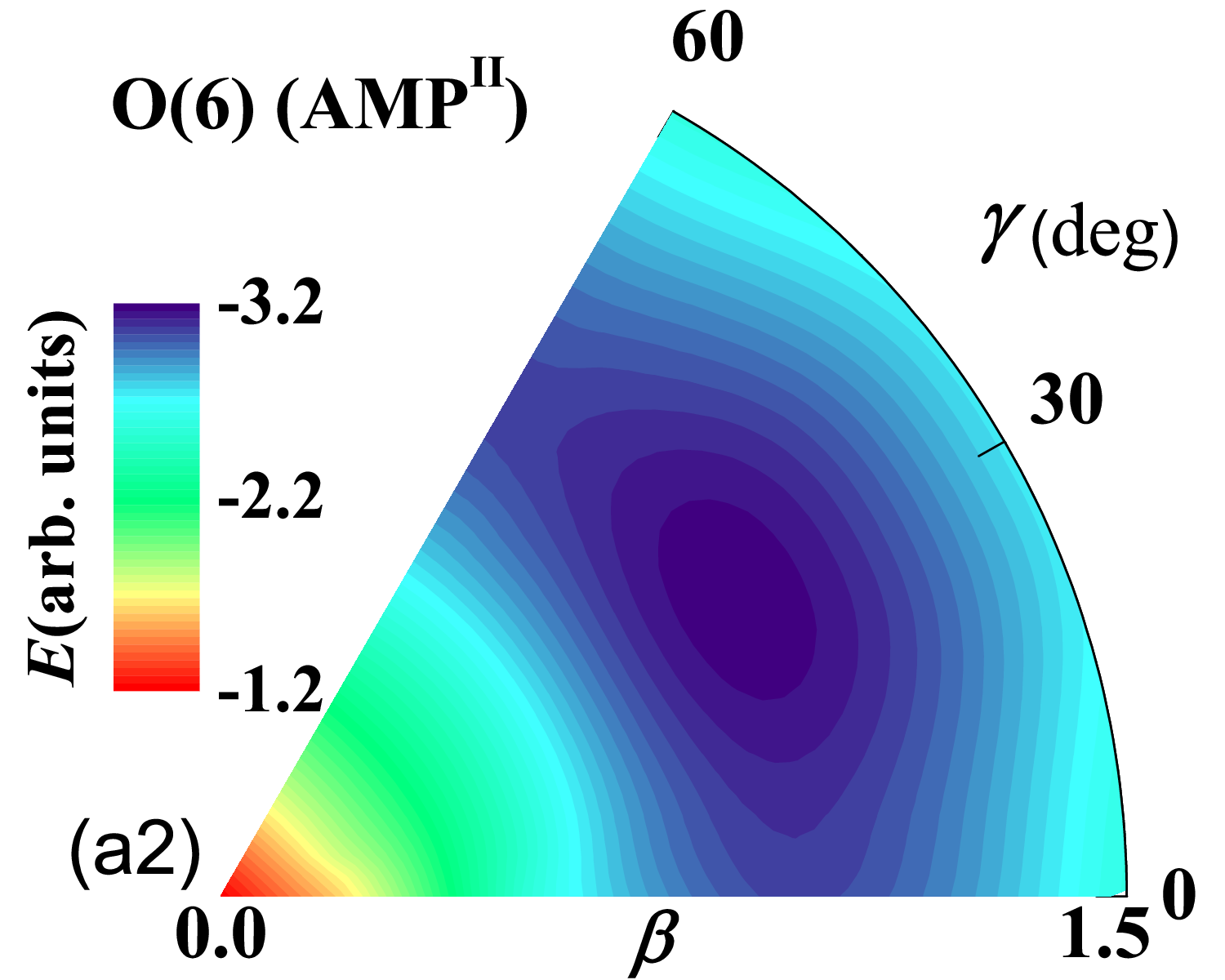}
\includegraphics[scale=0.15]{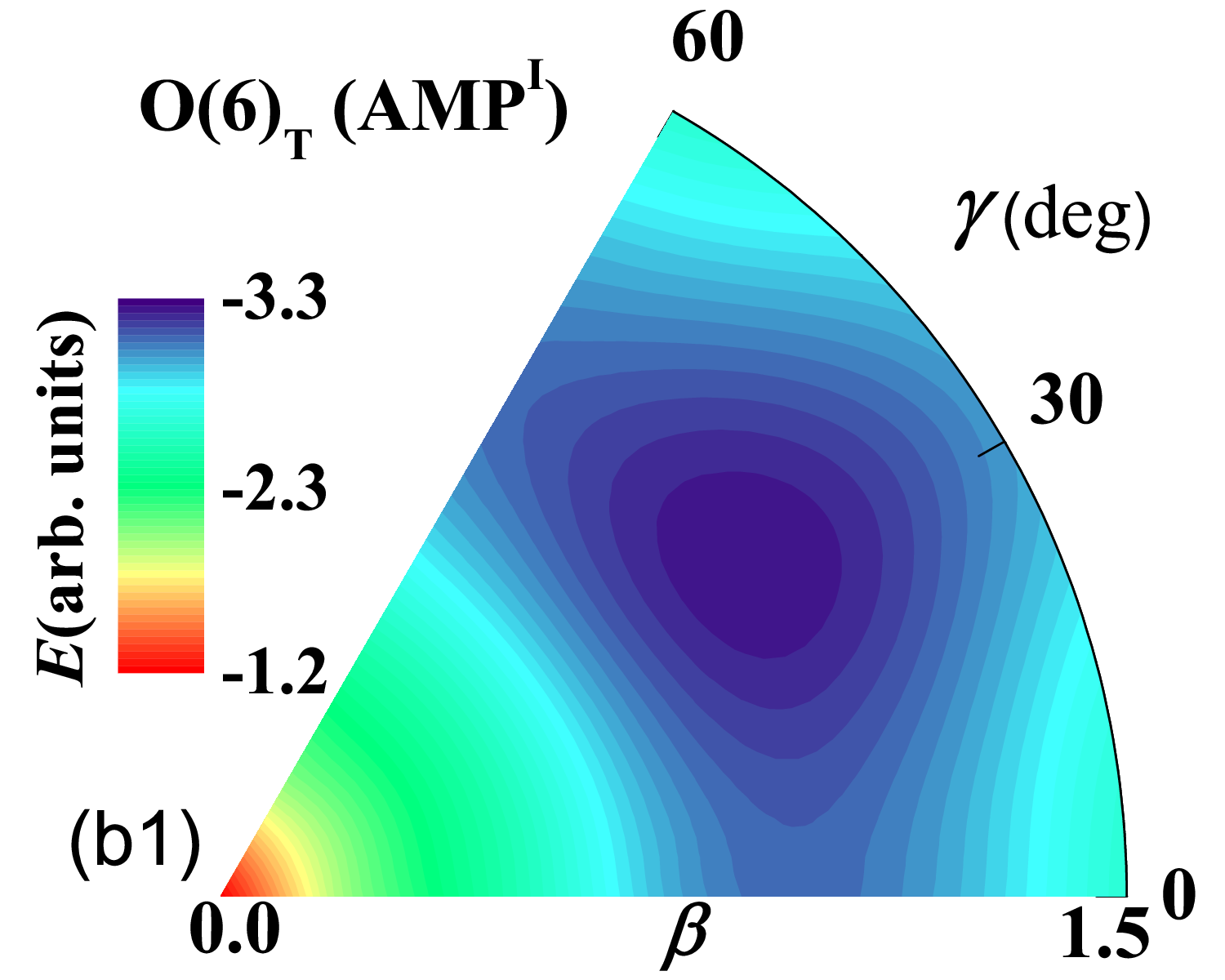}
\includegraphics[scale=0.15]{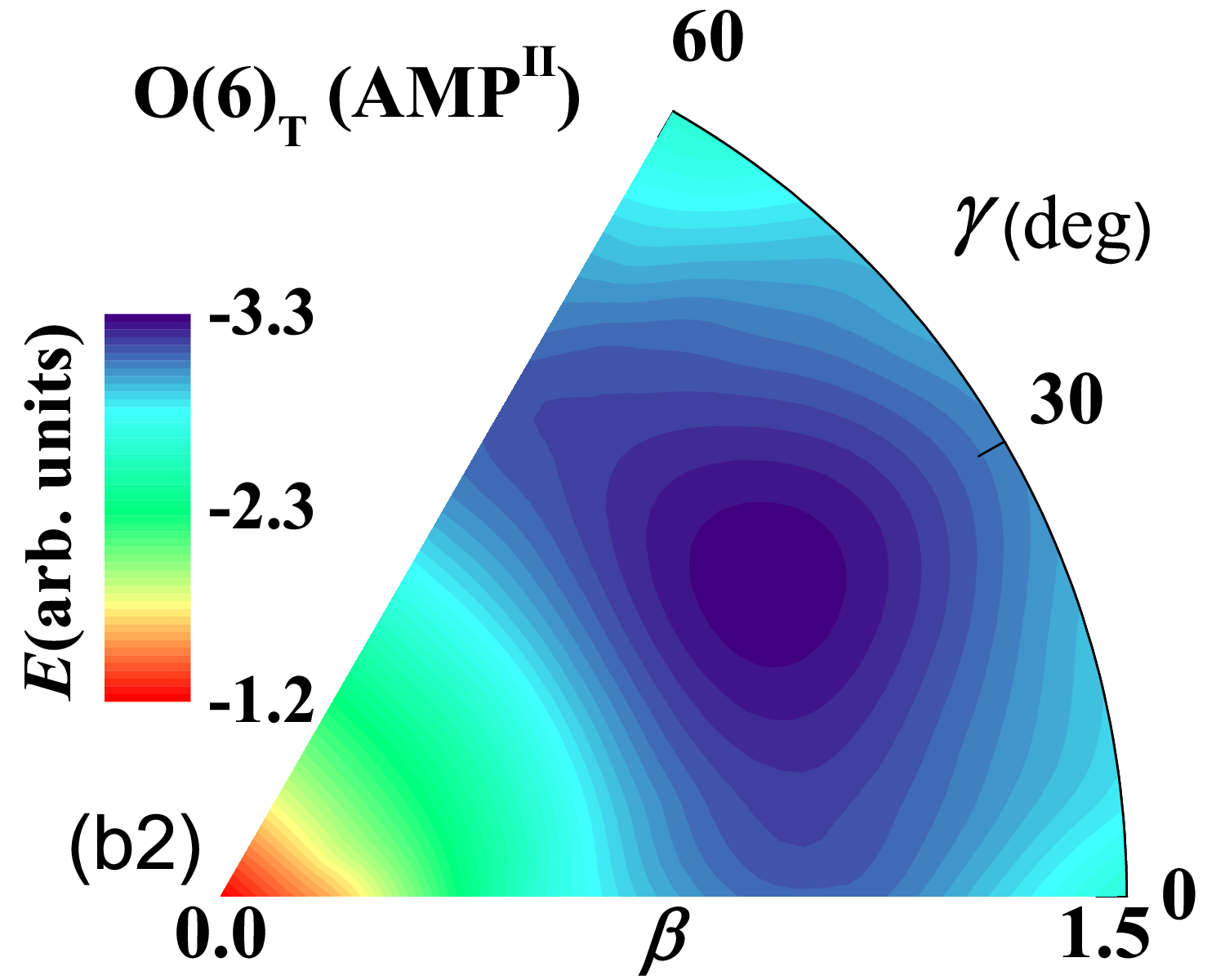}
\caption{(Color online) Contour plots of the potential energy surfaces (in arbitrary units) for $J=4$, computed using the two AMP methods, are presented for the cases labeled (e) and (f) in Table~\ref{T1}, denoted here as O(6) and O(6)$_\mathrm{T}$, respectively. \label{F3}}
\end{center}
\end{figure}

To further check AMP calculations, the O(6) limit and its triaxial extension (denoted $\mathrm{O}(6)_\mathrm{T}$), which correspond to maximal triaxiality at the mean-field level, are employed as illustrative cases for comparing the potential surfaces generated by the two AMP methods. Note that the parameters used for the two cases are identical to those specified in entries (e) and (f) of Table~\ref{T1}. As shown in Fig.~\ref{F3}, the resulting equilibrium $\beta$ and $\gamma$ deformations, as well as the global topographies of the potential energy surfaces, exhibit excellent agreement between the $\mathrm{AMP}^\mathrm{I}$ and $\mathrm{AMP}^\mathrm{II}$ calculations. This consistency supports the validity of the $K=0$ projection approximation for extracting quadrupole deformations in yrast states with $J>0$, a procedure previously employed in the AMP analysis of the IBM~\cite{Kuyucak1987,Kuyucak1987II,Zhang2021,Liu2006,Mu2005}.

\begin{figure}
\begin{center}
\includegraphics[scale=0.28]{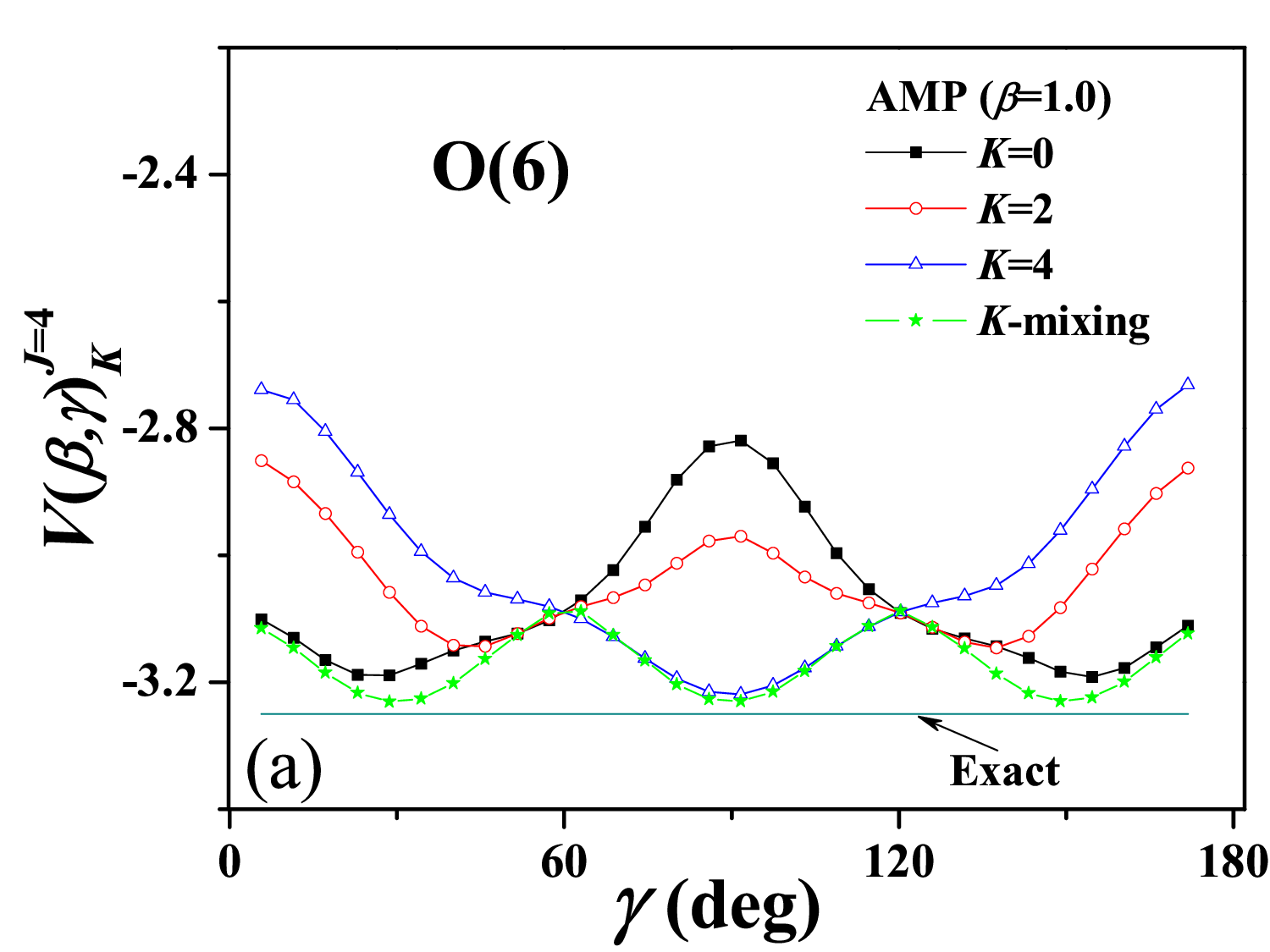}
\includegraphics[scale=0.28]{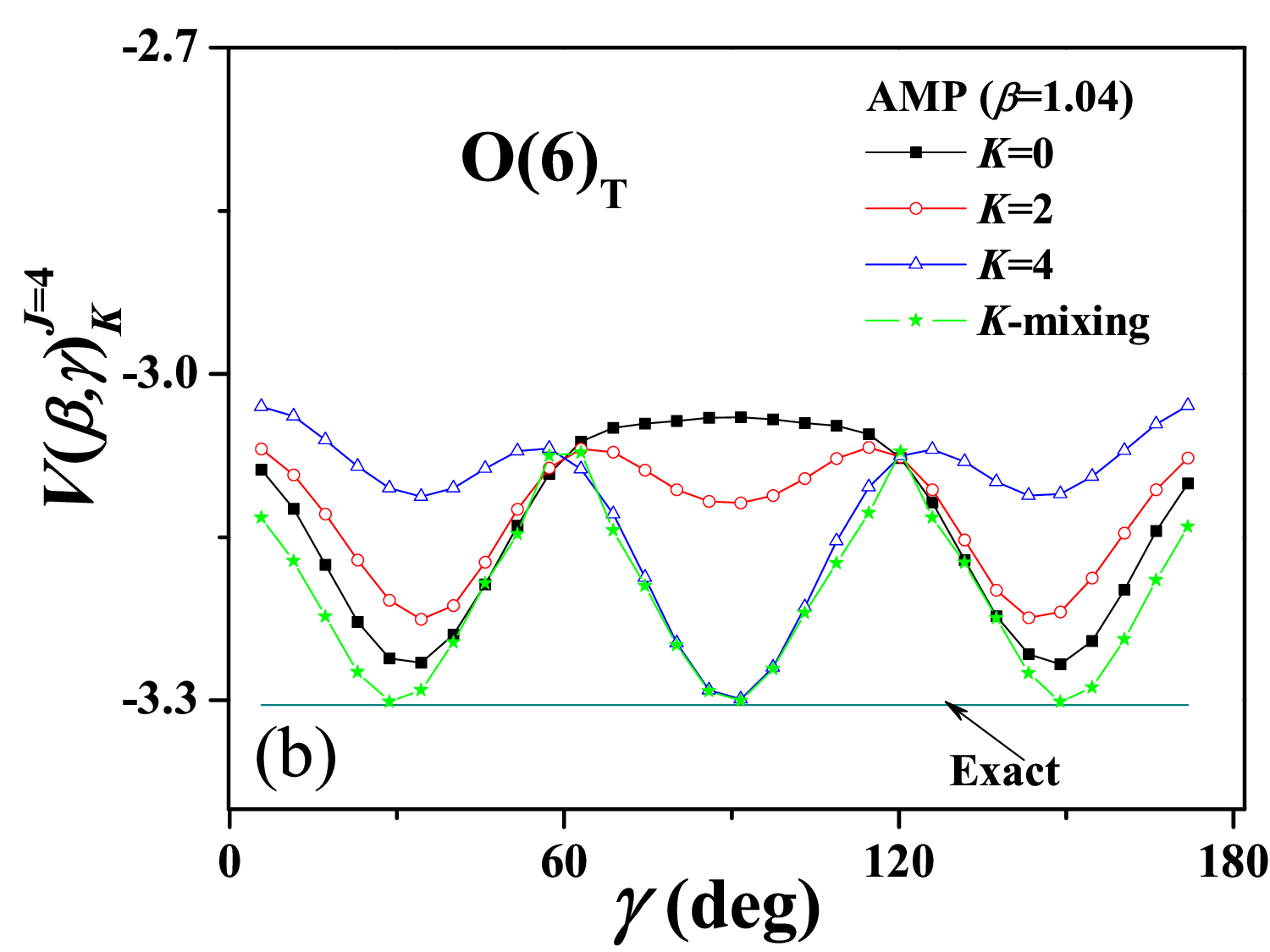}
\caption{(Color online) The $J=4$ potential curves (in arbitrary units) obtained from the $\mathrm{AMP}^\mathrm{I}$ calculations with $K=0,~2$, and $4$ are shown for both the O(6) limit and the O(6)$_\mathrm{T}$ system. For comparison, the corresponding $\mathrm{AMP}^\mathrm{II}$ results are included. In all calculations, $\beta$ is fixed at the equilibrium value $\beta_\mathrm{e}$ determined from the $K=0$ projection. "Exact" refers to the corresponding values listed in Table~\ref{T1}. \label{F4}}
\end{center}
\end{figure}

To examine $\mathrm{AMP}^\mathrm{I}$ for different $K$, the $J=4$ potential energy curves for the O(6) and $\mathrm{O}(6)_\mathrm{T}$ cases are plotted as functions of $\gamma$ in Fig.~\ref{F4}. These curves are obtained from $\mathrm{AMP}^\mathrm{I}$ calculations with $K=0$, $K=2$ and $K=4$, respectively. For comparison, the corresponding results solved from $\mathrm{AMP}^\mathrm{II}$ with $K$-mixing are provided in parallel. In the calculations, the deformation parameter $\beta$ is fixed at its equilibrium value determined from the $K=0$ projection in each case. As shown in Fig.~\ref{F4}, the potential curves with $K=0,~2$, and $4$ exhibit distinct evolutionary behaviors as functions of $\gamma$, yet share the same periodicity $\gamma_\mathrm{Peri.}=90^\circ$. In contrast, the $K$-mixing results retain the $\gamma_\mathrm{Peri.}=60^\circ$ periodicity characteristic of the underlying mean-field potential function given in (\ref{V}), a feature guaranteed by its explicit cos($3\gamma$) dependence.
Moreover, it is shown the global minima of the potential curves with different $K$ occur at different $\gamma$ values. For instance, in the O(6) limit, the minima for $K=0$ and $K=4$ are located at $\gamma=30^\circ$ and $\gamma=90^\circ$, respectively, as observed from Fig.~\ref{F4}(a). Nevertheless, the depths of these global minima are close across all $K$ values. It means that, within the IBM framework, rotational energy spectra obtained from different $K$ projections yield approximately equivalent results.
This observation is consistent with recent shell-model AMP studies~\cite{Gao2022,Lu2025}, which demonstrated that, in the systems studied, different $K$ projections can yield the same results, matching those obtained from the $K$-mixed projection. In the present study, $\mathrm{AMP}^\mathrm{II}$ calculations incorporating $K$-mixing consistently yield modest improvements over the $K$-fixed results, although these improvements remain quantitatively small, as indicated in Fig.~\ref{F4} and Table~\ref{T1}. Nevertheless, $K$-mixing significantly increase computational cost, particularly for systems with large angular momentum $J$ or large boson number $N$. In contrast, adopting a fixed-$K$ approximation in the AMP calculation, as defined by Eq.~(\ref{AMPI}), eliminates numerous intermediate steps and thereby substantially reduces computing time. Moreover, $K$-fixed projection circumvents the numerical instabilities~\cite{Wang2022} commonly encountered when constructing the orthonormalized basis via Eq.~(\ref{basis}). Therefore, the fixed-$K$ projection offers a computationally efficient while sufficiently accurate framework for AMP-based deformation analysis in the IBM.

\begin{figure}
\begin{center}
\includegraphics[scale=0.36]{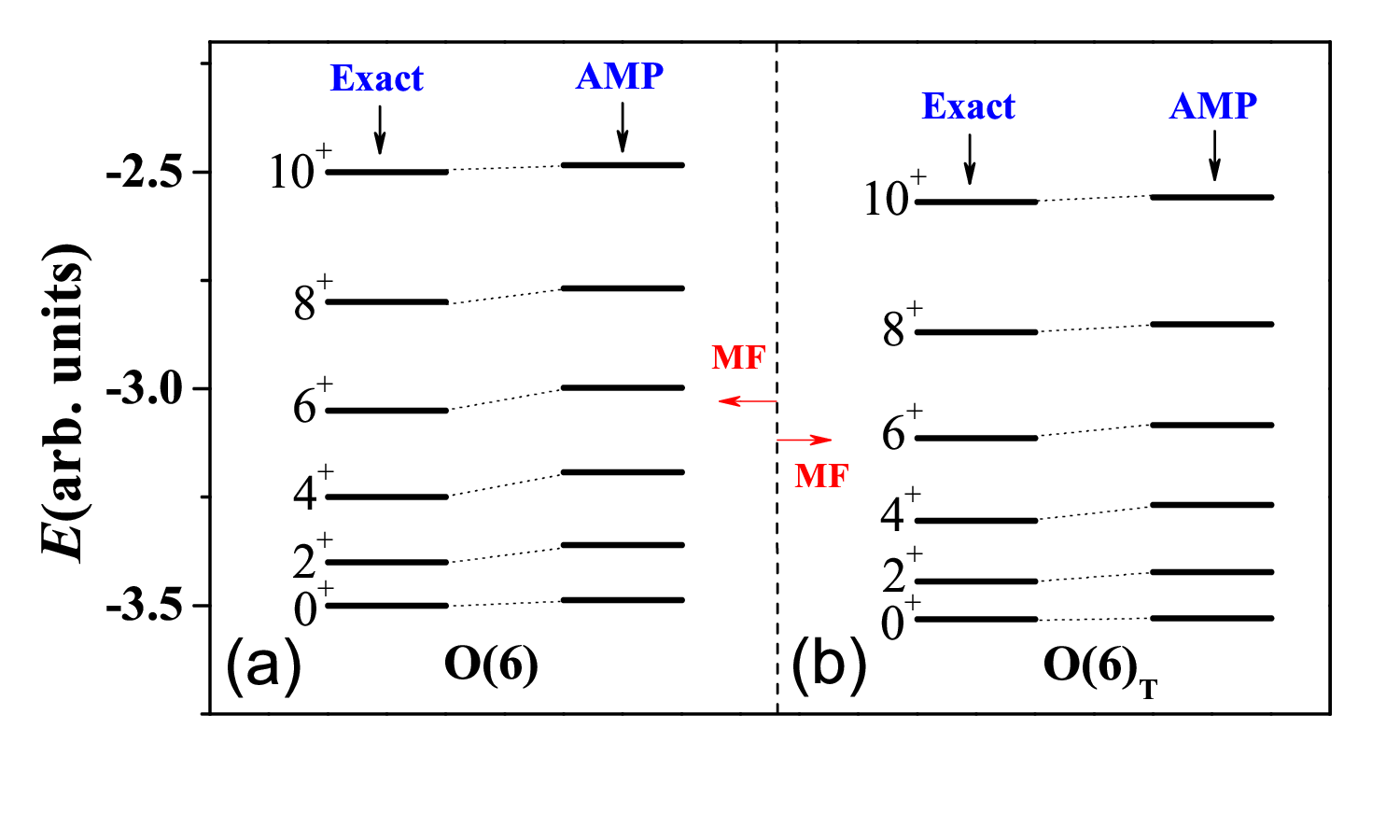}
\caption{(Color online) Yrast level energies (in arbitrary units), obtained from the AMP calculations with $K=0$, are presented for comparison with exact diagonalization results for both the O(6) limit and the O(6)$_\mathrm{T}$ system. The horizontal arrows mark the mean-field (MF) potential minimal values. \label{F5}}
\end{center}
\end{figure}

As further revealed in Fig.~\ref{F4}, $K=0$ consistently emerges as the optimal choice in the AMP$^\mathrm{I}$ calculations, provided our analysis is restricted to understanding the quadrupole geometry within the range of $\gamma\in[0^\circ,60^\circ]$. This is because the resulting $\gamma_\mathrm{e}$ and even $\beta_\mathrm{e}$ (see also Fig.~\ref{F3}) are consistent with those extracted from the $\mathrm{AMP}^\mathrm{II}$ calculations. To assess the overall performance of the $K=0$ projection in reproducing yrast level energies, results for both the O(6) limit and the $\mathrm{O}(6)_\mathrm{T}$ case are presented in Fig.~\ref{F5}, alongside the exact energies obtained from Hamiltonian diagonalization. As shown in Fig.~\ref{F5}, the AMP calculations with $K=0$ accurately reproduce both the ground-state energies, which lie significantly below their unprojected (mean-field) values, and excitation energies in both cases. This agreement confirms that the geometry configurations of the yrast states derived from $K=0$ projection remain physically robust, even in systems displaying triaxiality at the mean-field level.
For all other parameter sets shown in Fig.~\ref{F2}, which display reduced triaxiality, the conclusions drawn from the O(6) and O(6)$_\mathrm{T}$ analyses are further strengthened. Consequently, unless explicitly stated otherwise, we consistently set $K=0$ in all subsequent AMP analyses.

\begin{figure*}
\begin{center}
\includegraphics[scale=0.15]{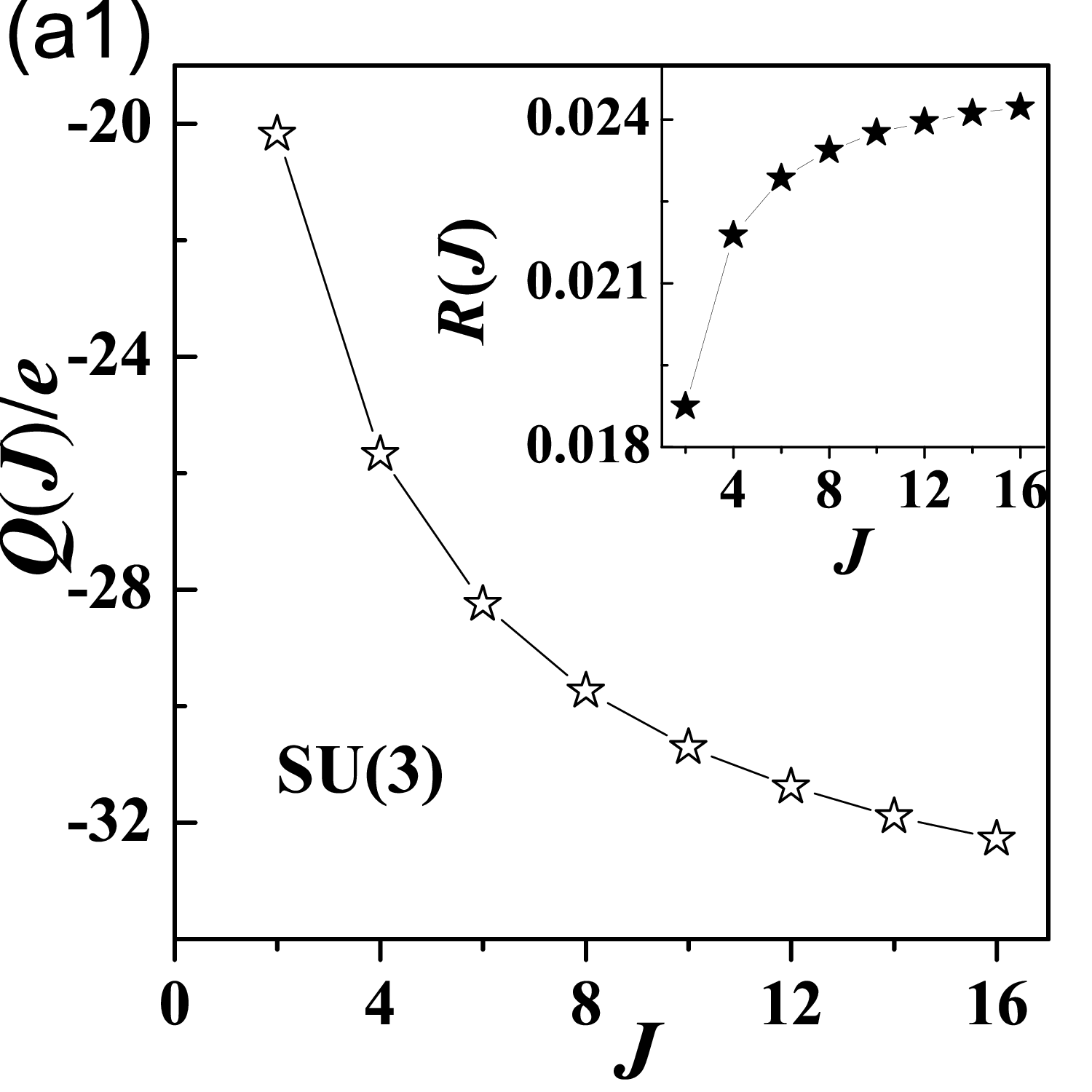}
\includegraphics[scale=0.15]{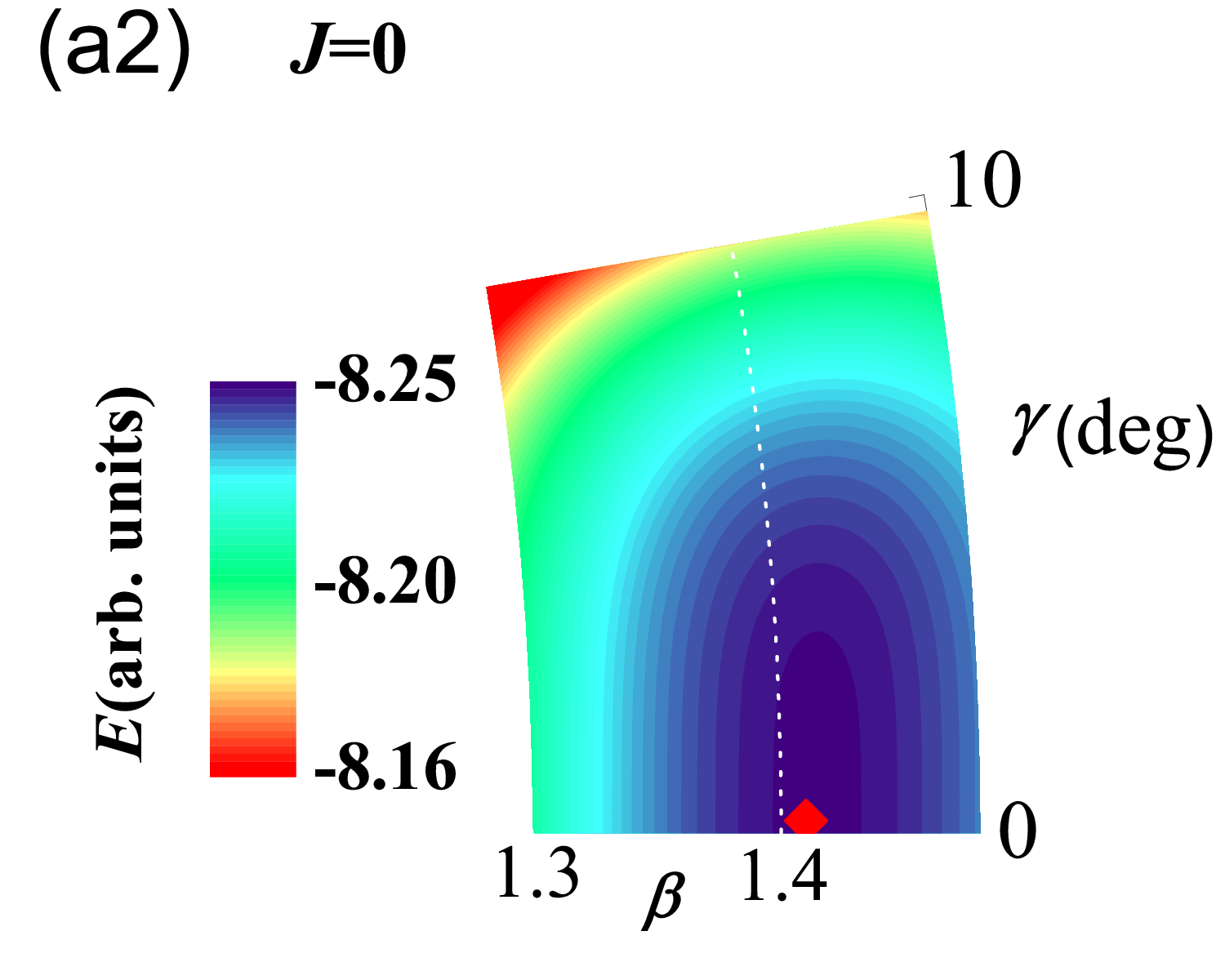}
\includegraphics[scale=0.15]{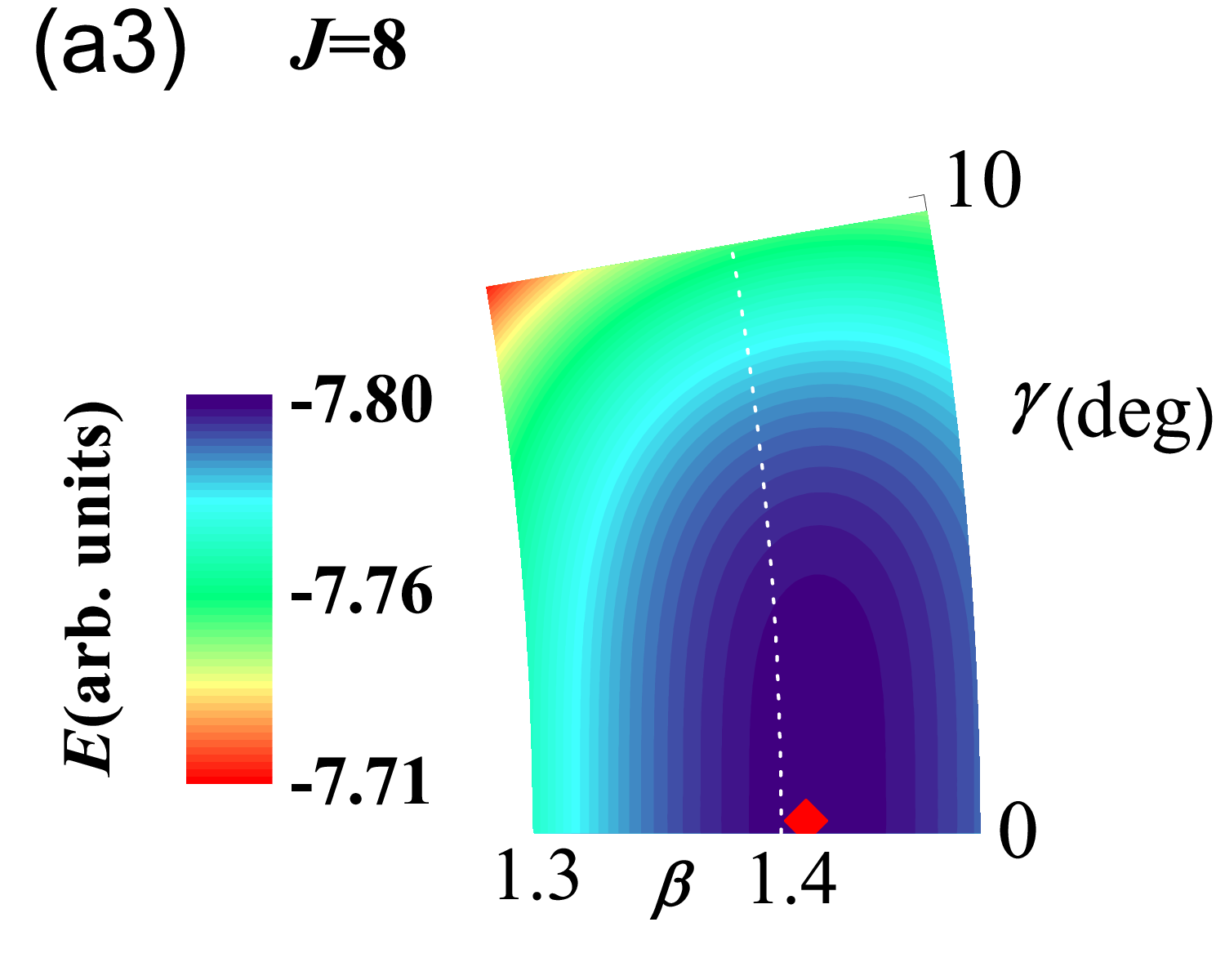}
\includegraphics[scale=0.15]{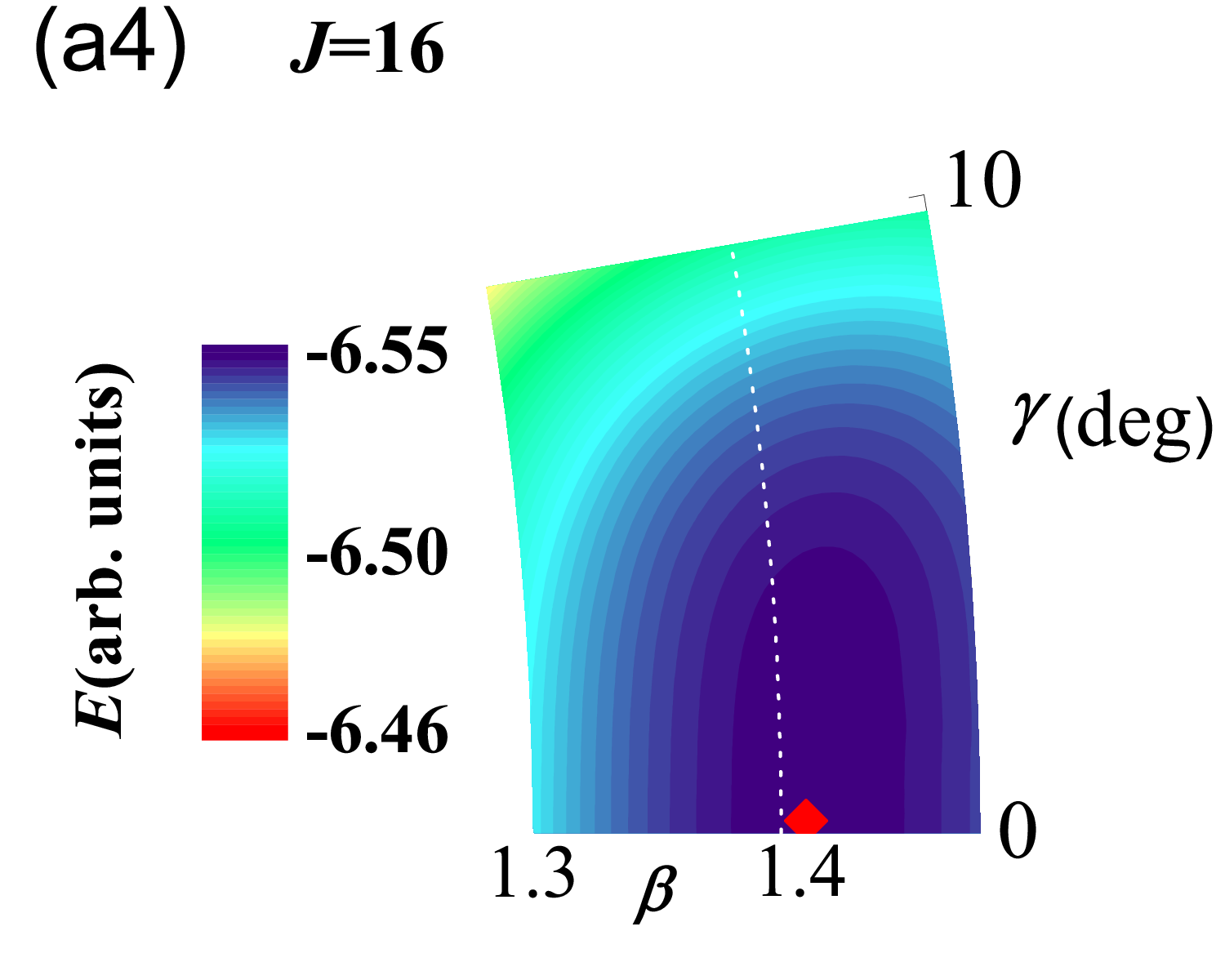}
\includegraphics[scale=0.15]{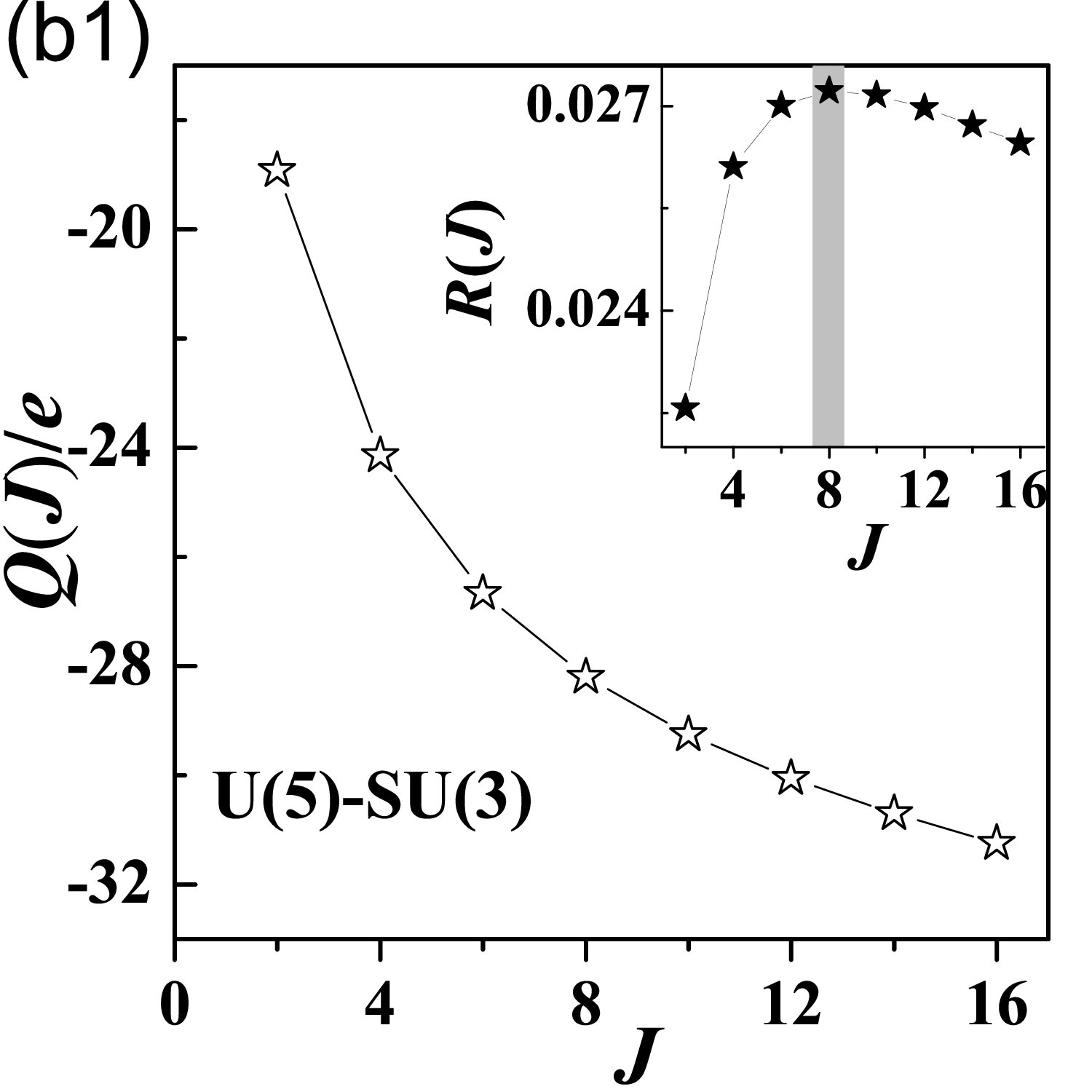}
\includegraphics[scale=0.15]{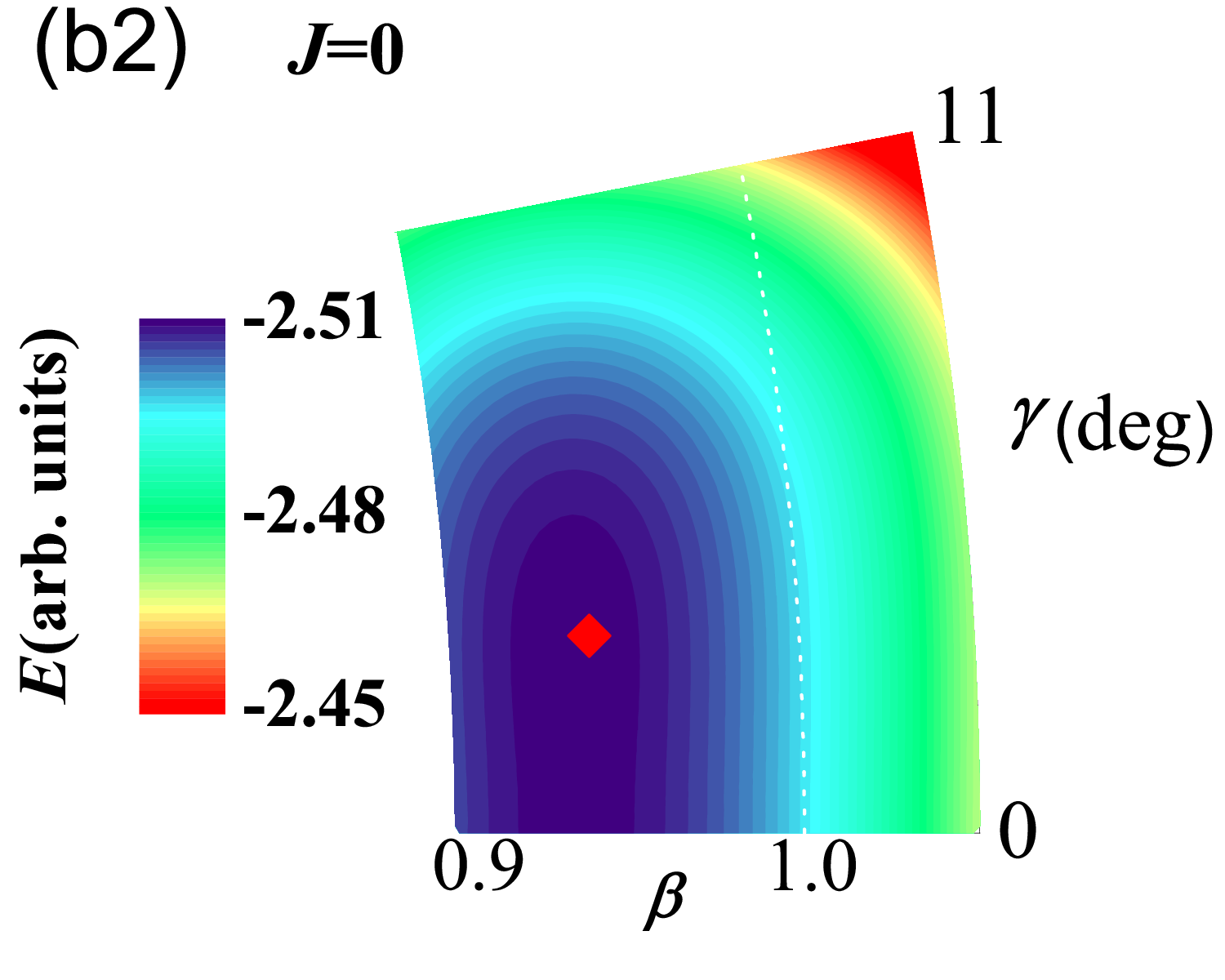}
\includegraphics[scale=0.15]{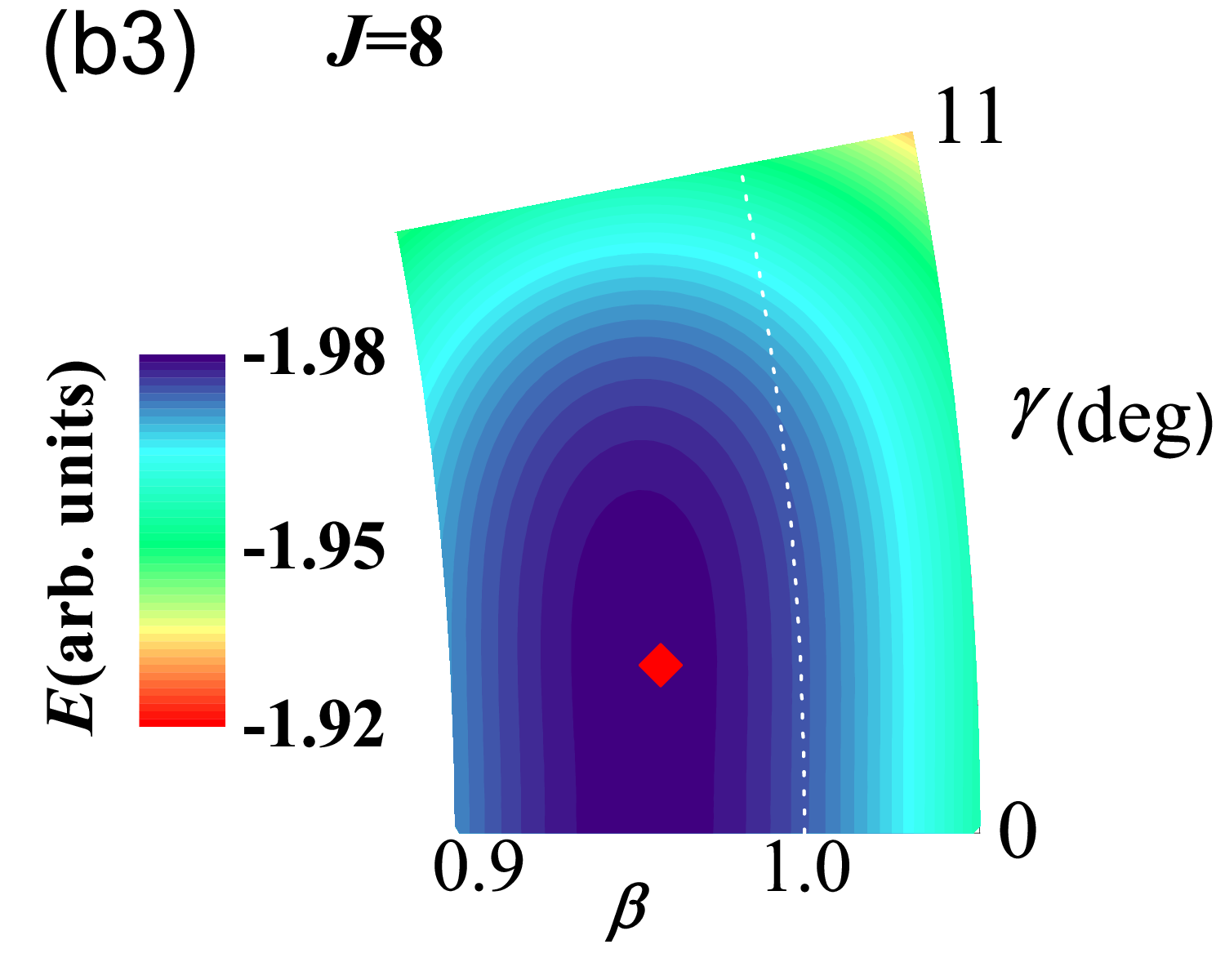}
\includegraphics[scale=0.15]{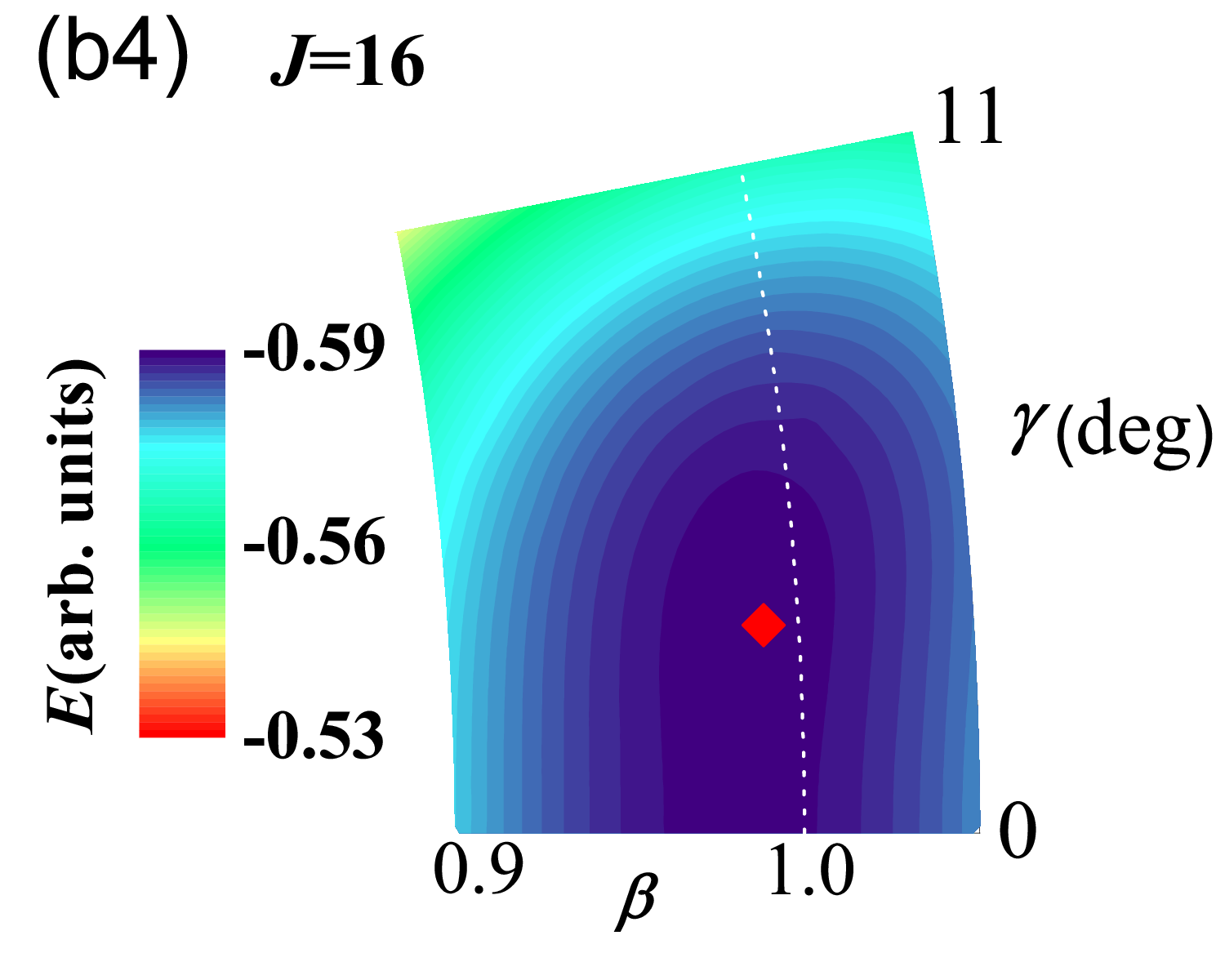}
\includegraphics[scale=0.15]{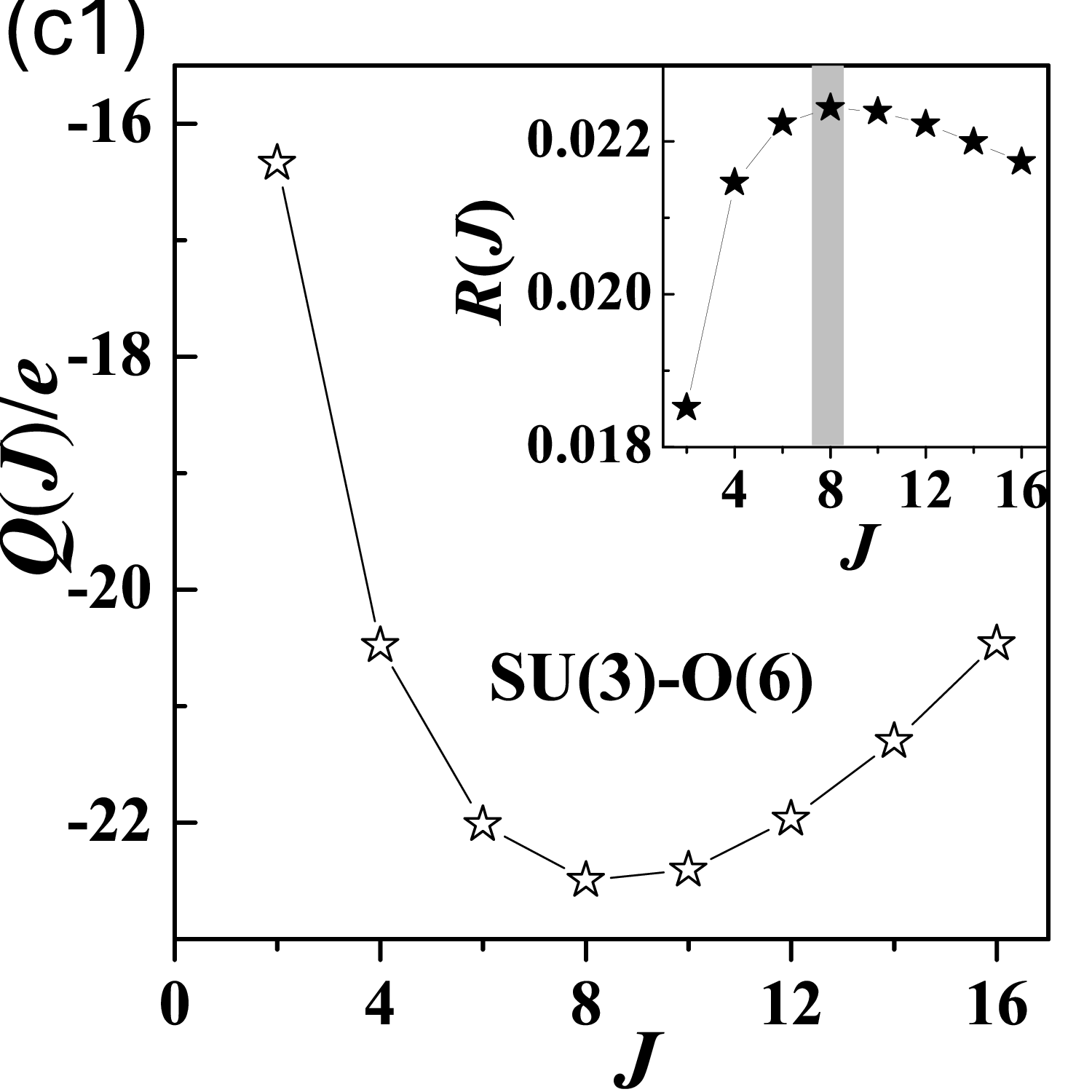}
\includegraphics[scale=0.15]{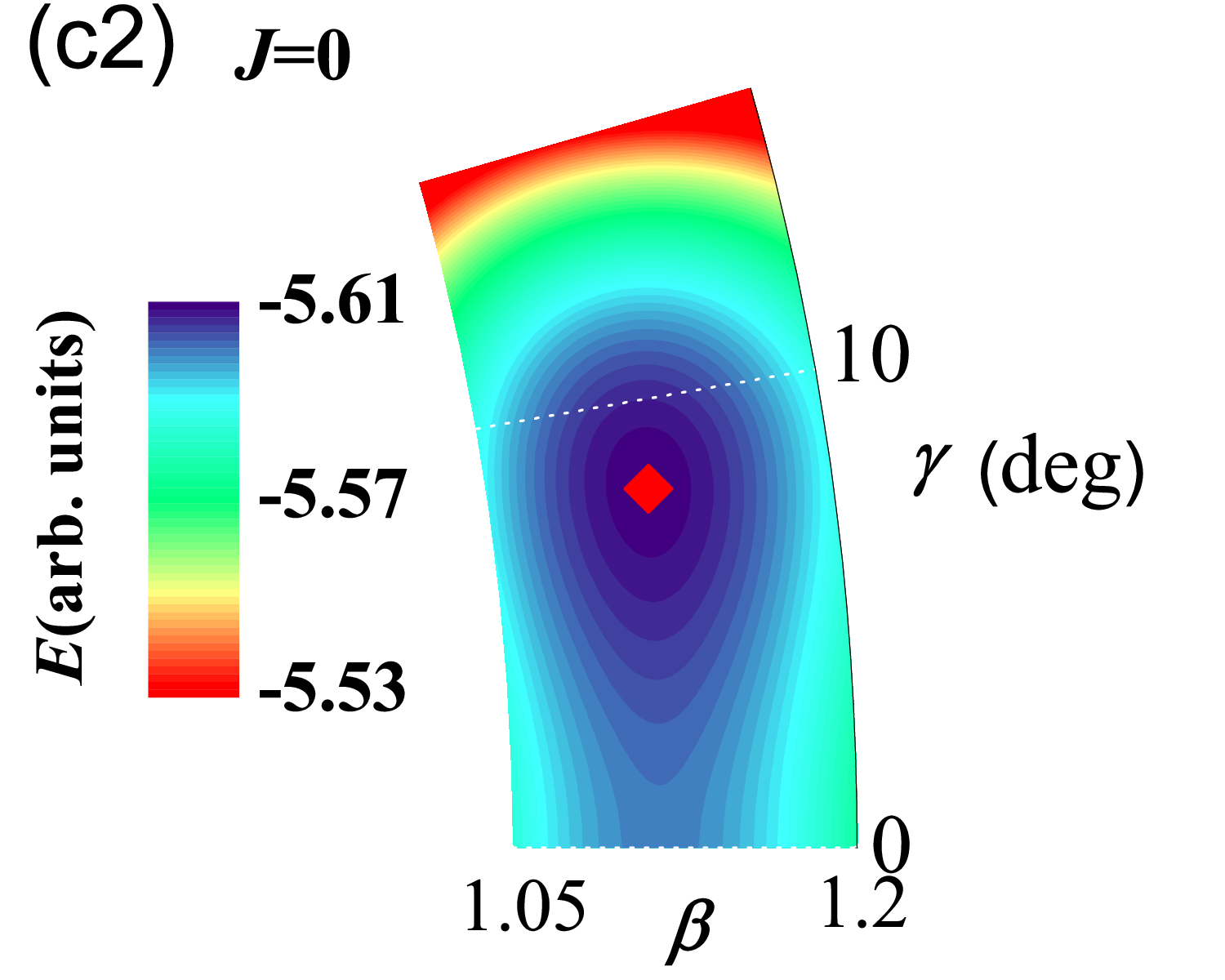}
\includegraphics[scale=0.15]{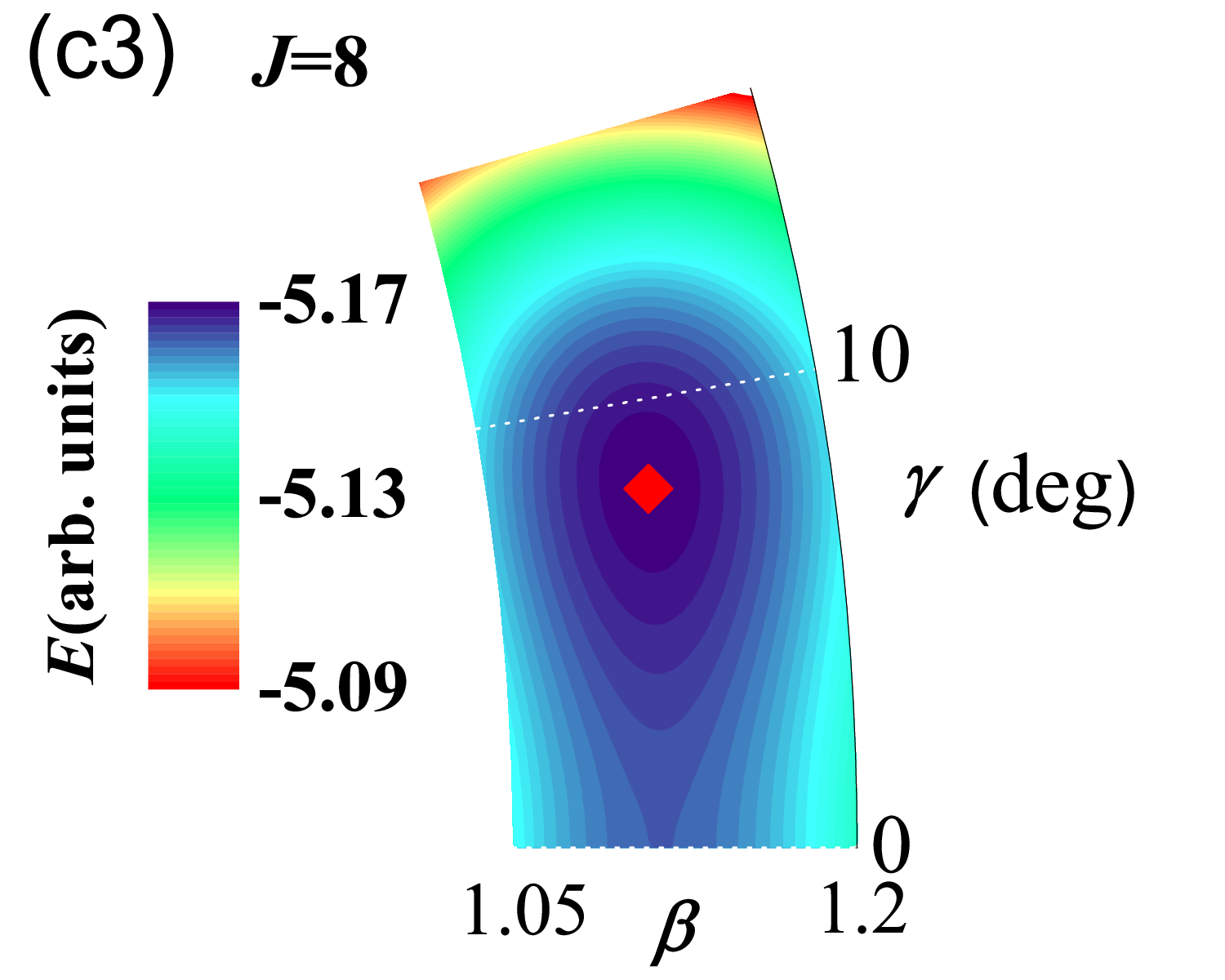}
\includegraphics[scale=0.15]{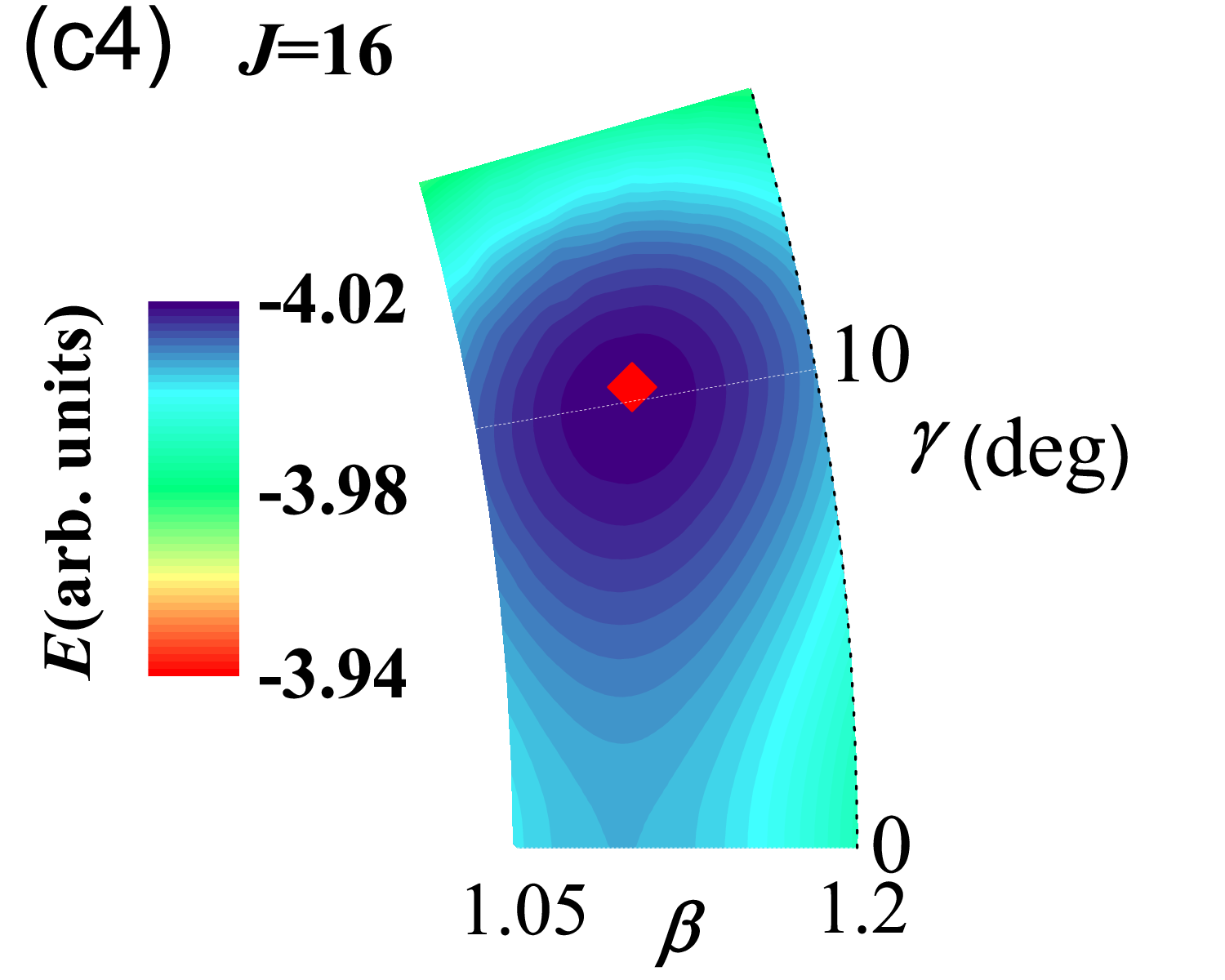}
\caption{(Color online) (a) The evolution of $Q(J)/e$ and $R(J)$ along the yrast line in the SU(3) limit ($N=15$) is shown, together with representative potential energy surfaces (in arbitrary units) for $J=0$, $8$, and $16$. Equilibrium deformations are indicated by red squares, with dotted lines indicating their possible shifts with increasing spin. (b) Same as in (a), but for the parameter point "b", located in the U(5)-SU(3) transitional region. (c) Same as in (b), but for the point "d", located in the SU(3)-O(6) transitional region. Gray bars mark where $R(J)$ evolution changes monotonicity with spin. \label{F6}}
\end{center}
\end{figure*}

\begin{center}
\vskip.2cm\textbf{C. Spin-Driven Deformation Evolution}
\end{center}\vskip.2cm

As established above, spin-dependent quadrupole deformations in the IBM can be extracted via AMP onto the coherent state. Of particular interest are systems exhibiting so-called Jacobi-type transitions,
as proposed in \cite{Zhang2017}. Analogous to the classical Jacobi transitions, from oblate to triaxial shapes, observed in hot rotating nuclei or in rapidly rotating nuclei at typical values of $J\sim60$~\cite{Alhassid1993,Alhassid1986,Ward2002,Mazurek2015,Shanmugam1995,Shanmugam2000}, the Jacobi-type transitions likewise describe a shape (deformation) evolutionary phenomenon, but occurring at significantly lower angular momentum ($J\sim4-10$) and with much smaller deformation amplitudes. Theoretically, Jacobi-type transitions are expected~\cite{Zhang2017,Zhang2021} to occur exclusively in transitional nuclei situated within the deformed region of the phase diagram depicted in Fig.~\ref{F1}. In the following, we select parameter points (b) and (d) to illustrate an AMP analysis for transitional systems.

As shown in Fig.~\ref{F1}, these two points correspond respectively to cases lying along the U(5)-SU(3) and SU(3)-O(6) legs of the phase diagram. For a comparison, the SU(3) limit is also included in the analysis. Specifically, we compute the normalized spectroscopic quadrupole moments $Q(J)/e$ and the E-GOS curves described by $R(J)=\frac{E(J)-E(J-2)}{J}$, a quantity originally introduced by Regan {\it~et~al.}~\cite{Regan2003} to discern structural transitions along the yrast line in vibrational nuclei. As $R(J)\propto\frac{1}{J}\frac{dE(J)}{dJ}$ exhibits greater sensitivity to subtle deformation variations within a rotational band than the excitation energy $E(J)$ alone, it was adopted~\cite{Zhang2017,Zhang2021} as a sensitive indicator of spectral evolution in deformed nuclei associated with Jacobi-type transitions. The resulting $J$-dependent behaviors are presented in Fig.~\ref{F6}. Moreover, to elucidate possible variations in quadrupole deformation, the potential energy surfaces obtained from the $\mathrm{AMP}^\mathrm{I}$ method with $K=0$ are provided for typical $J$ values. To examine the local structure near the minima in greater detail, the contour plots are restricted to a finite region centered at $\beta_\mathrm{e}$ and $\gamma_\mathrm{e}$. The corresponding potential energy values (obtained by MAP) $V(\beta_\mathrm{e},\gamma_\mathrm{e})$ are required to accurately reproduce the exact energy spectrum derived from Hamiltonian diagonalization (see Table~\ref{T2}). In the calculations, the boson number is set to $N=15$, consistent with the earlier study of Jacobi-type transitions reported in~\cite{Zhang2021}, where analogous analyses were performed on the evolution of E-GOS curves and the absolute amplitudes of the quadrupole moments, $|Q(J)|$.

As shown in Fig.~\ref{F6}(a1), the spectroscopic quadrupole moments of the yrast states in the SU(3) limit are all negative and decrease monotonically with increasing $J$, whereas the corresponding E-GOS curve increases monotonically (see the inset). These features are fully consistent with the prolate-rotor geometry conventionally associated with the SU(3) limit. As further observed from Fig.~\ref{F6}(a2)-(a4), the results indicate that the quadrupole deformation for different $J$ stabilizes at $\beta_\mathrm{e}\approx1.4$ and $\gamma_\mathrm{e}=0^\circ$, further confirming the prolate-rotor picture anticipated in the SU(3) limit.

Upon moving to the transitional system at the parameter point "b", Fig.~\ref{F6}(b1) reveals that $Q(J)/e$ values continue to decrease monotonically with increasing $J$ like in the SU(3) limit. In contrast, the corresponding E-GOS curve exhibits a non-monotonic behavior near $J=8$, signaling the onset of a Jacobi-type transition. As further observed from Fig.~\ref{F6}(b2)-(b4), the results indicate that the equilibrium deformations in the projected potentials stabilize within $\beta_\mathrm{e}\approx0.94-0.96$ for $J\leq8$, then increase to nearly $\beta_\mathrm{e}\approx1.0$ at $J=16$. Throughout this evolution, the triaxial parameter is confined around $\gamma_\mathrm{e}\sim3.0^\circ$. At the parameter point (d), Fig.~\ref{F6}(c1) shows that both $Q(J)/e$ and $R(J)$
display non-monotonic behaviors as functions of $J$, implying a qualitatively different manifestation of Jacobi-type transition mechanism. Consistent with this, Fig.~\ref{F6}(c2)-(c4) demonstrates that the $\gamma$ deformations evolve from $\gamma_\mathrm{e}\approx8^\circ$ ($J\leq8$) to $\gamma_\mathrm{e}\approx10^\circ$ ($J=16$), while $\beta_\mathrm{e}$ remains essentially unchanged. The enhancement in axial asymmetry at high angular momentum aligns quantitatively with the observed reduction in the absolute magnitude of $Q(J)$, as illustrated in Fig.~\ref{F6}(c1). Collectively, these findings imply that Jacobi-type transitions in the IBM systems may be closely related to rotational stretching of the quadrupole deformation, namely $\beta$-stretching in the U(5)-SU(3) transitional region and $\gamma$-stretching in the SU(3)-O(6) transitional region. Crucially, $\beta$ and $\gamma$ simultaneously serve as the classical order parameters for the respective ground-state shape transitions across these two regions~\cite{CJC2010,CJ2009}. This further strengthes the connection between Jacobi-type transitions and the associated ground-state shape transitions~\cite{Zhang2017,Zhang2021}.

Since Jacobi-type transitions are theoretically confined to the deformed region of the triangle phase diagram~\cite{Zhang2021}, their rotational stretching behavior can be qualitatively understood based on a simple rotor-model assumption~\cite{Li2016}. The triaxial rotor Hamiltonian can be written as
\begin{eqnarray}\label{rot}
\hat{H}_{\mathrm{rot}}=\sum_{k=1,2,3}\frac{\hat{J}_k^2}{2\Gamma_k}\, ,
\end{eqnarray}
where $\hat{J}_k$ with $k=1,~2,~3$ (or $x,~y,~z$) represent the projection of the angular momentum onto the body-fixed $k$ axis, and $\Gamma_k$ denotes the corresponding moment of inertia (MOI) defined by~\cite{Bohrbook}
\begin{eqnarray}\label{MOI}
\Gamma_k=\Gamma_0\beta^2\mathrm{sin}^2\Big(\gamma-\frac{2}{3}k\pi\Big)\, ,
\end{eqnarray}
with $\Gamma_0$ serving as an overall scale parameter.
Within the rotor model framework, stretching in either $\beta$ or $\gamma$ induces a direct modification of the MOIs. This becomes evident upon substituting $\beta=\beta_\mathrm{e}$ and $\gamma=\gamma_\mathrm{e}$ into Eq.~(\ref{MOI}). Notably, the scaling discrepancy in the deformation parameter $\beta$ between the geometric collective model and the IBM can be absorbed into the scale factor $\Gamma_0$, which is, in turn, determined by the experimentally measured or theoretically predicted $E(2_1^+)$ value. Consequently, an increase in $\beta$ leads only to a uniform rescaling of the MOIs, leaving the rotor wave functions unchanged for a given $J$. This feature provides a natural explanation for the pattern observed in Fig.~\ref{F6}(b1): the spectroscopic quadrupole moments $Q(J)$ continue to decrease monotonically, like in the SU(3) limit, whereas $R(J)$ exhibits a non-monotonic variation beyond $J=8$, reflecting the enhanced MOIs at higher angular momentum. In contrast, an increase in $\gamma$ alters both the magnitude and the anisotropic distribution of the MOIs, thereby modifying the rotor wave functions. This insight helps interpret the evolution shown in Fig.~\ref{F6}(c): for $J>8$, an enhancement of $\gamma$ deformation, i.e., increased triaxiality, leads to a reduction in the absolute magnitudes of both $Q(J)$ and $R(J)$.

\begin{table}
\scriptsize
\caption{The excitation energies (in arbitrary units) $E(J_1^+)-E(0_1^+)$ for yrast states, obtained via exact diagonalization of the IBM Hamiltonian ($N=15$) at parameter point "b" (U(5)-SU(3) transitional region) and parameter point "d" (the SU(3)-O(6)) transitional region), both labeled in Fig.~\ref{F6}, are tabulated for comparison with those predicted by the AMP and the rotor model calculations. Rotor$^\mathrm{I}$ and Rotor$^\mathrm{II}$ denote results obtained using variable and fixed MOIs, respectively (see the parameter illustration in the text).}\label{T2}
\begin{tabular}{cccccc}\hline\hline
U(5)-SU(3)&$J=2$&$J=6$&$J=8$&$J=14$&$J=16$ \\ \hline
Exact&0.045&0.311&0.529&1.499&1.922\\
AMP&0.045&0.311&0.529&1.497&1.921\\
Rotor$^\mathrm{I}$&0.045&0.309&0.523&1.452&1.858\\
Rotor$^{\mathrm{II}}$&0.045&0.315&0.542&1.579&2.045\\
$(\beta_\mathrm{e},\gamma_\mathrm{e})$&$(0.938,3.0^\circ)$&$(0.948,3.0^\circ)$&$(0.955,3.1^\circ)$&$(0.979,3.5^\circ)$&$(0.985,3.6^\circ)$\\
 \hline
SU(3)-O(6)&$J=2$&$J=6$&$J=8$&$J=14$&$J=16$ \\ \hline
Exact&0.037&0.256&0.436&1.234&1.582\\
AMP&0.037&0.256&0.436&1.235&1.582\\
Rotor$^\mathrm{I}$&0.037&0.258&0.442&1.270&1.621\\
Rotor$^{\mathrm{II}}$&0.037&0.258&0.442&1.273&1.641\\
$(\beta_\mathrm{e},\gamma_\mathrm{e})$&$(1.12,8.0^\circ)$&$(1.12,8.0^\circ)$&$(1.12,8.1^\circ)$&$(1.12,9.7^\circ)$&$(1.12,10.4^\circ)$\\
\hline\hline
\end{tabular}
\end{table}

In fact, a quantitative assessment of the stretching effects can also be achieved under the assumption of the rotor model by converting changes in quadrupole deformations into corrections to the excitation energies. Using the rotor Hamiltonian (\ref{rot}) with deformation parameters $\beta=\beta_\mathrm{e}(J)$ and $\gamma=\gamma_\mathrm{e}(J)$ extracted from the projected potentials,
one can compute the yrast energies $E(J_1^+)$ for the two transitional cases illustrated in Fig.~\ref{F6}. For a quantitative comparison, the $E(J_1^+)$ values solved from the IBM via exact diagonalization and the AMP method ($K=0$) are listed in Table~\ref{T2}, alongside corresponding rotor model calculations. To isolate the influence of deformation stretching effects, two distinct rotor model implementations are presented: Rotor$^\mathrm{I}$ employs equilibrium deformation parameters $(\beta_\mathrm{e},\gamma_\mathrm{e})$, extracted from the AMP calculations, for each individual state in the MOIs expressions; Rotor$^\mathrm{II}$ instead fixes the deformation parameters $(\beta_\mathrm{e},\gamma_\mathrm{e})$ in MOIs at the ground-state values, identical to those determined for the $J=2_1^+$ state, as displayed in Table~\ref{T2}. Furthermore, the overall scaling factor in MOIs is determined such that the calculated energy spacing $E(2_1^+)-E(0_1^+)$ reproduces the exact IBM results (see Table~\ref{T2}): this yields $\Gamma_0=101.3$ for parameter point "b" (U(5)-SU(3)) and $\Gamma_0=89.5$ for parameter point "d" (SU(3)-O(6)).

As shown in Table~\ref{T2}, the exact diagonalization results for both cases are reproduced with high fidelity by the AMP calculations, confirming that the extracted equilibrium deformation parameters $(\beta_\mathrm{e},\gamma_\mathrm{e})$ provide a physically meaningful geometric description of the yrast states in the IBM. Moreover, the table reveals that rotor model calculations also yield
a globally accurate description of the excitation energies $E(J_1^+)$, while incorporating deformation-stretching into MOIs clearly improve the rotor model's agreement with the IBM spectra, especially for states with $L>8$. This supports that the yrast level sequences in the present cases are reasonably approximated by a rotor mode, despite that the presence of deformation-stretching already implies a certain degree of intrinsic softness in the rotating IBM systems. The $R(J)$ values corresponding to the calculated $E(J_1^+)$ can be directly extracted from Table~\ref{T2}. The stretching effects associated with the Jacobi-type transitions depicted in Fig.~\ref{F6} can then be quantified via the difference $\Delta R=\frac{R(16)-R(8)}{R(8)-R(2)}$. In the strict SU(3) limit, it is straightforward to deduce that $\Delta R$ should be strictly positive, as no deformation stretching occurs in this case. This is consistent with the trend observed in the inset of Fig.~\ref{F6}(a1). In contrast, for parameter points "b" and "d", Rotor$^{\mathrm{I}}$ calculations, employing variable MOIs, yield negative values, giving $\Delta R=-0.32$ and $\Delta R=-0.24$, respectively. These values are in reasonable agreement with the exact IBM results, $\Delta R=-0.17$ and $-0.19$. By comparison, Rotor$^{\mathrm{II}}$ calculations, where $\beta$ and $\gamma$ in MOIs have been fixed at their $J=0$ values, yield instead two positive values, $\Delta R=+0.13$ and $\Delta R=+0.5$, deviate markedly from the IBM results. Such differences in rotor model calculations actually reflet that the deformation-stretching effects indeed influence spectral evolution, even when the associated deformation change appears relatively small, e.g., $\Delta\gamma=2^\circ\sim3^\circ$ at the parameter point "d" (see Table~\ref{T2}). Collectively, these results demonstrated that rotational stretching of the quadrupole deformation provides a physically transparent explanation for the Jacobi-type transitional behavior in the yrast spectra observed in Fig.~\ref{F6}.

Similarly, the stretching effects on the spectroscopic quadrupole moments $Q(J)$ can also be quantified within the rotor model framework, in which the $E2$ transitional operator is defined as~\cite{Li2016}
\begin{eqnarray}
\hat{T}(E2)^{\mathrm{rot}}=e^\prime\beta[\mathrm{cos}(\gamma)D_{u,0}^{2}+\frac{1}{\sqrt{2}}\mathrm{sin}(\gamma)(D_{u,2}^{2}+D_{u,-2}^{2})]\, ,
\end{eqnarray}
where $e^\prime$ denotes the effective charge. Analogous to the quantity $\Delta R$, we define $\Delta Q=\frac{Q(16)-Q(8)}{Q(8)-Q(2)}$ as a dimensionless measure of rotational stretching effects. Rotor$^{\mathrm{I}}$ calculations yield values of opposite sign at parameter points "b" and "d", namely $\Delta Q=+0.257$ and $\Delta Q=-0.276$, respectively. These values are in good agreement with the IBM results, $\Delta Q=+0.328$ and $-0.329$, extracted from Fig.~\ref{F6}(b1) and Fig.~\ref{F6}(c1), respectively. In contrast, Rotor$^{\mathrm{II}}$ calculations yield instead $\Delta Q=+0.263$ and $\Delta Q=+0.097$ for the two cases. This difference reaffirms the pivotal role of stretching effects in distinguishing and understanding the two classes of Jacobi-type transitions.

We emphasize that deformation stretching discussed here is not an independent phenomenon, but rather a dynamic manifestation of intrinsic deformation characterized by a small degree of softness, i.e., reduced rigidity.
In other words, the rotational-stretching effects are physically meaningful only when all yrast states can be adequately described within a collective rotational framework, as exemplified by the cases shown in Fig.~\ref{F6}. In such cases, dynamical effects of rotation can be qualitatively estimated using a rotor (or rotor-like) model, as outlined above. By contrast, for intrinsically soft systems, such as the parameter point "a" shown in Fig.~\ref{F1} and Fig.~\ref{F2}, the concept of deformation stretching provides no particular advantage over the well established notions of $\gamma$-softness or $\beta$-softness, even if the equilibrium deformation varies with angular momentum. Therefore, the above analysis of Jacobi-type shape transitions applies exclusively to nuclear systems possessing a very deformed ground-state configuration at the mean-field level.

\begin{center}
\vskip.2cm\textbf{IV. Application to Relevant Phenomena}
\end{center}\vskip.2cm

\begin{figure*}
\begin{center}
\includegraphics[scale=0.16]{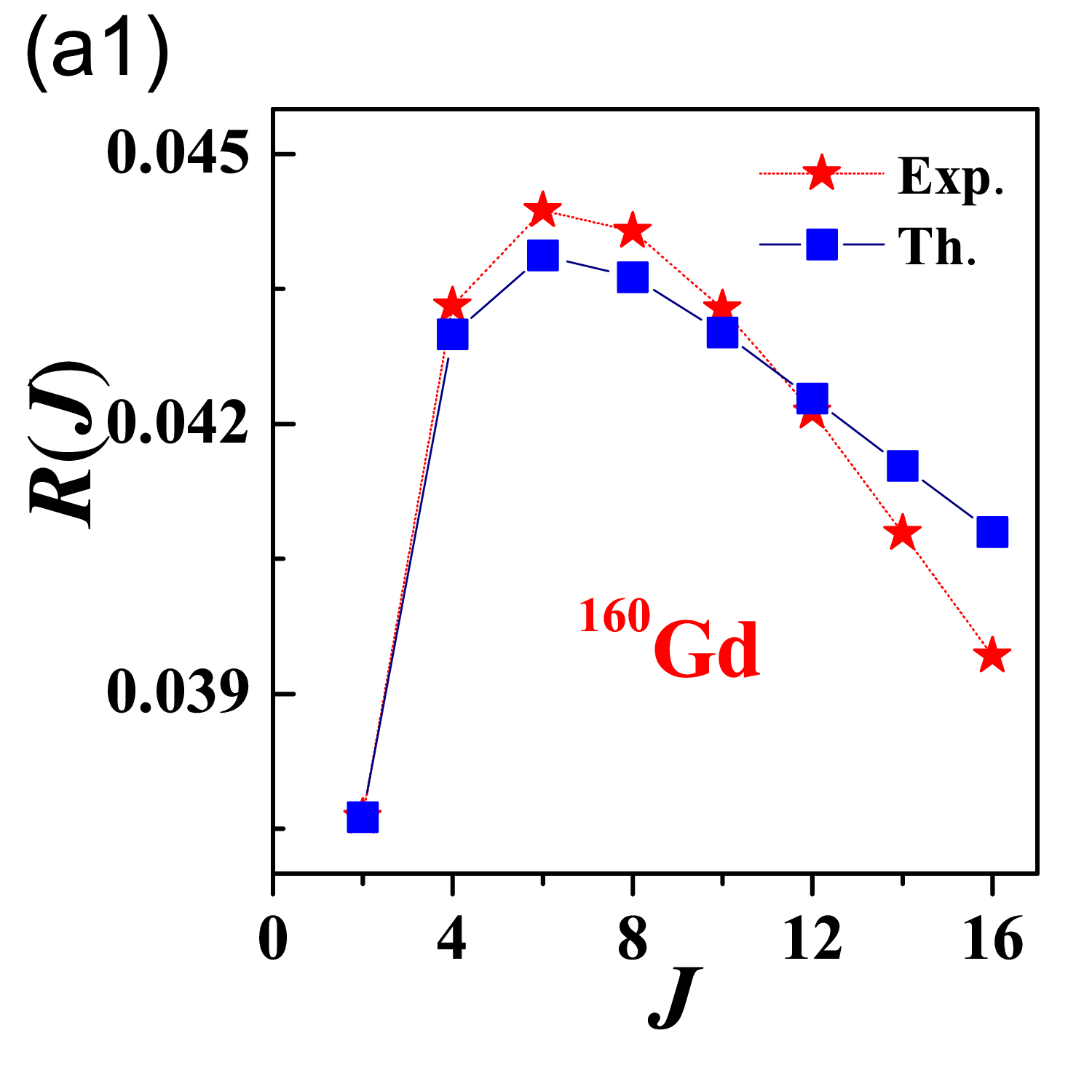}
\includegraphics[scale=0.16]{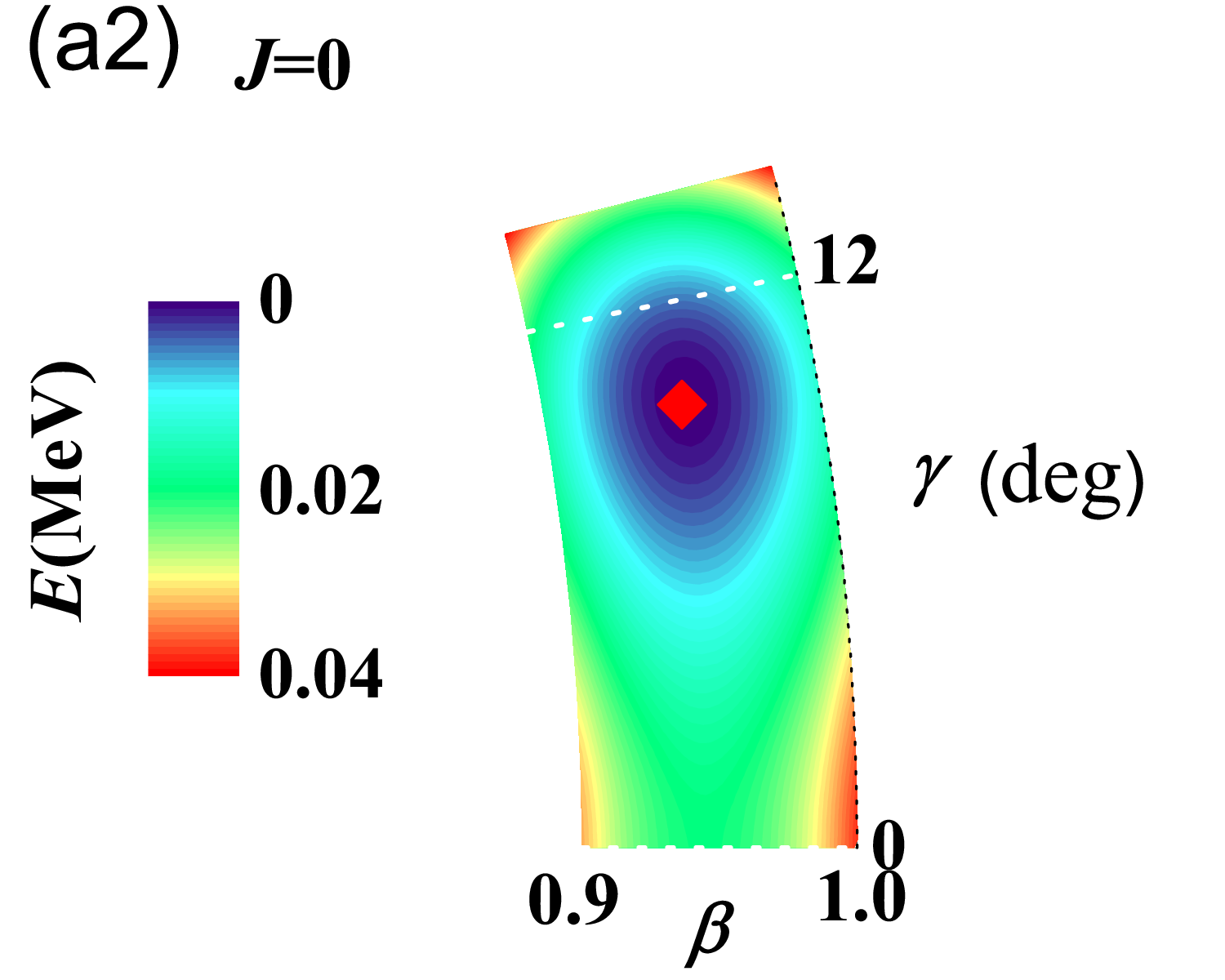}
\includegraphics[scale=0.16]{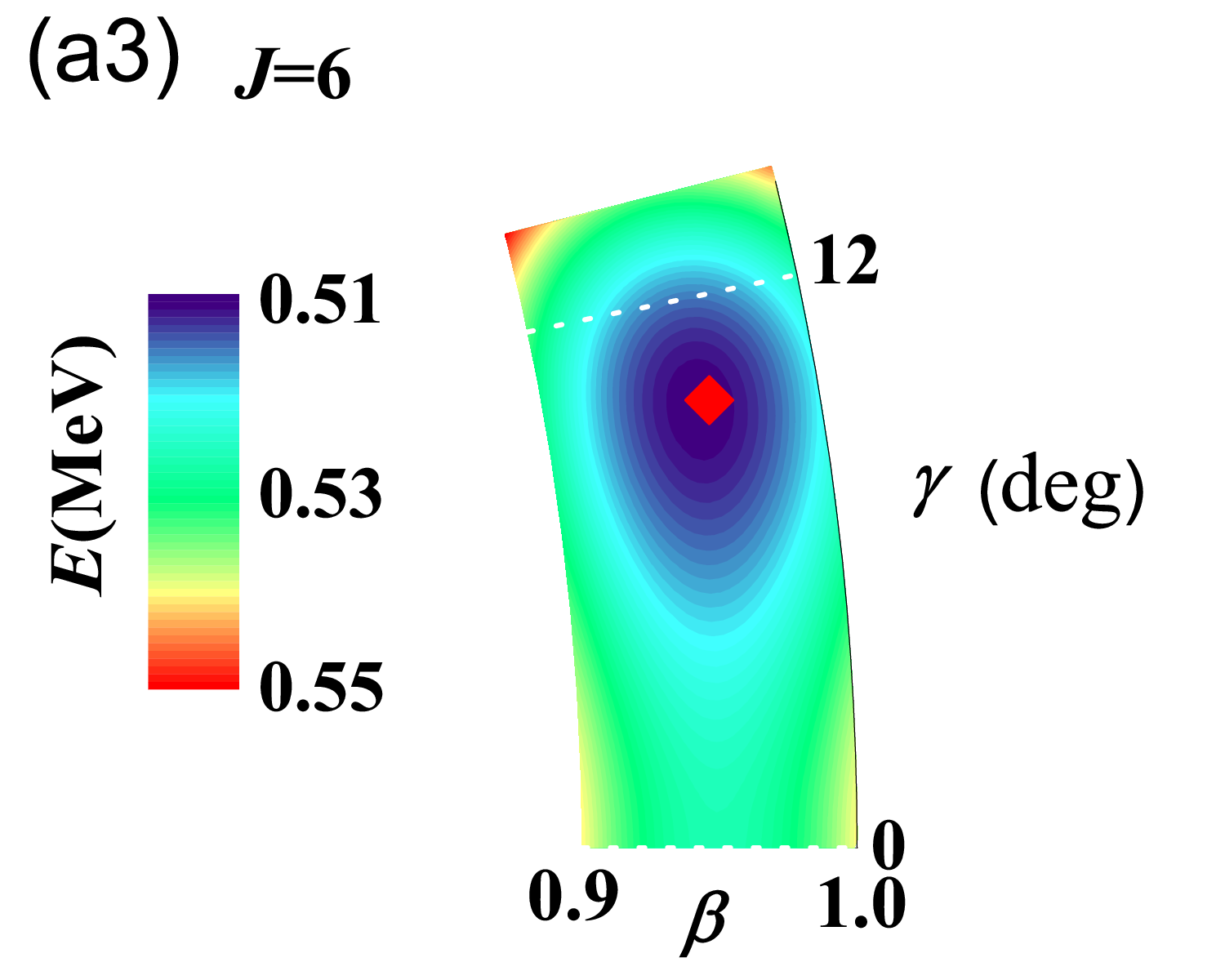}
\includegraphics[scale=0.16]{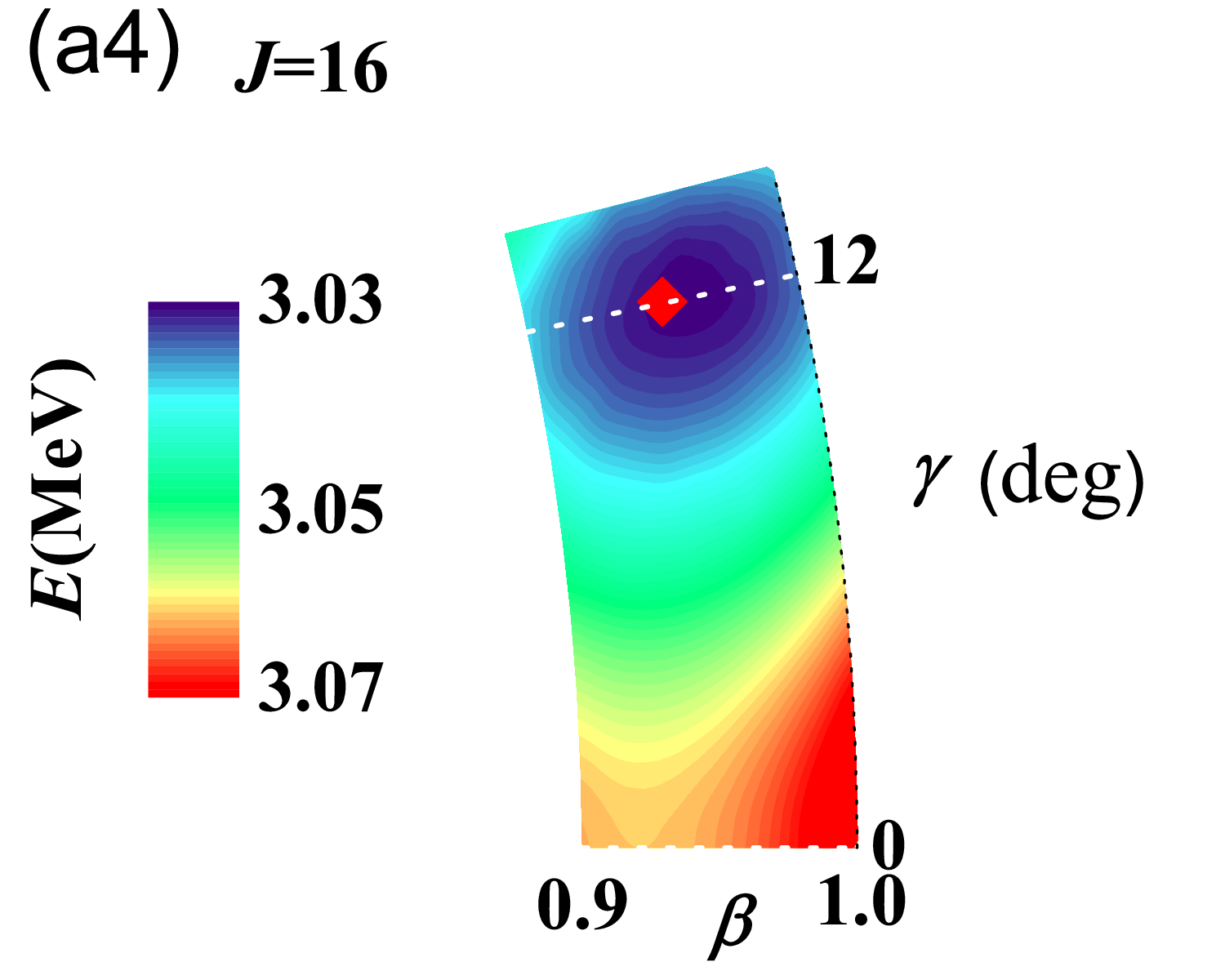}
\includegraphics[scale=0.16]{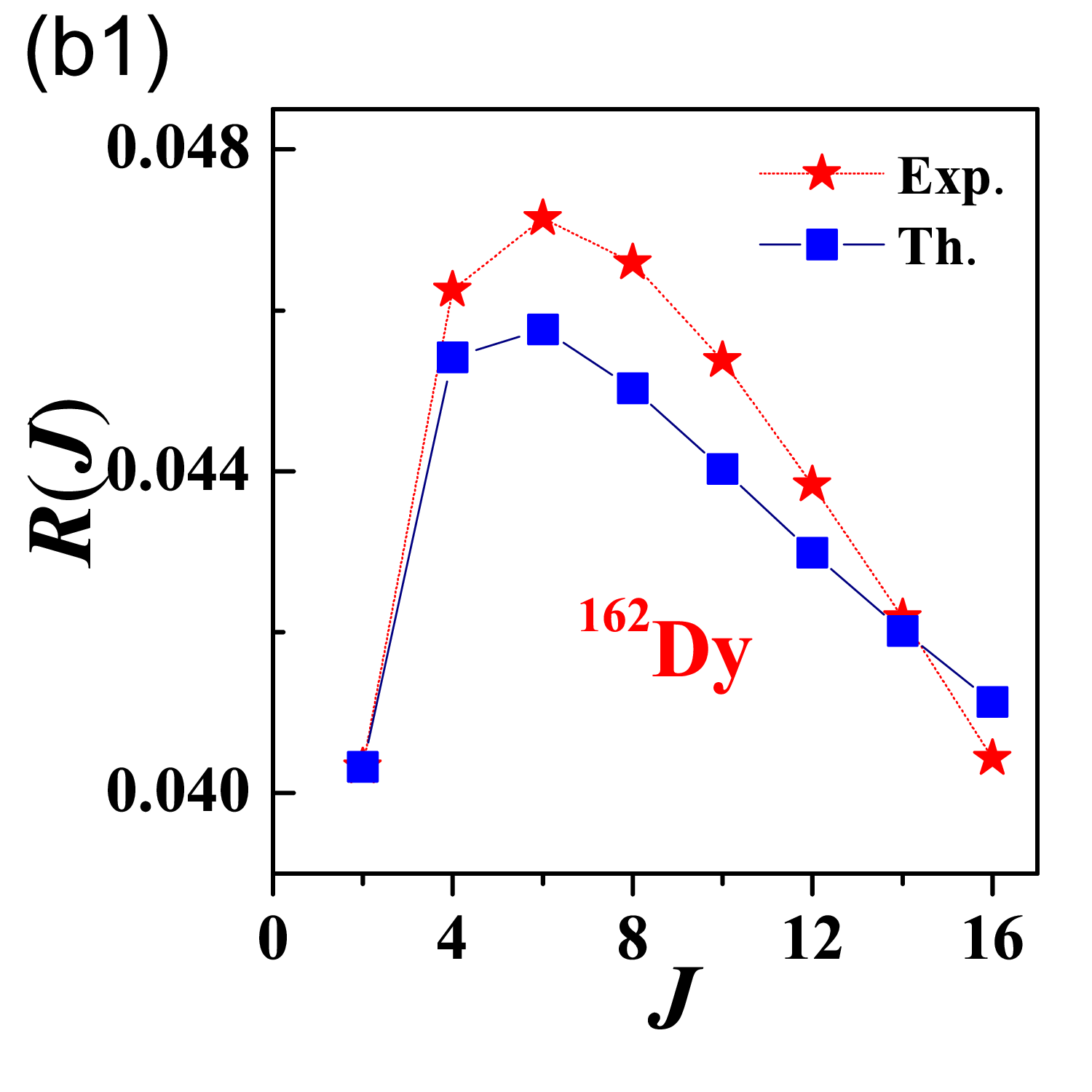}
\includegraphics[scale=0.16]{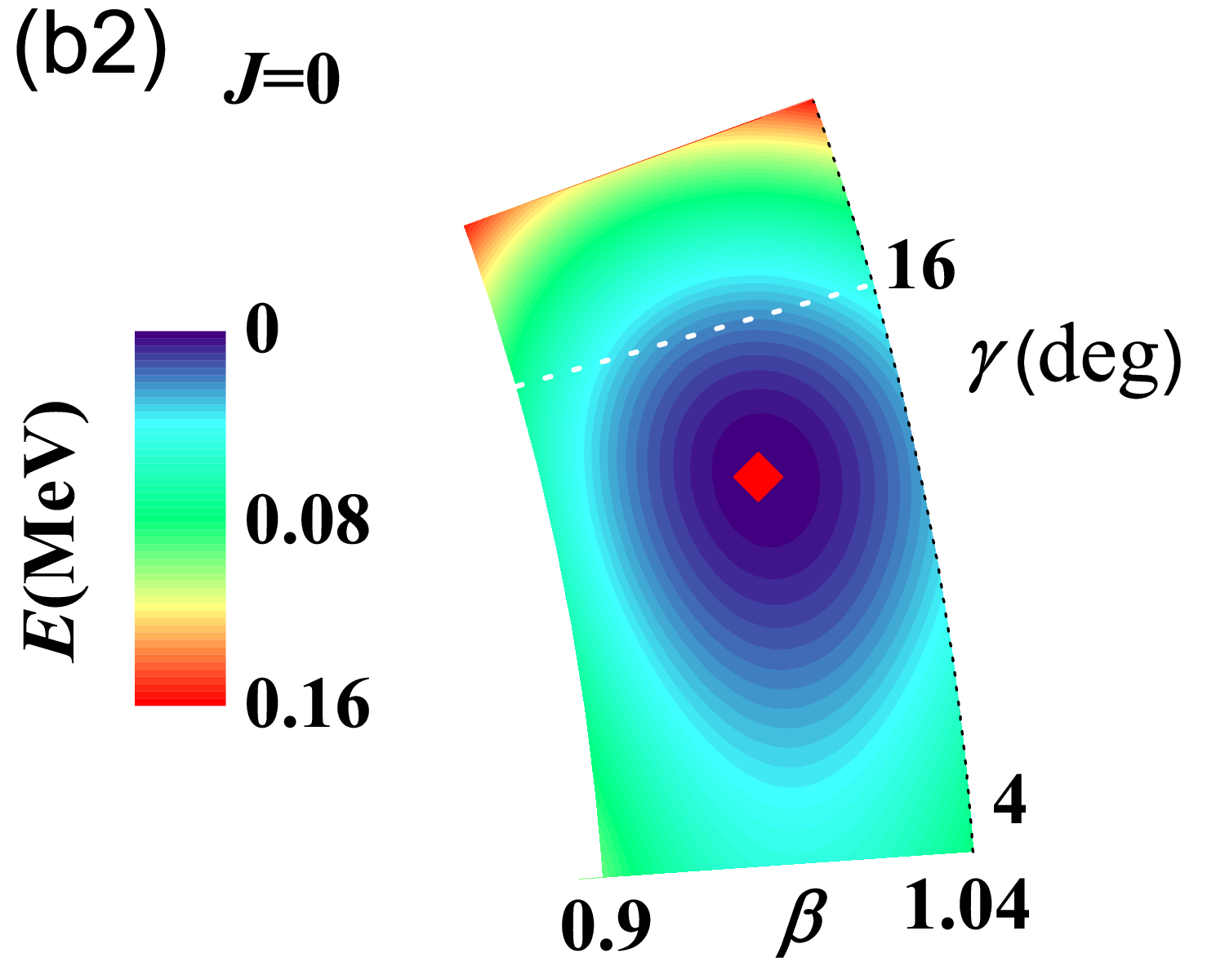}
\includegraphics[scale=0.16]{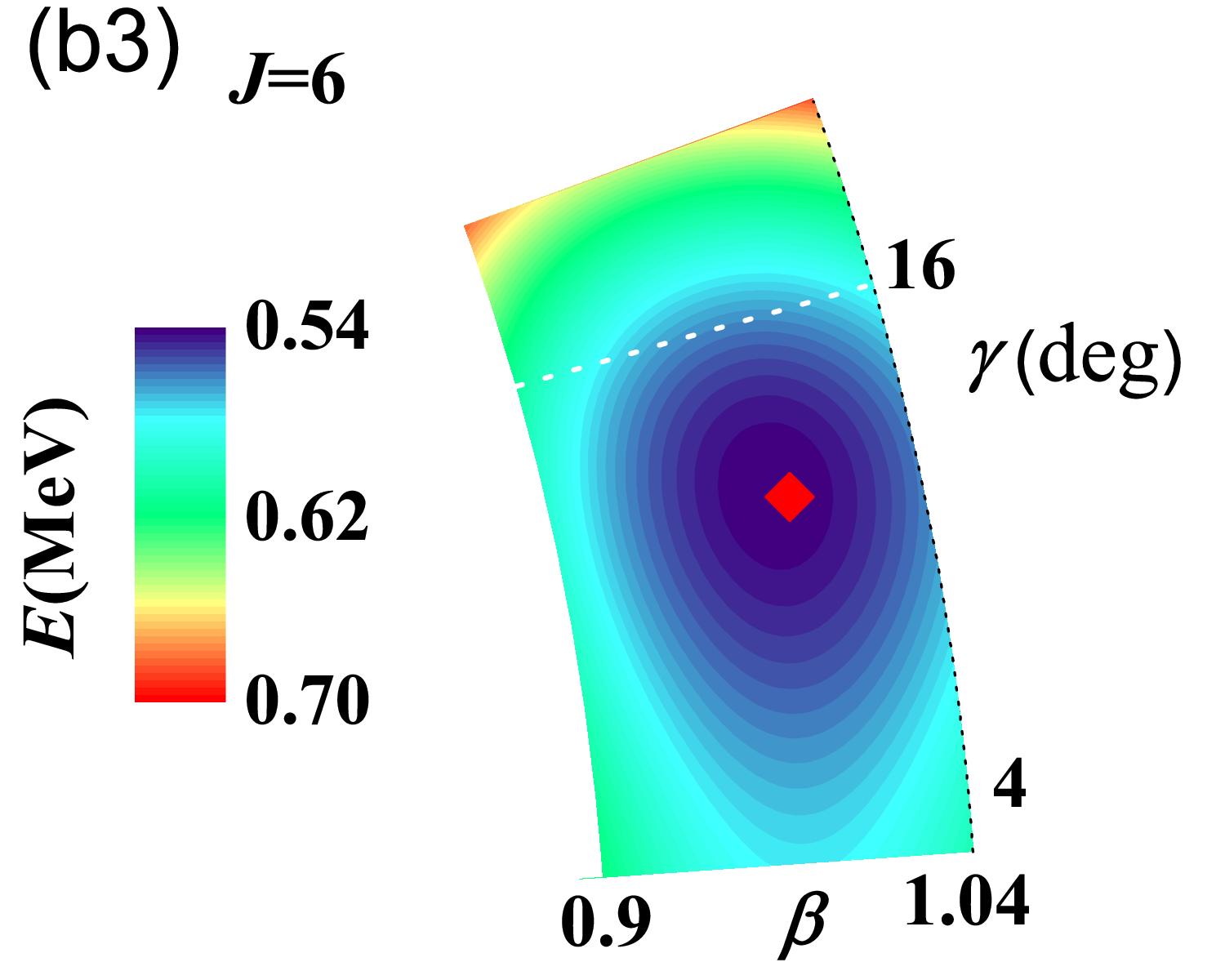}
\includegraphics[scale=0.16]{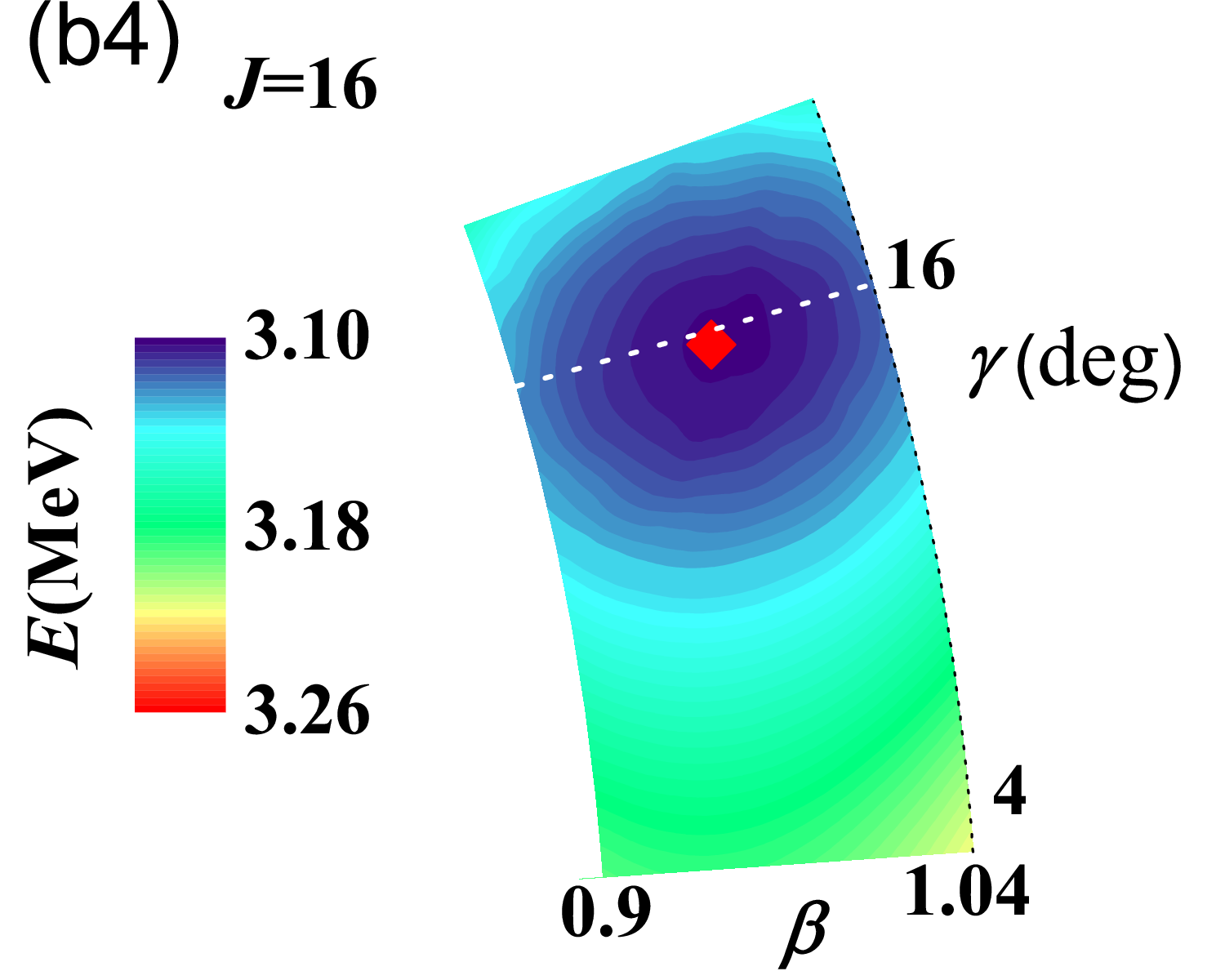}
\caption{(Color online) (a1) $R(J)$ values (unit in MeV/$\hbar$) extracted from yrast-level data in $^{\mathrm{160}}$Gd~\cite{Reich2005} are compared with theoretical predictions of the consistent-$Q$ Hamiltonian (\ref{CQ}), using parameters from \cite{McCutchan2004}. (a2) Using the same parameter set employed for $^{\mathrm{160}}$Gd, the potential energy surface at $J=0$, evaluated in the vicinity of the equilibrium deformation ($\beta_\mathrm{e}$ and $\gamma_\mathrm{e}$), with the red square marking its location. (a3) Same as in (a2), but for $J=6$. (a4) Same as in (a2), but for $J=16$. (b1)-(b4) Same as in (a1)-(a4), but for $^{\mathrm{162}}$Dy~\cite{Reich2007}. All potential energy surfaces are referenced to the minimum of the $J=0$ potentials, with $V_{\mathrm{min}}^{J=0}=0$. \label{F7}}
\end{center}
\end{figure*}

As discussed above, the AMP method can be employed to analyze the quadrupole geometry of yrast states in transitional nuclei. This approach enables the interpretation of anomalous features in yrast spectra by examining the spin-dependent evolution of the potential energy surfaces. Conversely, a pronounced variation in quadrupole deformations with increasing spin must inevitably manifest itself in the associated excitation spectrum. In this section, we present an analysis of two distinct phenomena linked to modifications in the spectral properties of yrast states.

\begin{center}
\vskip.2cm\textbf{A. Jacobi-types transitions in $^{160}$Gd and $^{162}$Dy}
\end{center}\vskip.2cm

In the preceding section, it has been demonstrated that a rotational stretching of the $\beta$ or $\gamma$ deformation parameters can modify MOIs, thereby accounting for
Jacobi-type transitions in nuclear systems. Such transitions have been experimentally observed across a broad range of deformed rare-earth nuclei~\cite{Zhang2021}. In Fig.~\ref{F7}, we present $^{160}$Gd and $^{162}$Dy as representative cases, showing both the calculated E-GOS curves for their yrast bands and the corresponding AMP potential energy surfaces at selected spin values. The calculations employ the consistent-$Q$ Hamiltonian defined in (\ref{CQ}), with dimensionless parameters for $^{160}$Gd ($^{162}$Dy) set to $\eta=0.84~(0.92)$ and $\chi=-0.53~(-0.31)$, respectively; the scale parameter is fixed at $\varepsilon$=1.612 (1.889) MeV, as determined by
reproducing the experimental $2_1^+$ excitation energy $E(2_1^+)=0.075$ (0.081) MeV. These parameter sets yield reasonable description of the low-lying spectroscopic properties of both $^{160}$Gd~\cite{Reich2005} and $^{162}$Dy~\cite{Reich2007}, as established in Ref.~\cite{McCutchan2004}. For example, the calculated values of two typical energy ratios, $R_{4/2}=E(4_1^+)/E(2_1^+)$ and $R_{2/2}=E(2_2^+)/E(2_1^+)$, for $^{160}$Gd ($^{162}$Dy) are $R_{4/2}^{\mathrm{Th.}}=3.29$ $(3.25)$ and $R_{2/2}^{\mathrm{Th.}}=13.52$ $(11.05)$, respectively, which agree well with the corresponding experimental values: $R_{4/2}^{\mathrm{Exp.}}=3.30$ $(3.29)$ and $R_{2/2}^{\mathrm{Exp.}}=13.13$ $(11.01)$. The large experimental $R_{4/2}$ values indicate that both rare-earth nuclei possess well-deformed ground-state configurations, and their IBM parameter points lie firmly within the deformed region of the triangle phase diagram, thereby supporting the applicability of the aforementioned Jacobi-type transition analysis to them.

As shown in Fig.~\ref{F7}(a1), the theoretical E-GOS curve successfully reproduces the characteristic Jacobi-type transitional behavior observed in $^{160}$Gd: $R(J)$ attains its maximum at $J=6$ and subsequently decreases up to $J=16$. This decrease yields a theoretical deformation-changing indicator $\Delta R=\frac{R(16)-R(6)}{R(6)-R(2)}=-0.49$, which, although smaller in magnitude, is consistent in sign and order of magnitude with the experimental value, $\Delta R=-0.73$. As evident from panels (a2)-(a4) in Fig.~\ref{F7}, the AMP potential energy surfaces exhibit a systematic increase in triaxiality: the equilibrium $\gamma$-deformation rises from $\gamma_\mathrm{e}\approx 10^\circ$ at $J\leq6$ to $\gamma_\mathrm{e}\approx 12^\circ$ at $J=16$, while the $\beta$ equilibrium deformation remains nearly constant at $\beta_\mathrm{e}\approx0.95$. At these equilibrium deformations, the $K=0$ AMP calculations reproduce the IBM exact solutions with high fidelity. When this $\gamma$-stretching trend is incorporated into the rotor model Hamiltonian (\ref{rot}), it yields $\Delta R=-0.34$. This value is in reasonable agreements with the aforementioned theoretical estimation of $\Delta R$, which is extracted directly from the IBM calculations presented in Fig.~\ref{F7}(a1). Consequently, the Jacobi-type transition observed in $^{160}$Gd, evidenced by the non-monotonic evolution of the E-GOS curve, may, to some extent, be attributed to the rotational stretching effect on the $\gamma$ deformation.

Similarly, as shown in Fig.~\ref{F7}(b1), the experimental $R(J)$ values exhibit a non-monotonic variation across $J=6$, signaling a Jacobi-type transition along the yrast line in $^{162}$Dy. The transition is well reproduced by the IBM calculations employing the parameters mentioned above. As illustrated in Fig.~\ref{F7}(b2)-(b4), the same parameters yield AMP potential energy surfaces whose equilibrium $\gamma$-deformation increases with spins, analogous to the behavior observed in $^{160}$Gd. Using the extracted equilibrium deformation parameters as input, rotor model calculations yield $\Delta R=\frac{R(16)-R(6)}{R(6)-R(2)}=-0.69$, which agrees well with the IBM prediction ($\Delta R=-0.85$) and the experimental value ($\Delta R=-0.99$) extracted from Fig.~\ref{F7}(b1). Therefore, the Jacobi-type transitional behavior observed in $^{162}$Dy can be understood, at least in part, as a consequence of rotational stretching of the $\gamma$ deformation. It should be mentioned that the parameters employed here were obtained from a global fit to the low-lying spectroscopic data for two rare-earth nuclei~\cite{McCutchan2004}. While a refined fit to yrast-band energies would undoubtedly improve the quantitative descriptions of $R(J)$ and $\Delta R$, such an optimization lies beyond the scope of the present study.

In addition, the present theoretical analysis indicates that $\gamma$-stretching effects can lead to a non-monotonic change in spectroscopic quadrupole moments $Q(J)$ on spin, suggesting maximal absolute values $|Q(J)|$ at $J=6$ for both $^{160}$Gd and $^{162}$Dy. This prediction imposes a more stringent constraint on current theoretical interpretations of Jacobi-type transitions in the two rare-earth nuclei, pending availability of $Q(J)$ measurements. Undoubtedly, the above geometric description of the yrast-state evolution rests on the assumption that the yrast structures in these deformed rare-earth nuclei are fully captured by quadrupole-rotor modes within the IBM framework. To solidify the conclusion drawn from the present AMP analysis, further theoretical investigations employing complementary models remain necessary, particularly given that different models may yield quantitatively distinct predictions for quadrupole deformations even for the same nucleus.

\begin{figure}
\begin{center}
\includegraphics[scale=0.20]{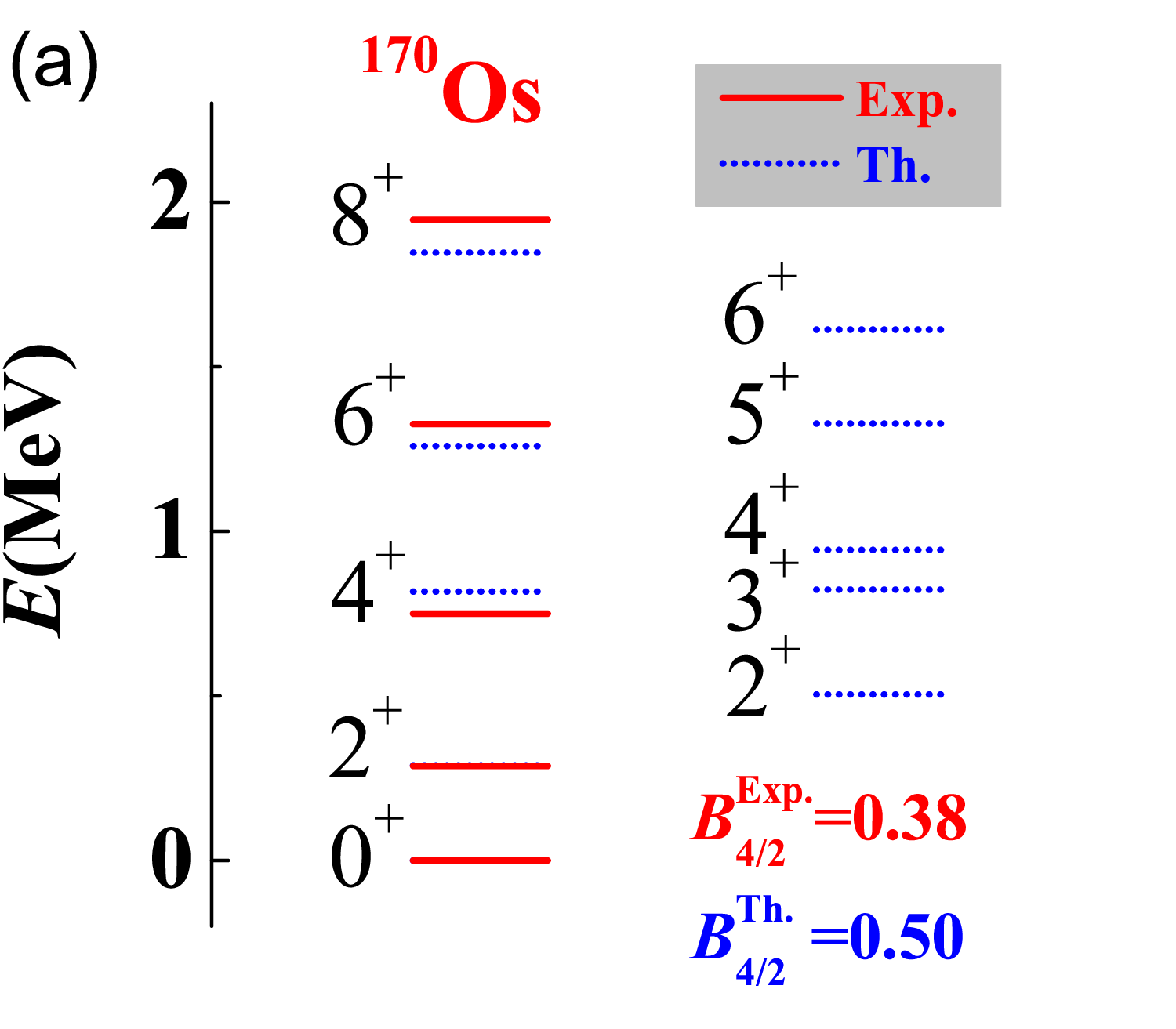}
\includegraphics[scale=0.16]{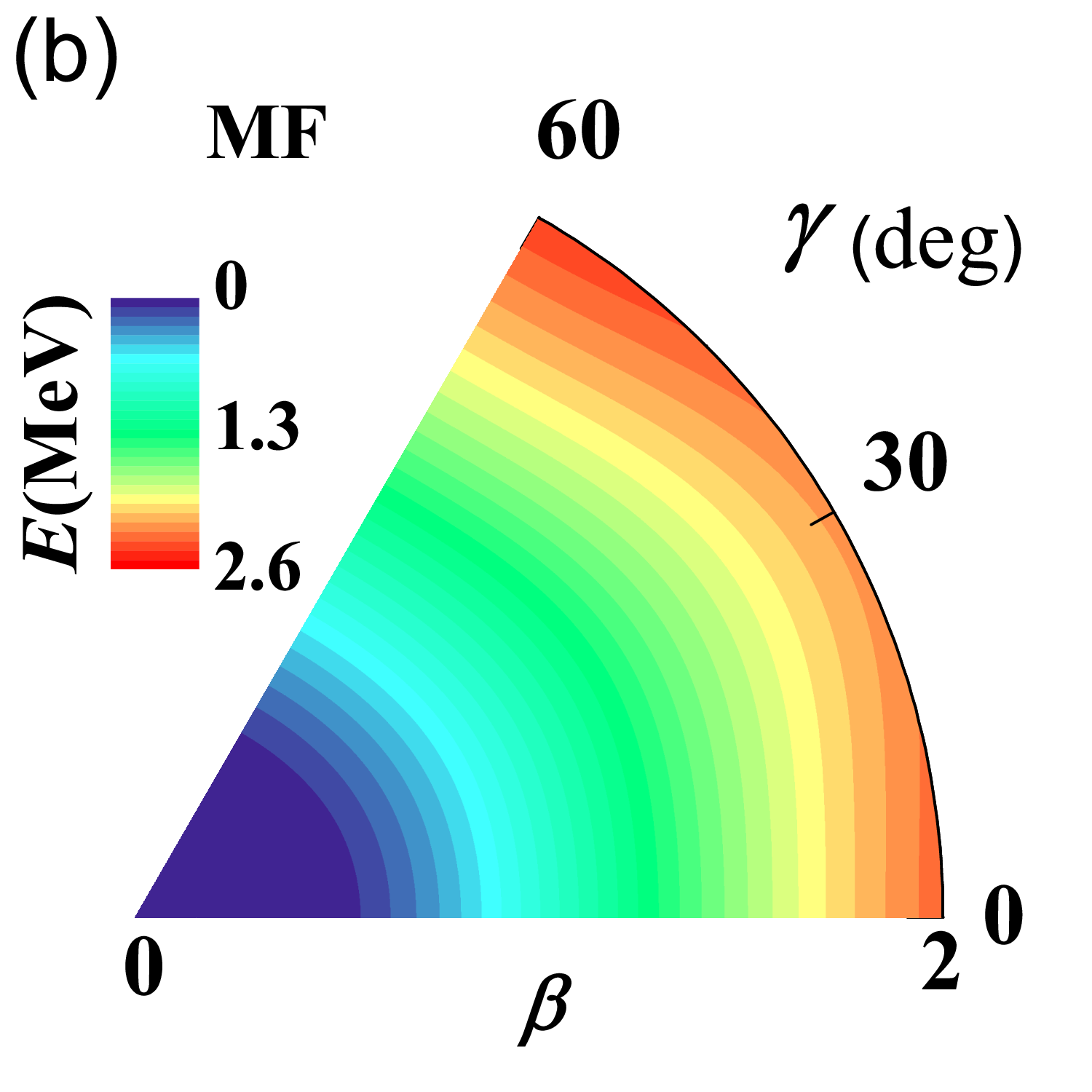}
\includegraphics[scale=0.16]{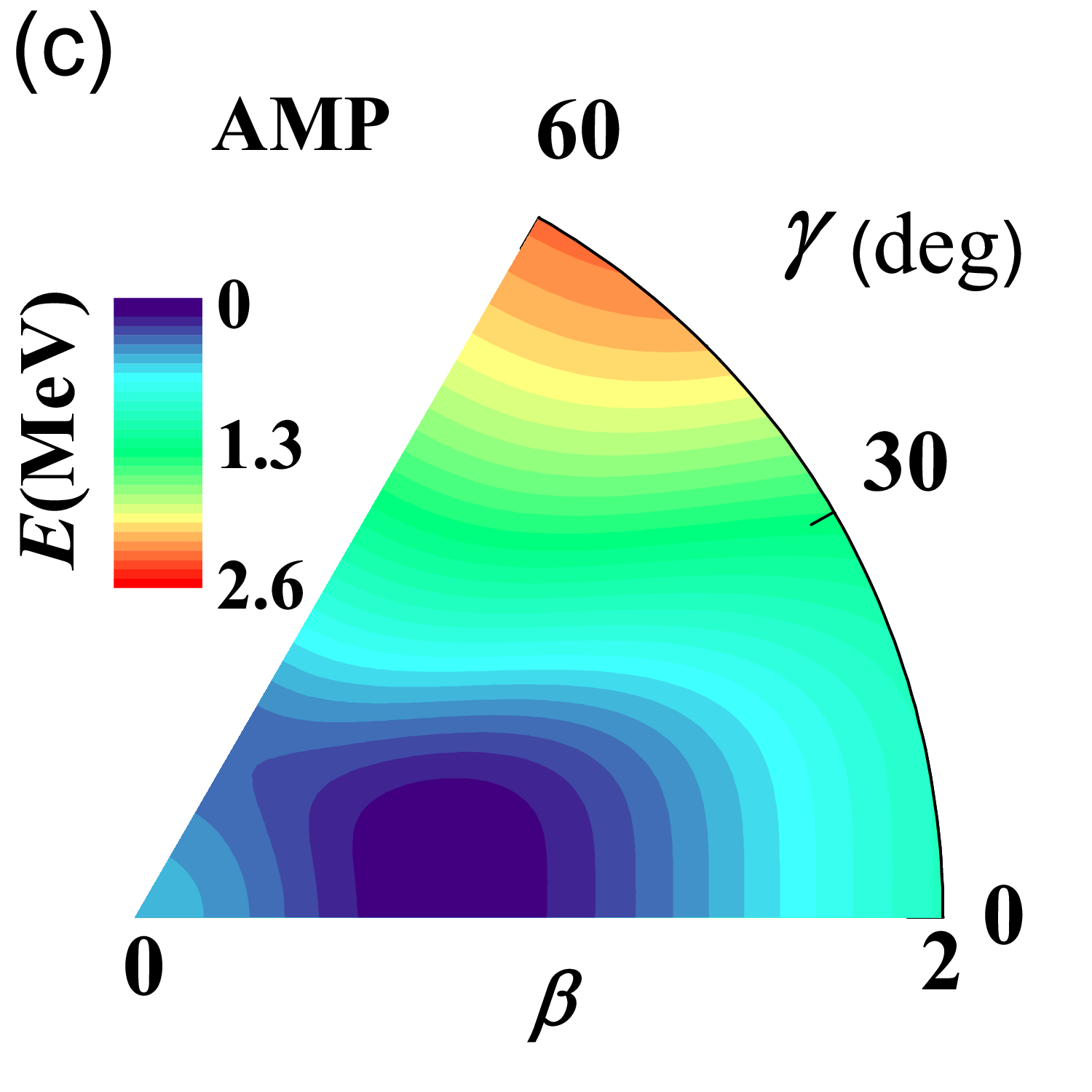}
\caption{(Color online) (a) Low-lying level scheme of $^{170}$Os calculated with model parameters from \cite{Teng2025} is compared to experimental data~\cite{Goasduff2019}. (b) Mean-field potential energy surface. (c) Projected potential energy surface at $J=0$. Each potential energy surface is referenced to its own potential minimum, with $V_{\mathrm{min}}=0$. \label{F8}}
\end{center}
\end{figure}

\begin{figure}
\begin{center}
\includegraphics[scale=0.18]{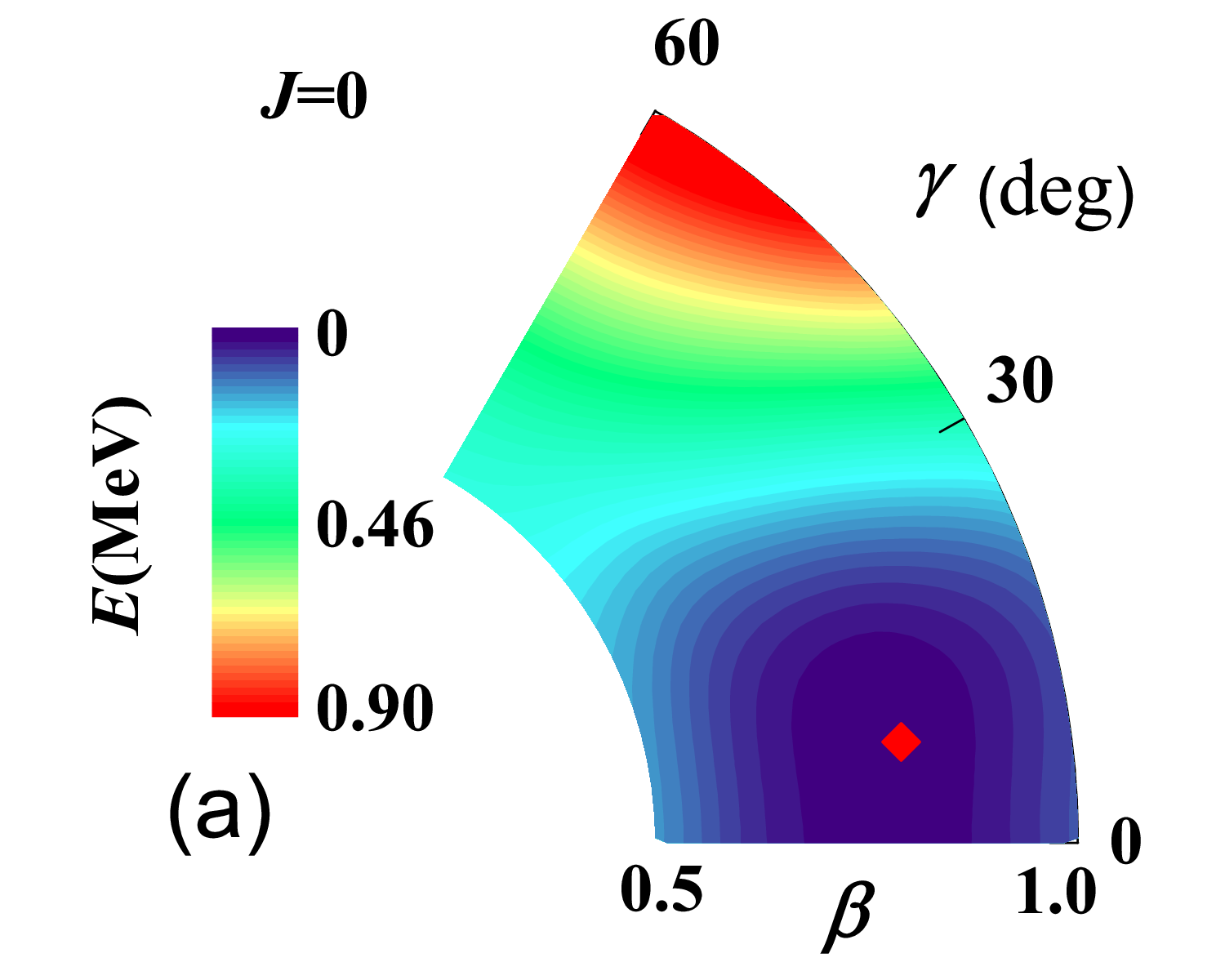}
\includegraphics[scale=0.18]{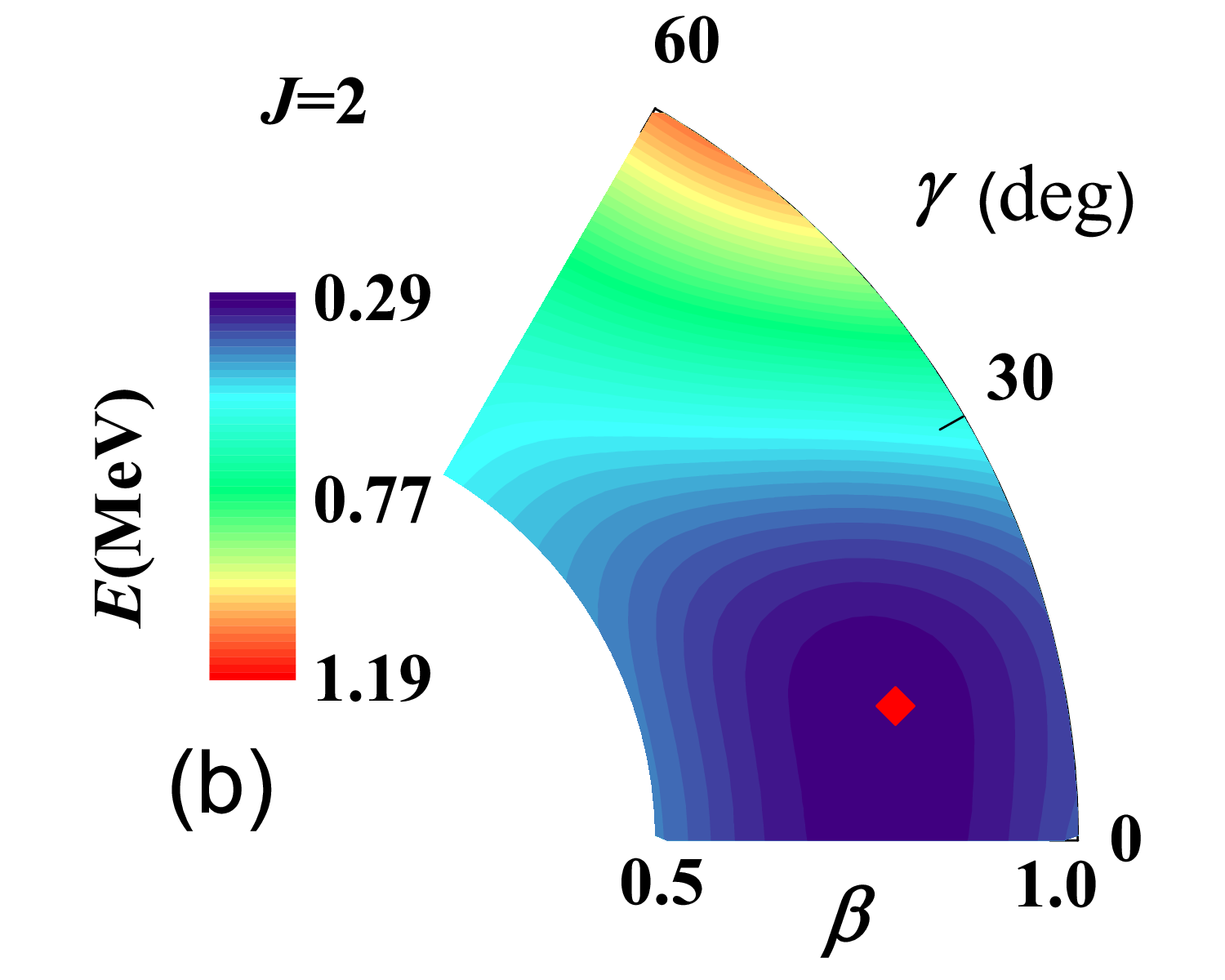}
\includegraphics[scale=0.18]{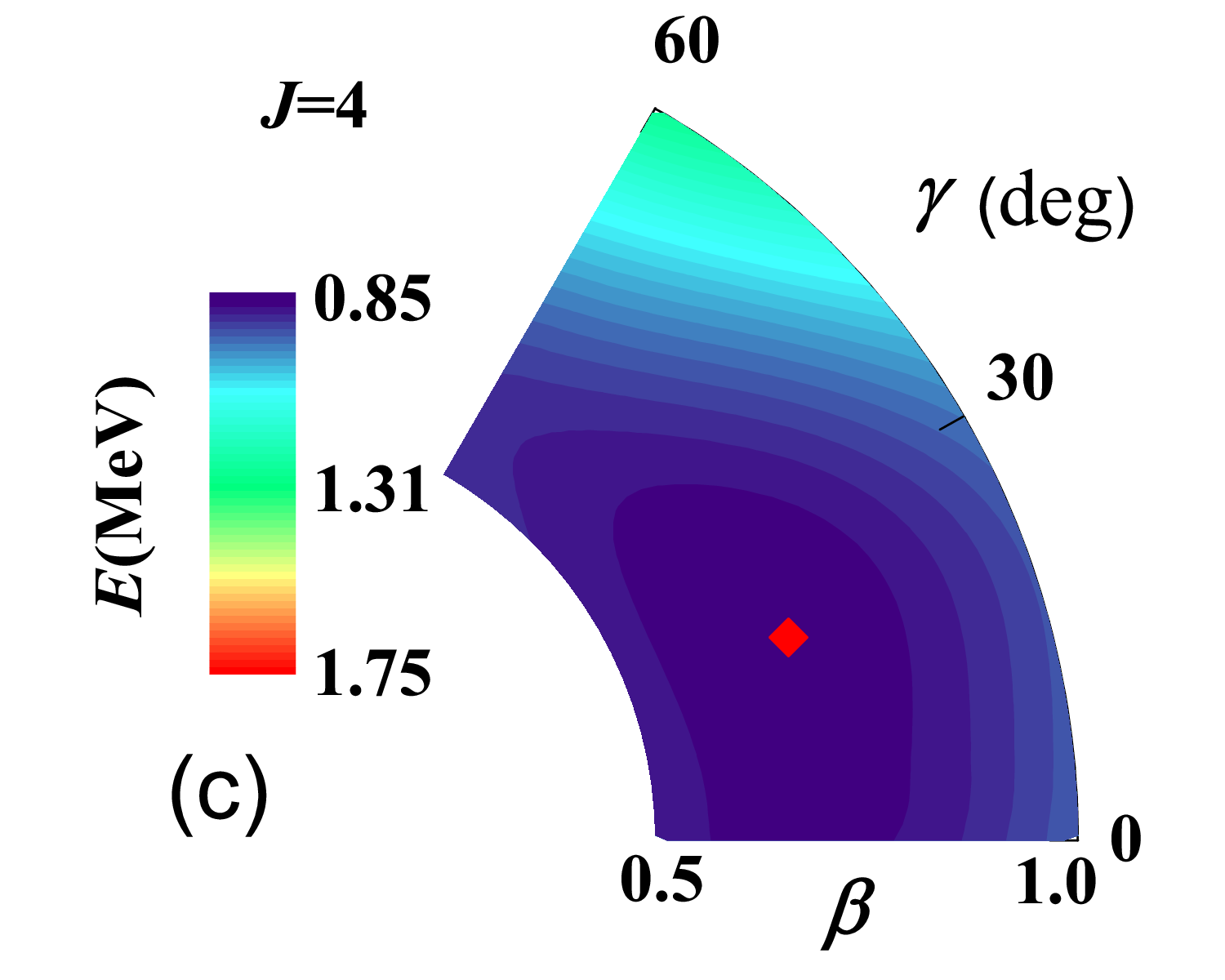}
\caption{(Color online)Using the parameters employed for $^{170}$Os, the projected potentials surfaces for $J=0$, 2 and 4 are displayed, with the red squares marking the equilibrium deformations $\beta_\mathrm{e}$ and $\gamma_\mathrm{e}$. All potential energy surfaces are referenced to the minimum of the $J = 0$ potentials, with $V_{\mathrm{min}}^{J=0}=0$. \label{F9}}
\end{center}
\end{figure}

\begin{center}
\vskip.2cm\textbf{B. $B(E2)$ anomaly in $^{170}$Os}
\end{center}\vskip.2cm

Recently, the $B(E2)$ anomaly, a phenomenon characterized by $B_{4/2}=B(E2;4_1^+\rightarrow2_1^+)/B(E2;2_1^+\rightarrow0_1^+)<1.0$ alongside $R_{4/2}=E(4_1^+)/E(2_1^+)>2.0$, has attracted considerable attention~\cite{Grahn2016,Saygi2017,Cederwall2018,Goasduff2019,Zhang2021,Zhang2022,Zhang2024,Pan2024,Wang2020,Zhang2025,Teng2025,Teng2025II,Teng2025III,Fu2025}, as it challenges conventional interpretations of collective nuclear structures: all standard collective models predict $B_{4/2}>1.0$. Until very recently, this anomaly has been successfully described within the framework of the IBM~\cite{Zhang2022,Zhang2024,Pan2024,Wang2020,Zhang2025,Teng2025,Teng2025II,Teng2025III} via a band-mixing mechanism arising from static or dynamic triaxiality~\cite{Zhang2025}. In particular, it was proposed~\cite{Teng2025II} that the effective $\gamma$ deformation in a $B(E2)$-anomaly system undergoes substantial change at low spins. Within the IBM, the effective $\gamma$ deformation can be quantified using the formula~\cite{Elliott1986}
\begin{eqnarray}\label{effgamma}
\mathrm{Cos}(3\gamma_\mathrm{a})=-\Big(\frac{7}{2\sqrt{5}}\Big)^{1/2}\frac{\langle(\hat{Q}^\chi\times\hat{Q}^\chi\times\hat{Q}^\chi)^{(0)}\rangle}{(\langle\hat{Q}^\chi\times\hat{Q}^\chi)^{(0)}\rangle^{3/2}}\, ,
\end{eqnarray}
where $\langle~\hat{A}~\rangle$ denotes the expectation value of operator $\hat{A}$ in the given state.
Based on the above discussions, the AMP method introduced here provides a natural framework to probe the spin dependence of $\gamma$ deformation in such low-spin anomalous phenomena. In what follows, we take $^{170}$Os as an example to analyze the quadrupole deformation in a $B(E2)$ anomaly nucleus. As demonstrated in Ref.~\cite{Teng2025}, the Hamiltonian given in Eq.~(\ref{H}) successfully reproduces the available experimental data, including the hallmark $B(E2)$ anomalies, across the neutron-deficient Os isotopes. For $^{170}$Os, the parameter set (in MeV) is $a_1=0.25,~a_2=-0.115/N,~a_3=0.368/N,~b_1=-0.902/N,~b_2=0.127/N$, with the boson number $N=9$ and the dimensionless parameter $\chi=-0.98$. We retain these parameters~\cite{Teng2025} here not to refine the fit to the limited experimental data, but rather to model a $B(E2)$ anomaly system for an AMP analysis.

With the given parameters, the calculated level scheme and the corresponding mean-field (unprojected) and projected ($J=0$) potential energy surfaces are presented in Fig.~\ref{F8}.
As shown in Fig.~\ref{F8}(a), the yrast level energies, along with the strongly suppressed $B_{4/2}$ ratio, in $^{170}$Os are well reproduced by the theoretical calculations, which further predict the existence of a low-energy side band. Fig.~\ref{F8}(b) shows that the mean-field potential surface exhibits a spherical-like deformation, whereas the $J=0$ projected potential surface in Fig.~\ref{F8}(c) displays a pronounced $\beta$-deformed geometry for the ground state. Such a dramatic change caused by AMP has not been observed in the cases illustrated in Fig.~\ref{F2}. This underscores the critical role of AMP in revealing the true ground-state geometry of the $B(E2)$ anomaly system modeled in the IBM. The effects originating from the rotor-like term $LQL$~\cite{Zhang2025}, which explicitly couples rotational motion to quadrupole deformation, are absent in the consistent-$Q$ Hamiltonian. Notably, the minimum of the projected potential in Fig.~\ref{F8}(c) aligns closely the exact solution solved from Hamiltonian diagonalization. The resulting equilibrium triaxiality, $\gamma_\mathrm{e}=9^\circ$, is in good agreement with the value $\gamma=10^\circ$ predicted for $^{170}$Os within the covariant density function theory with the point-coupling interaction PC-PK1~\cite{Zhao2010}.

To understand the deformation evolution in this $B(E2)$ anomaly system at low spins, Fig.~\ref{F9} presents the projected potentials of $J=0,~2$ and $4$, obtained from the AMP$^\mathrm{I}$ calculations. It is shown that the quadrupole deformation of the $2_1^+$ state closely resembles that of the $0_1^+$ state, indicating that the two states share a similar collective configuration and hence exhibit a relatively strong $E2$ transition between them. This is quantitatively confirmed by the large experimental value, $B(E2;2_1^+\rightarrow0_1^+)\approx97$W.u.~\cite{Goasduff2019}. In contrast, the $J=4$ potential surface exhibits pronounced $\gamma$ softness and enlarged equilibrium triaxiality $\gamma_\mathrm{e}$, reflecting a significant reduction in the overlap between the collective wavefunctions of the $2_1^+$ and $4_1^+$ states, potentially resulting in $B_{4/2}<1.0$ as observed in Fig.~\ref{F8}(a). In addition, the substantial $\gamma$ softness renders a rigid-rotor assumption inadequate in this case, implying that the observed $B(E2)$ anomaly cannot be ascribed to stretching effects and is thus fundamentally distinct from Jacobi-type transitions analyzed above.

It should be noted that, unlike the cases for $J=0$ and $J=2$, the $K=0$ projection for $J=4$ within the present scheme yields a global minimum at $\gamma_\mathrm{e}\approx145^\circ$, which lies outside the range $\gamma\in[0^\circ,60^\circ]$. To ensure a consistent description of the $\gamma$ deformation across different $J$, we require that the global minimum of each projected potential $V(\beta,\gamma)_K^J$ must accurately reproduce the exact diagonalization results within $\gamma\in[0^\circ,60^\circ]$. Accordingly, the $J=4$ potential surface displayed in Fig.~\ref{F9}(c) corresponds to the $K=4$ projection, which satisfies this criterion and reproduces well the exact solution over this $\gamma$ interval. This behavior can be partially understood from the results shown in Fig.~\ref{F4}, where it is evident that varying the $K$ value in the AMP calculations leads to the global minima located at distinct $\gamma$ values. Since all potential surfaces presented in Fig.~\ref{F9} reproduce the exact solutions accurately, as clearly demonstrated by the displayed energy scales, the quadrupole deformations extracted from these projected potentials are theoretically reliable. Specifically, the resulting equilibrium deformations are $\gamma_\mathrm{e}\approx9^\circ$, $12^\circ$, and $20^\circ$ for $J=0$, $J=2$, and $J=4$, respectively. These values agree closely with the effective $\gamma$ deformation $\gamma_\mathrm{a}\approx10^\circ,~10^\circ$, and $20^\circ$ obtained directly from the wavefunctions via Eq.~(\ref{effgamma}). Furthermore, the enhanced $\gamma$ softness observed for $J=4$ is also consistent with prior IBM analysis~\cite{Teng2025II}, which predicted relatively large fluctuation in $\gamma$ for this nucleus at low spins.

\begin{center}
\vskip.2cm\textbf{V. Summary}
\end{center}\vskip.2cm

In summary, we have performed a systematic analysis of the AMP method within the framework of the IBM. The results demonstrate that AMP with $K$-fixed projection yields predictions in agreement with those from the more computationally demanding $K$-mixied AMP, thereby supporting recent shell-model based assessments of AMP accuracy~\cite{Gao2022,Lu2025}.
By analyzing the spin dependence of the quadrupole geometries in deformed systems, we demonstrate that incorporating deformation-stretching effects provides a self-consistent explanation for the Jacobi-type transitional behavior, manifested both in the E-GOS curve and in the evolution of spectroscopic quadrupole moments, offering an intuitively geometric perspective on this phenomenon. A concrete illustration of such stretching effects is provided by the AMP analysis of $^{160}$Gd~\cite{Reich2005} and $^{162}$Dy~\cite{Reich2007}. Moreover, AMP calculations reveal that pronounced low-spin deformation changes may occur in an IBM system with $B(E2)$ anomaly, as exemplified by $^{170}$Os~\cite{Goasduff2019}. This finding opens a new angle for investigating this exotic phenomenon, which remains incompletely understood from a microscopic perspective~\cite{Teng2025III,Fu2025}. Collectively, these results demonstrate that the AMP method serves as a sensitive probe of the geometric structure underlying exotic nuclear behaviors, motivating its extension to explore other nuclear deformation effects including shape and phase coexistence~\cite{Iachello1998}. Related work is in progress.

\bigskip

\begin{acknowledgments}
\begin{acknowledgments}
We gratefully acknowledge Z. C. Gao for inspiring this work and for providing valuable guidance on the angular momentum projection method.
This work is supported by the National Natural Science Foundation of China (Grant No. 12375113).
\end{acknowledgments}
\end{acknowledgments}



\end{document}